\documentclass[apj, twocolappendix, numberedappendix, appendixfloats]{emulateapj}
\pdfoutput=1 
\usepackage{amsmath,amstext}
\usepackage[T1]{fontenc}
\usepackage{apjfonts}
\usepackage[breaklinks,colorlinks,citecolor=blue]{hyperref} 
\usepackage[figure,figure*]{hypcap}

\defcitealias{Caplar:2015aa}{C15}
\defcitealias{Weigel:2016aa}{W16}
\defcitealias{Ajello:2012aa}{A12}
\defcitealias{Mauch:2007aa}{MS07}
\newcommand{\Vmax}{$V_\mathrm{max}$}
\newcommand{\Ndraw}{$N_\mathrm{draw}$}
\newcommand{\loglambda}{$\log \lambda$}
\newcommand{\logMBH}{$\log M_\mathrm{BH}$}
\newcommand{\logLbol}{$\log L_\mathrm{bol}$}

\newcommand{\Mmin}{$\log (M_\mathrm{min}/M_\odot) = 9$}
\newcommand{\Mmax}{$\log (M_\mathrm{max}/M_\odot) = 12$}

\newcommand{\kbolX}{-1}
\newcommand{\kbolR}{-3}
\newcommand{\ERDFmin}{-8}
\newcommand{\ERDFmax}{1}
\newcommand{\fitLminX}{41.5}
\newcommand{\fitLmaxX}{45.6}
\newcommand{\fitLminR}{20.4}
\newcommand{\fitLmaxR}{26.4}
\newcommand{\logMstarbluegreen}{10.67}

\newcommand{\logMstarred}{10.77}
\newcommand{\alphabred}{-0.46}
\newcommand{\breakAjello}{43.71}

\newcommand{\breakMaucherg}{40.74}
\newcommand{\mmmu}{-2.75}
\newcommand{\mmsigma}{0.3}
\newcommand{\mmrho}{38.2}

\newcommand{\deltaXa}{$0.47^{+ 0.20}_{- 0.42}$}

\newcommand{\massdeperdfmbinsize}{0.1}
\newcommand{\lambdastarSchechterX}{$-1.47^{+0.12}_{-0.11}$}
\newcommand{\alphaSchechterX}{$-1.64^{+0.09}_{-0.08}$}
\newcommand{\lambdastarlognormalR}{$-7.95^{+0.42}_{-0.52}$}
\newcommand{\sigmalognormalR}{$1.72^{+0.13}_{-0.11}$}

\shorttitle{Two mass independent ERDFs for BHs at z$\sim$0}
\shortauthors{Weigel et al.}

\begin{document}

\title{AGN and their host galaxies in the local Universe: two mass independent Eddington ratio distribution functions characterize black hole growth}

\author{
	Anna K. Weigel \altaffilmark{1},
	Kevin Schawinski \altaffilmark{1}, 
	Neven Caplar \altaffilmark{1},
	O. Ivy Wong \altaffilmark{2, 3},\\
	Ezequiel Treister \altaffilmark{4, 5, 6} and 
	Benny Trakhtenbrot \altaffilmark{1}
	}

\altaffiltext{1}{Institute for Astronomy, Department of Physics, ETH Zurich, Wolfgang-Pauli-Strasse 27, CH-8093 Zurich, Switzerland}
\altaffiltext{2}{International Centre for Radio Astronomy Research, The University of Western Australia M468, 35 Stirling Highway, Crawley, WA 6009, Australia}
\altaffiltext{3}{ARC Centre for Excellence for All-Sky Astrophysics (CAASTRO), Australia}
\altaffiltext{4}{Instituto de Astrof\'isica, Facultad de F\'isica, Pontificia Universidad Cat\'olica de Chile, 306, Santiago 22, Chile}
\altaffiltext{5}{EMBIGGEN Anillo, Concepcio\'n, Chile}
\altaffiltext{6}{Departamento de Astronom\'ia Universidad de Concepcio\'n, Casilla 160-C, Concepc\'ion, Chile}

\begin{abstract}
We use a phenomenological model to show that black hole growth in the local Universe ($z \lesssim 0.1$) can be described by two separate, mass independent Eddington ratio distribution functions (ERDFs). We assume that black holes can be divided into two independent groups: those with radiatively efficient accretion, primarily hosted by optically blue and green galaxies, and those with radiatively inefficient accretion, which are mainly found in red galaxies. With observed galaxy stellar mass functions as input, we show that the observed AGN luminosity functions can be reproduced by using mass independent, broken power law shaped ERDFs.  We use the observed hard X-ray and 1.4 GHz radio luminosity functions to constrain the ERDF for radiatively efficient and inefficient AGN, respectively. We also test alternative ERDF shapes and mass dependent models. Our results are consistent with a mass independent AGN fraction and AGN hosts being randomly drawn from the galaxy population. We argue that the ERDF is not shaped by galaxy-scale effects, but by how efficiently material can be transported from the inner few parsecs to the accretion disc. Our results are incompatible with the simplest form of mass quenching where massive galaxies host higher accretion rate AGN. Furthermore, if reaching a certain Eddington ratio is a sufficient condition for maintenance mode, it can occur in all red galaxies, not just the most massive ones.\\
\end{abstract}


\section{Introduction}
An active area of research in astrophysics today is the existence and the extent of the galaxy-black hole connection. Assuming there is a causal link, are galaxies dictating the growth of black holes? Or is the energy of active galactic nuclei (AGN) effective enough to impact galaxy evolution? 

Local scaling relations between black holes and their host galaxies (e.g. \citealt{Kormendy:1995aa, Magorrian:1998aa, Ferrarese:2000aa, Gebhardt:2000aa, Tremaine:2002aa, Marconi:2003aa, Gultekin:2009aa, Kormendy:2013aa, Savorgnan:2016ac, Graham:2016aa}) are often attributed to galaxy-black hole co-evolution (see e.g. \citealt{Silk:1998aa, Fabian:1999aa, King:2003aa, Di-Matteo:2005aa}), but they could also have a non-causal origin, for instance galaxy-galaxy mergers \citep{Peng:2007aa, Jahnke:2011aa}. Besides the local scaling relations, the similar redshift evolution of the star formation rate density and the black hole accretion rate density points towards a possible galaxy-black hole connection (e.g. \citealt{Boyle:1998aa, Heckman:2004aa, Hasinger:2005aa, Silverman:2009aa, Mullaney:2012ab}). Furthermore, the effect of AGN on their host galaxies in the form of feedback is associated with the quenching of star formation \citep{Sanders:1988aa, Di-Matteo:2005aa, Cattaneo:2009ab,  Fabian:2012aa} and could explain the observed bi-modality in color-magnitude and color-mass space (\citealt{Bell:2003aa, Baldry:2004aa, Faber:2007aa, Martin:2007aa, Schawinski:2014aa}). The galaxy-black hole connection hence seems complex and multifaceted, especially if we consider additional aspects such as galaxy morphology and redshift evolution.

Our aim is to understand the effect of black holes on their host galaxies and vice versa on a global scale. To do so we introduce a model that allows us to put observations into context and to estimate and interpret their impact.

For galaxies large surveys, such as SDSS \citep{York:2000aa, Abazajian:2009aa} or zCOSMOS \citep{Lilly:2007aa}, allow us to examine the population in great detail. We are able to constrain the luminosity and stellar mass functions (at $z \sim 0$ e.g.: \citealt{Blanton:2003aa, Baldry:2012aa, Perez-Gonzalez:2008aa, Moustakas:2013aa}; \citealt{Weigel:2016aa} hereafter \citetalias{Weigel:2016aa}; \citealt{Moffett:2016aa}) and to study the evolution of the specific star formation rate (sSFR) distribution, which links the two \citep{Sargent:2012aa, Bernhard:2014aa, Ilbert:2014aa}. 

For black holes, the equivalent to the sSFR distribution is the Eddington ratio distribution function (ERDF)\footnote{Note that this is only approximately true. The sSFR direcly corresponds to the stellar mass e-folding time. However, deriving the black hole mass e-folding time from the Eddington ratio requires the assumption of the radiative efficiency.} The ERDF is the distribution of normalized black hole accretion rates ($\lambda = L_{\rm bol}/L_{\rm Edd}$)\footnote{Throughout the paper we will be referring to $\lambda$ as the Eddington ratio.}. While $\lambda$ is an indicator of the small scale accretion process, the ERDF describes the distribution of one of the fundamental black hole properties and represents a powerful population probe. In analogy to galaxies, the ERDF links the black hole mass function to the AGN luminosity function. While observationally constraining the AGN luminosity function is possible for a range of wavelengths, redshifts and luminosities (e.g. \citealt{Richards:2006aa, Ueda:2003aa, Hasinger:2005aa, Bongiorno:2007aa, Hopkins:2007aa, Aird:2015aa, Miyaji:2015aa}), measuring reliable  black hole masses and their distribution for a large sample is more complex \citep{Trakhtenbrot:2012aa, Shen:2013aa, Peterson:2014aa, Mejia-Restrepo:2016aa}.

In previous work, the black hole mass function and the corresponding Eddington ratio values have been measured observationally (\citealt{Heckman:2004aa, McLure:2004aa, Kollmeier:2006aa, Kauffmann:2009aa, Schulze:2010aa, Trump:2011aa, Shen:2012aa, Kelly:2012aa, Nobuta:2012aa, Bongiorno:2012aa, Lusso:2012aa, Aird:2012aa, Schulze:2015aa, Jones:2016aa, Bongiorno:2016aa, Aird:2017aa}) and have been studied using a phenomenological approach (\citealt{Merloni:2004aa, Shankar:2004aa, Yu:2004aa, Merloni:2008aa, Hopkins:2009aa,  Shen:2009aa, Shankar:2009aa, Cao:2010aa, Li:2011aa, Conroy:2013aa, Shankar:2013ab, Novak:2013aa, Veale:2014aa, Aversa:2015aa, Caplar:2015aa, Tucci:2016aa}).

We present the results of a phenomenological model that links the galaxy and black hole populations in the local Universe ($z \lesssim 0.1$). The purpose of this analysis is not to infer from the data what the ERDFs properties are, as some studies have done (e.g. \citealt{Kollmeier:2006aa, Shen:2008aa, Kauffmann:2009aa, Schulze:2010aa, Aird:2012aa, Jones:2016aa}), but rather to forward model the process: if we assume a certain, simple ERDF shape, how far can we go in characterizing the AGN population. Our aim is to make the simplest and most straightforward assumptions possible. These assumptions might not be exactly true, but they do allow us to make inferences about black hole growth on a global scale. For example, we assume that AGN can be separated into two independent populations: AGN that accrete radiatively efficiently and are primarily found in optically blue and green galaxies and AGN with radiatively inefficient accretion that are mostly found in optically red, quiescent galaxies \citep{Fabian:2012aa, Kormendy:2013ab, Heckman:2014aa}. Starting from the galaxy stellar mass function, we predict the black hole mass function and AGN luminosity function by assuming an ERDF. This ERDF is assumed to have a broken power law shape and to be mass independent. We test if the observed AGN luminosity functions, in the X-rays for radiatively efficient AGN and in the radio for radiatively inefficient AGN, can be reproduced with this simple model and constrain the corresponding ERDFs. We discuss the physical implications of our results in the context of  mass quenching \citep{Sanders:1988aa, Di-Matteo:2005aa, Cattaneo:2009ab, Peng:2010aa, Fabian:2012aa, Peng:2012aa}, maintenance mode \citep{Croton:2006aa, Bower:2006aa,  Springel:2006aa,  Somerville:2008aa, Fabian:2012aa} and black hole growth in general.

The paper is structured as follows. In Section \ref{sec:model} we introduce our model. After having discussed our assumptions, we establish our method in Section \ref{sec:method}. Section \ref{sec:app_results} summarizes the application of the method to the observations and our results. We examine the implications of these results in Section \ref{sec:implications}. In Section \ref{sec:discussion} we discuss variations and caveats of our model and compare our results to previous work. We conclude this paper with a summary in Section \ref{sec:summary}.

Throughout this paper, we denote the logarithm to base 10 as $\log$ and assume a $\Lambda$CDM cosmology with $h_0 = 0.7$, $\Omega_m = 0.3$ and $\Omega_\Lambda = 0.7$ \citep{Komatsu:2011aa}. \\
\section{Model and assumptions}\label{sec:model}
\begin{figure}
	\includegraphics[width=\columnwidth]{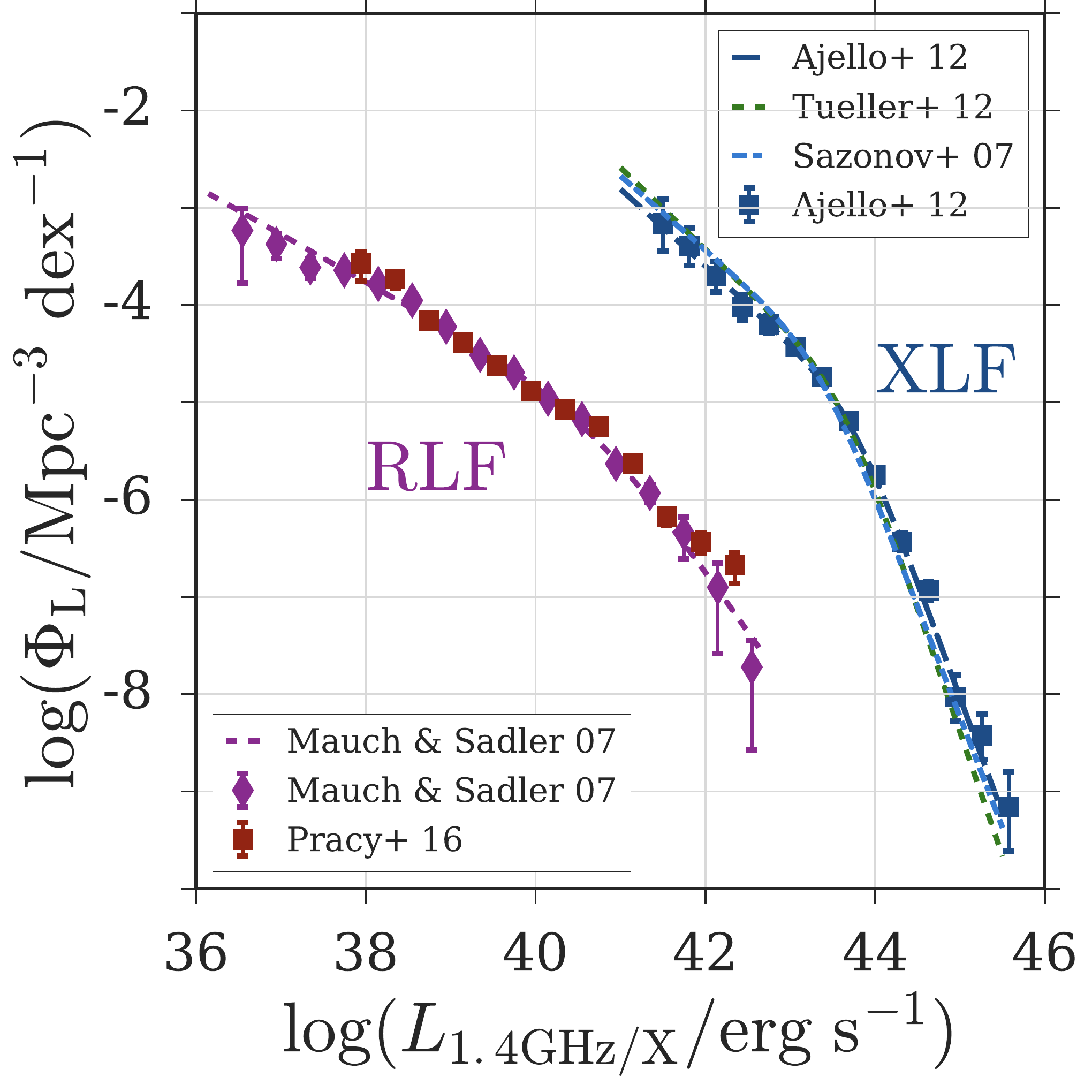}
	\caption{\label{fig:rlf_xlf_comp}AGN radio and X-ray luminosity functions. Shown on the right-hand side of this figure are the hard X-ray luminosity functions (XLF) by \cite{Ajello:2012aa} (15 - 55 keV),  \cite{Tueller:2008aa} (14 - 195 keV) and \cite{Sazonov:2007aa} (17 - 60 keV). The 1.4 GHz radio luminosity functions (RLF) by \citetalias{Mauch:2007aa} and \cite{Pracy:2016aa} are illustrated on the left-hand side. To allow for an easier comparison between the RLF and the XLF we converted the radio luminosities from $\rm W\ Hz^{-1}$ to $\rm erg\ s^{-1}$. Compared to the RLF, the XLF is significantly steeper and has a pronounced break.}
\end{figure}

The ERDF, $\xi(\lambda)$, describes the distribution of Eddington ratios of a black hole population. It shows what fraction of black holes have Eddington ratios within a certain range and thus links the black hole mass function to the AGN luminosity function. As we have discussed above, the black hole population is linked to the galaxy population through local scaling relations. By correlating the stellar mass and black hole mass functions, the shape of the AGN luminosity function can hence be traced back to the ERDF and the stellar mass function (\citealt{Caplar:2015aa}, hereafter \citetalias{Caplar:2015aa}). 

Our aim is to test if the observed AGN luminosity function can be reproduced by a simple model. Instead of starting with the AGN luminosity function and inferring the ERDF, we use a forward modelling approach and base our model on the underlying galaxy population and its stellar mass function. We assume that the ERDF is mass independent and broken power law shaped, test if this ERDF allows us to reproduced the observed AGN luminosity function and constrain the ERDFs parameters.

In Figure \ref{fig:rlf_xlf_comp} we show the hard X-ray luminosity (XLF) function by \cite{Ajello:2012aa} (hereafter \citetalias{Ajello:2012aa}) and the 1.4 GHz radio luminosity function (RLF) by \cite{Mauch:2007aa} (hereafter \citetalias{Mauch:2007aa}) which we use to compare our predictions to the observations. For the RLF we converted the 1.4 GHz radio luminosity from $\rm{W\ Hz^{-1}}$ to $\rm{erg\ s^{-1}}$ to allow for an easier comparison to the XLF. The figure shows that compared to the XLF, the RLF is significantly shallower. The shape of the AGN luminosity function depends on the shape of the stellar mass function and the ERDF. Yet even if we used different stellar mass functions as input, due to their significantly different shapes we are unable to reproduce both the XLF and the RLF with a single ERDF. Different ERDF shapes indicate that in our model radio and X-ray AGN cannot be considered as being part of the same black hole growth mode. Therefore our first assumption is the following:

\newcounter{counter}
\begin{enumerate}
	\item{In the local Universe AGN can be separated into radiatively efficient and inefficient AGN, detected in the hard X-rays and at 1.4 GHz, respectively. The two groups show little overlap, can be treated separately and each have a characteristic ERDF.}
	\setcounter{counter}{\value{enumi}}
\end{enumerate}

We discuss the need for two separate ERDFs in more detail in Section \ref{sec:two_erdfs}. We make additional assumptions to link the black hole and galaxy populations and to be able to compare our predicted AGN luminosity functions to the observations. We stress that these assumptions are intended to describe the galaxy and black hole populations on a global scale. They might not be able to capture all underlying complexities, but are meant to be as simple as possible, broadly true and sufficient to infer general relations. We assume 

\begin{enumerate}
	\setcounter{enumi}{\value{counter}}
	\item{that radiatively efficient AGN are primarily found in optically blue and green galaxies,}
	\item{a constant bolometric correction for the hard X-rays of $k_{\rm bol, X} =  \log(L_{\rm X}/L_{\rm bol}) = \kbolX$,}
	\item{that radiatively inefficient AGN are mostly hosted by optically red, quiescent galaxies,}
	\item{a constant bolometric correction for 1.4 GHz of $k_{\rm bol, R} =  \log(L_{\rm R}/L_{\rm bol}) = \kbolR$,}
	\item{a constant $M_{\rm BH} - M_{\rm host}$ scaling relation to convert stellar into black hole masses ($\log(M_{\rm BH}/M) = \mmmu$, scatter \mmsigma\ dex),}
	\item{and that both, the ERDF for radio AGN and the ERDF for X-ray AGN, are mass independent and broken power law shaped.}
\end{enumerate}

We will now discuss each of these assumptions in more detail. 
\subsection{Two accretion modes}\label{sec:acc_modes}
We assume that active black holes can be observed in two distinct accretion modes: radiative and radio mode. In the radiative mode, AGN accrete via a  geometrically thin, optically thick accretion disc (e.g. \citealt{Shakura:1973aa}). The Eddington ratio and the radiative efficiency of these AGN are of the order of $10\%$, i.e. the potential energy of the infalling matter is efficiently converted to radiation. On the contrary, AGN that are in the radio mode have a low radiative efficiency and low Eddington ratios. Their accretion flows are geometrically thick and optically thin, which results in a radiative cooling time that is longer than the infall time \citep{Narayan:1994aa, Narayan:2008aa}. Radio or jet mode AGN are often found in massive, red elliptical galaxies (e.g. \citealt{Best:2005aa}), whereas radiative or quasar mode AGN tend to be hosted by galaxies with bluer colors and higher SFRs (see discussion below and e.g. \citealt{Nandra:2007aa, Hickox:2009aa, Treister:2009ab, Schawinski:2009ab, Netzer:2009aa, Griffith:2010aa, Goulding:2014aa}). 

The two AGN populations show an overlap in the form of high excitation radio galaxies. Based on optical emission lines, radio galaxies can be split into high and low excitation radio galaxies (HERGs \& LERGs, \citealt{Hardcastle:2006aa,Hardcastle:2007aa, Smolcic:2009aa, Heckman:2014aa, Ching:2017aa}). The distinction is linked to different AGN accretion modes (\citealt{Hardcastle:2006aa, Smolcic:2009aa}): HERGs accrete via radiatively efficient accretion, whereas LERGs accrete via radiatively inefficient, advection dominated accretion. HERGs, which are similar to Seyferts with low levels of radio emission, can thus also be detected using selection methods based on, for example, X-ray  \citep{Wong:2016aa} or infrared \citep{Hardcastle:2013aa} emission. Bright HERGs are often referred to as radio loud quasars. 

\cite{Heckman:2014aa} show that below $\log (P_{1.4\ \mathrm{GHz}}/\mathrm{W\ Hz}^{-1}) = 26$ the space density of LERGs is significantly higher than the space density of HERGs (also see \citealt{Best:2012aa, Gendre:2013aa, Heckman:2014aa, Pracy:2016aa}). While HERGs become dominant at high radio luminosities, their space density is several orders of magnitude lower than the space density of LERGs at, for example, $\log (P_{1.4\ \mathrm{GHz}}/\mathrm{W\ Hz}^{-1}) = 22$. The number of radio AGN that accrete via radiatively efficient accretion is thus negligible. 

For our model we neglect this overlap between the radiatively efficient and inefficient AGN populations and consider X-ray and radio AGN separately. We furthermore assume that they are detected in the hard X-rays and at 1.4 GHz, respectively.
\subsection{The host galaxies of radiatively efficient AGN}\label{sec:blue_Xray_AGN}
In the local Universe, the bimodality that we observe in color-mass and color-magnitude space is well established \citep{Bell:2003aa, Baldry:2004aa, Faber:2007aa, Martin:2007aa, Schawinski:2014aa}. On one hand, star-forming galaxies, which are predominantly late-type/disc dominated, inhabit the `blue cloud'. On the other hand, quiescent galaxies, mostly early type and/or bulge dominated galaxies, are found on the `red sequence'. Between these two populations lies the transition zone, the `green valley' \citep{Bell:2004aa, Martin:2007aa, Fang:2012aa, Schawinski:2014aa}. Previous work has shown that X-ray selected AGN are found in galaxies that show signs of star formation, so either in the blue cloud or the green valley \citep{Silverman:2009aa, Treister:2009ab, Koss:2011aa,Schawinski:2009ab, Hickox:2009aa, Rosario:2013ab, Goulding:2014aa}. For our simple model we assume that most of the X-ray selected AGN are hosted by blue and green galaxies and use the stellar mass function of these galaxies as input when constraining the ERDF relative to the XLF.

\subsection{Bolometric correction for hard X-ray selected AGN}\label{sec:bol_corr_xray}
To be able to compare our predicted to the observed XLF, we need to convert from bolometric to hard X-ray luminosities. Although the SEDs of AGN show a high degree of uniformity from the X-rays to the infrared \citep{Elvis:1994aa, Richards:2006ab}, previous studies have suggested that the X-ray bolometric correction might be luminosity or Eddington ratio dependent \citep{Marconi:2004aa, Hopkins:2007aa, Vasudevan:2009aa, Lusso:2010aa, Lusso:2012aa}. We consider the simplest model and assume a constant bolometric correction. We discuss the effect that a luminosity dependent bolometric correction would have on our results in Section \ref{sec:lum_dep_kbol}.

\cite{Rigby:2009aa} determine $L_{17 - 60\ \mathrm{keV}}/L_\mathrm{2 - 10\ \mathrm{keV}}$ to be $1.34$. We use the same value to convert from 15 - 55 keV to 2 - 10 keV. Averaging over a wide range of Eddington ratios, we assume $L_\mathrm{bol}/L_\mathrm{2 - 10\ \mathrm{keV}} = 20$ based on the results by \cite{Vasudevan:2009aa}. We conclude:
\begin{equation}
\begin{aligned}
\log L_{15 - 55\ \mathrm{keV}} =& \log L_\mathrm{bol} - \log k_\mathrm{soft - bol} - \log k_\mathrm{hard - soft}\\
\simeq & \log L_\mathrm{bol} - \log 20 + \log 1.34\\
k_\mathrm{bol, X} =& \log \left(\frac{L_{15 - 55\ \mathrm{keV}}/\mathrm{erg\ s^{-1}}}{L_\mathrm{bol}/\mathrm{erg\ s^{-1}}}\right) = -1.
\end{aligned}
\end{equation}
\subsection{The host galaxies of radiatively inefficient AGN}\label{sec:red_radio_AGN}
In our model we assume that the 1.4 GHz radio luminosity function is produced by radiatively inefficient AGN which are hosted by red, quiescent galaxies. The fact that low Eddington ratio radio AGN are primarily found in massive ellipticals is well established \citep{Matthews:1964aa, Yee:1987aa, Best:2005aa, Hickox:2009aa}. The HERG and LERG distinction has been introduced more recently  \citep{Hardcastle:2006aa,Hardcastle:2007aa, Smolcic:2009aa, Heckman:2014aa}. As we discussed above, the RLF is dominated by LERGs which accrete via radiatively inefficient accretion. These LERGs are predominantly hosted by optically quiescent, red galaxies  \citep{Smolcic:2009aa, Smolcic:2009ab, Janssen:2012aa, Ching:2017aa}. 
\subsection{Bolometric correction for radio AGN}\label{sec:bol_corr_radio}
To convert our predicted bolometric luminosity functions to 1.4 GHz, we use a constant bolometric correction of 
\begin{equation}
k_\mathrm{bol, R} = \log \left(\frac{L_{1.4\ \mathrm{GHz}}/\mathrm{erg\ s}^{-1}}{L_\mathrm{bol}/\mathrm{erg\ s}^{-1}}\right) = \kbolR.
\end{equation}
This is a simplified assumption since the connection between the AGN bolometric luminosity and the 1.4 GHz core emission is complex. First of all, radio loudness $\mathcal{R}$, i.e. the ratio of radio to optical emission, is a function of the Eddington ratio  \citep{Woo:2002aa, Ho:2002aa}; the higher $\lambda$, the lower the radio loudness.  Depending on whether the total or only the core radio luminosity is taken into account, there is furthermore evidence for \citep{Sikora:2007aa} and against \citep{Broderick:2011aa} a radio loud/radio quiet dichotomy in $\mathcal{R}-\lambda$ space. In addition, radio AGN release energy in the form of mechanical jet power and so only a fraction of their total energy is emitted in the form of radiation \citep{Falcke:1995aa, Birzan:2004aa, Birzan:2008aa, Kording:2008aa, Cattaneo:2009aa, Cavagnolo:2010aa, Plotkin:2012aa, Turner:2015aa, Godfrey:2016aa, Mingo:2016aa}.

To constrain the shape of $\xi(\lambda)$, we need to propose a $k_\mathrm{bol, R}$ value. Motivated by the fact that radio AGN, specifically LERGs, tend to have low Eddington ratios compared to radiatively efficient AGN or HERGs \citep{Ho:2002aa, Evans:2006aa, Hardcastle:2007aa, Merloni:2008aa, Hickox:2009aa, Smolcic:2009aa, Alexander:2012aa, Fabian:2012aa, Best:2012aa}, we choose $k_\mathrm{bol, R} = \kbolR$. 

We stress that we do not claim $k_\mathrm{bol, R}$ to be precisely \kbolR. This value represents an assumption which allows us to test if the observed RLF is consistent with a mass independent ERDF and a constant bolometric correction. While summarizing all complexities that affect $k_\mathrm{bol, R}$ into a single constant allows us to constrain the shape of the radio ERDF, we do not claim to be able to determine the absolute value of the ERDF break, as we will discuss in more detail below.
\subsection{The local $M_\mathrm{BH} - M_\mathrm{host}$ relation}\label{sec:MM}
To convert stellar to black hole masses, we use the local $M_\mathrm{BH} - M_\mathrm{host}$ scaling relation. We parametrize the relation as
\begin{equation}
\log \left(\frac{M_\mathrm{BH}}{M_\odot}\right) = \mu + \beta \times \log \left(\frac{M_\mathrm{host}}{M_\odot} \right)																	
\end{equation}
and choose $\mu = -2.75$, $\beta = 1$ and $\sigma = 0.3$ dex. This $\mu$ value lies between the results by \citeauthor{Haring:2004aa} (\citeyear{Haring:2004aa}, $\mu = -2.8$), \citeauthor{Jahnke:2009aa} (\citeyear{Jahnke:2009aa}, $\mu = -2.75$), \citeauthor{Kormendy:2013ab} (\citeyear{Kormendy:2013ab}, $\mu = -2.31$), \citeauthor{McConnell:2013aa} (\citeyear{McConnell:2013aa}, $\mu = -2.54$), \citeauthor{Marleau:2013aa} (\citeyear{Marleau:2013aa}, $\mu = -3.02$)  and \citeauthor{Reines:2015aa} (\citeyear{Reines:2015aa}, $\mu = -3.55$). We do not take into account variations in the relation that might arise due to different levels of star formation, different morphological classifications or due to using bulge instead of total stellar mass \citep{Sani:2011aa, Reines:2015aa, Savorgnan:2016ab, Terrazas:2016aa} and use stellar mass $M$ for the host mass $M_\mathrm{host}$. 
\subsection{A broken power law shaped ERDF}
While alternative ERDF shapes have been discussed previously \citep{Kollmeier:2006aa, Kauffmann:2009aa, Hopkins:2009aa, Cao:2010aa, Aird:2012aa, Bongiorno:2012aa, Nobuta:2012aa, Conroy:2013aa, Aird:2013ab, Veale:2014aa, Hickox:2014aa, Schulze:2015aa, Trump:2015aa, Jones:2016aa, Bongiorno:2016aa}, we assume a broken power law shaped ERDF for both radiatively efficient and inefficient AGN. A broken power law shaped ERDF solves the apparent discrepancy between a Schechter function shaped stellar mass function and an observed  broken power law shaped luminosity function. Furthermore, a broken power law shaped ERDF allows us to change both the high and the low Eddington ratio end slopes. Besides a functional form for the ERDF, we assume that the ERDF is mass independent. Mass dependent ERDFs have been proposed \citep{Schulze:2015aa, Bongiorno:2016aa}, yet not introducing a mass dependence constitutes a straightforward first assumption. In Sections \ref{sec:erdf_mass} and \ref{sec:erdf_shapes} we discuss a possible mass dependence and alternative ERDF shapes, respectively. 

\section{Method}\label{sec:method}
\begin{figure*}
	\includegraphics[width=\textwidth]{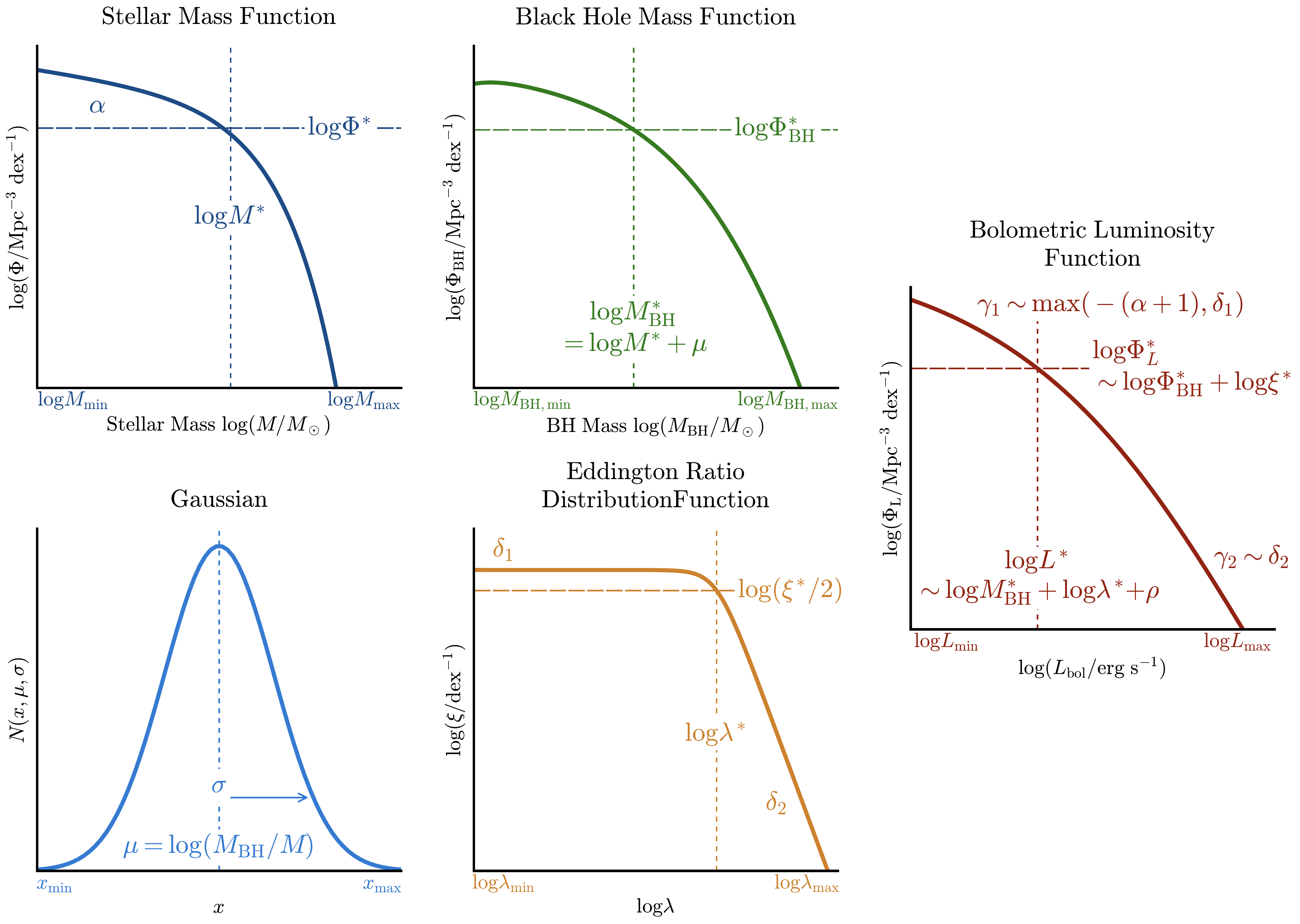}
	\caption{\label{fig:cartoon_para}Overview over the most important parameters and variables used in our model. The stellar mass function is described as a standard single or double Schechter function with break $\log M^{*}$, slope $\alpha$ and normalization $\Phi$ (see equations \ref{eq:single_schechter} and \ref{eq:double_schechter}). We construct the black hole mass function $\Phi_\mathrm{BH}$ from the stellar mass function by assuming a constant $\log M_\mathrm{BH}$ to $\log M$ ratio $\mu$ with log-normal scatter $\sigma$ (see Sections \ref{sec:random_draw} and \ref{sec:convolution}). We assume a broken power law Eddington ratio distribution function $\xi( \lambda)$ with break $\log \lambda^{*}$ and the slopes $\delta_1$ and $\delta_2$.  We describe the luminosity function as a broken power law with break $\log L^{*}$, normalization $\Phi^{*}_{L}$, faint end slope $\gamma_1$ and bright end slope $\gamma_2$ (see equation \ref{eq:broken_pl_lf}). In Section \ref{sec:pred_lf} we discuss how $\log  L^{*}$, $\gamma_1$ and $\gamma_2$ are related to the stellar mass function and Eddington ratio distribution function parameters.}
\end{figure*}

\begin{figure*}
	\includegraphics[width=\textwidth]{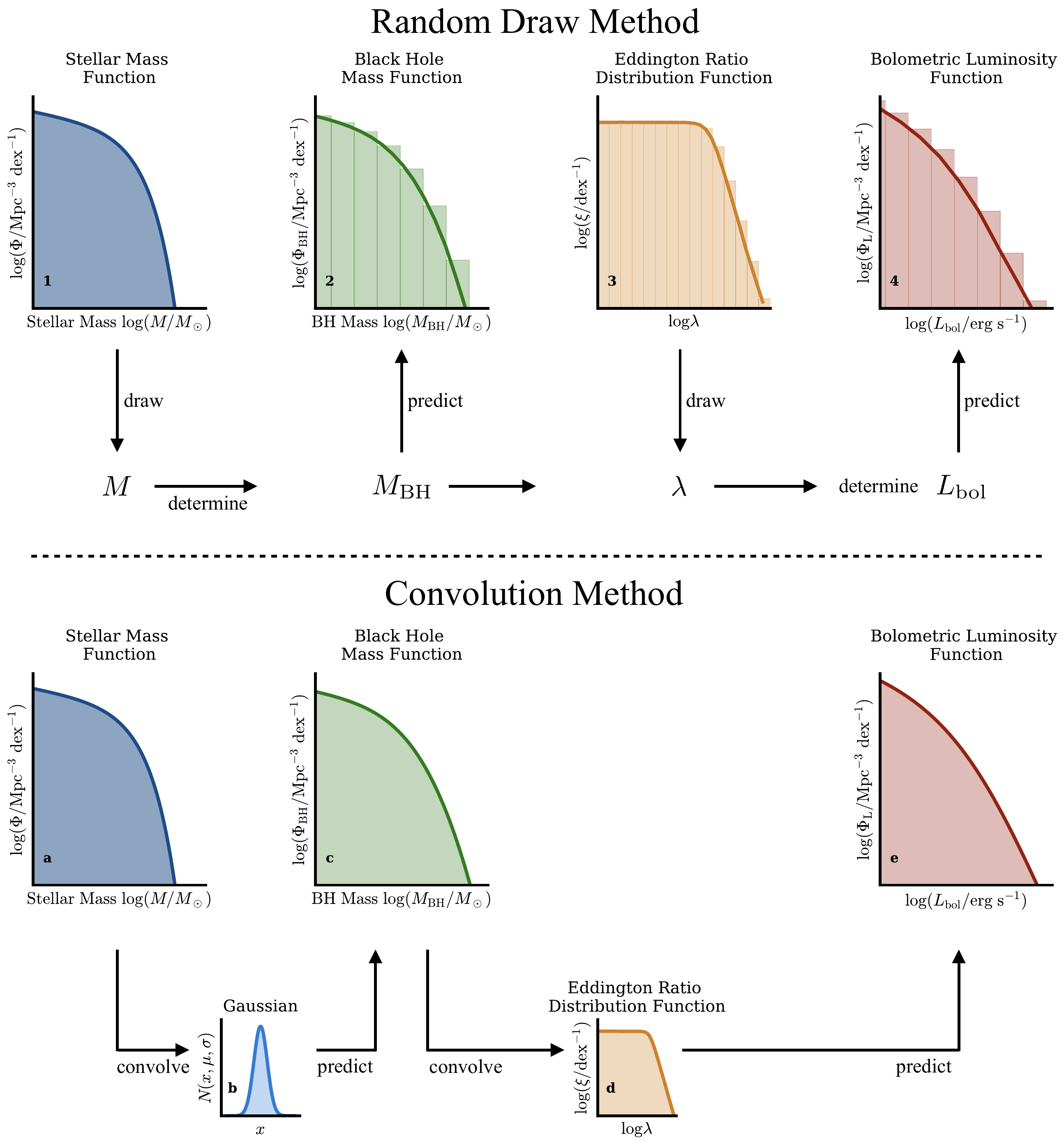}
	\caption{\label{fig:cartoon} Schematic illustrating the two equivalent methods to predict the bolometric luminosity function shape. For both techniques we use an observed stellar mass function as input and assume an Eddington ratio distribution function. By applying constant bolometric corrections our predicted luminosity functions can then be compared to observed X-ray and radio luminosity functions. Our first method, which is based on a random draw (see Sec. \ref{sec:random_draw}), is illustrated in the upper panels. We randomly draw stellar mass values from the input stellar mass function (panel 1) and convert these to black hole masses, assuming a constant $\log M_\mathrm{BH}$ to $\log M$ ratio $\mu$ with log-normal scatter $\sigma$. Furthermore, we randomly draw Eddington ratio values from the input Eddington ratio distribution function $\xi(\lambda)$ (panel 3). This allows us to not only predict the the black hole mass function (panel 2), but also the shape of the bolometric luminosity function (panel 4). The second technique is summarized in the lower panels. It is computationally less expensive since it is based on convolutions (see Sec. \ref{sec:convolution}). We convolve the input stellar mass function (panel a) with a normal distribution with mean $\mu$ and standard deviation $\sigma$ (panel b) to predict the black hole mass function (panel c). To determine the shape of the bolometric luminosity function (panel e), we convolve $\Phi_\mathrm{BH}$ with the Eddington ratio distribution function (panel d).}
\end{figure*}

After introducing our assumptions in the previous section, we now discuss our method. Our method relies on an input stellar mass function and an assumed ERDF. We parametrize the single Schechter \citep{Schechter:1976aa} stellar mass function in the following functional way: 
\begin{equation}\label{eq:single_schechter}
\Phi(M) = \frac{dN}{d\log M} = \ln(10) \Phi^{*} \left(\frac{M}{M^{*}}\right)^{\alpha + 1} \exp\left(-\frac{M}{M^{*}}\right).
\end{equation}
Here, the factor of $\ln(10)$ and the $+1$ in the power law exponent are due to the conversion from $dM$ to $d\log M$. Similarly the double Schechter function is given by:  
\begin{equation}\label{eq:double_schechter}
\begin{aligned}
\Phi(M) &= \frac{dN}{d\log M}\\
& = \ln(10) \exp\left(-\frac{M}{M^{*}}\right) \left[\Phi^{*}_1 \left(\frac{M}{M^{*}}\right)^{\alpha_1 + 1}  + \Phi^{*}_2 \left(\frac{M}{M^{*}}\right)^{\alpha_2 + 1}\right].
\end{aligned}
\end{equation}
As we assume that the ERDF is broken power law shaped and mass independent, we parametrize the broken power law ERDF in the following way 
\begin{equation}\label{eq:broken_pl_erdf}
\xi(\lambda) = \frac{dN}{d\log \lambda} = \xi^{*} \times \left[\left(\frac{\lambda}{\lambda^{*}}\right)^{\delta_1} + \left(\frac{\lambda}{\lambda^{*}}\right)^{\delta_2} \right]^{-1}
\end{equation}
and define $\delta_1$ and $\delta_2$ as the low and high Eddington ratio slopes, respectively. With $\Phi(M)$ and $\xi(\lambda)$ as input, we can predict the AGN bolometric luminosity function, $\Phi_\mathrm{L}(L)$, which is commonly parametrized as a broken power law of the form:
\begin{equation}\label{eq:broken_pl_lf}
\Phi_\mathrm{L}(L) = \frac{dN}{d\log L} = A \times \left[\left(\frac{L}{L^{*}}\right)^{\gamma_1} + \left(\frac{L}{L^{*}}\right)^{\gamma_2} \right]^{-1}.
\end{equation}
$\gamma_1$ represents the faint, $\gamma_2$ the bright end of the luminosity function. With the appropriate bolometric corrections, we can then compare our prediction to the observed XLF or RLF.

We introduce two manifestations of our method: the first is based on random draws (see Section \ref{sec:random_draw}), whereas for the second we employ multiple convolutions (see Section \ref{sec:convolution}). In Section \ref{sec:pred_lf} we discuss the properties of the predicted luminosity function. To constrain the ERDFs of radiatively efficient and inefficient AGN we use a Markov chain Monte Carlo (MCMC) which we introduce in Section \ref{sec:mcmc}. We also discuss our model in the context of key terms such as `variability', `occupation fraction', `duty cycle' and `active black hole fraction' (Section \ref{sec:flickering}). Figures \ref{fig:cartoon_para} and \ref{fig:cartoon} summarize the random draw and the convolution method and all relevant parameters.

The results of the two methods, which we introduce below, are equivalent, but differ in their computational expense. The random draw approach is computationally more expensive, but more easily adjusted. For instance, we use the random draw method to demonstrate the effects of a mass dependent ERDF (see Section \ref{sec:erdf_mass}). The convolution approach is computationally less expensive and thus used to constrain the ERDF shape with a MCMC.
\subsection{The random draw method}\label{sec:random_draw}
Our first method to predict the bolometric luminosity function is based on random draws and is illustrated in the top panels of Figure \ref{fig:cartoon}. We use a stellar mass function and an ERDF as input and perform the following steps: 

\begin{itemize}
	\item {
		\textit{Determine $N_\mathrm{draw}$}:
		Before drawing from the stellar mass function, we need to determine the number of values to be drawn, \Ndraw. \Ndraw\ is a simple scaling factor which determines the lowest number densities that we can probe with the random draw method. \Ndraw\ can be given a physical meaning by linking it to the stellar mass function. For example \Ndraw\ can be defined as:  
		\begin{equation}\label{eq:nr_draws}
		N_\mathrm{draw} = V(z_\mathrm{min}, z_\mathrm{max}, \Omega) \times \int_{\log M_\mathrm{min}}^{\log M_\mathrm{max}} \Phi(M) d\log M
		\end{equation}
		Here, $\log M_\mathrm{min}$ and $\log M_\mathrm{max}$ denote the minimum and maximum stellar mass values that we consider in the draw, respectively. $V(z_\mathrm{min}, z_\mathrm{max}, \Omega)$ can be chosen to represent the comoving volume (e.g. \citealt{Hogg:1999aa}) of a specific survey with solid angle $\Omega$.
	}
	\item {
		\textit{Construct the cumulative distribution function (CDF) of the stellar mass function and draw  stellar mass values}:
		After having determined \Ndraw, we construct the CDF of the stellar mass function by computing: 
		\begin{equation}\label{eq:cdf}
		\mathrm{CDF}(\log M) = \frac{\int_{\log M_\mathrm{min}}^{\log M} \Phi(M) d\log M}{\int_{\log M_\mathrm{min}}^{\log M_\mathrm{max}} \Phi(M) d\log M}.
		\end{equation}
		We randomly draw \Ndraw\ numbers between 0 and 1 from a uniform distribution and by inverting $\mathrm{CDF}(\log M)$, assign the corresponding stellar mass values.
	}
	\item{
		\textit{Convert all stellar masses to black hole masses by assuming a conversion factor $\mu$ with scatter $\sigma$}:
		We assume the local $M_\mathrm{BH} - M_\mathrm{host}$ relation (see Section \ref{sec:MM}) and draw \Ndraw\ conversion factors from a  normal distribution with mean $\mu = \log M_\mathrm{BH}/M = \mmmu$ and scatter $\sigma = \mmsigma$. \logMBH\ is then given by the sum of these conversion factors and the previously drawn stellar mass values. Having determined \logMBH, allows us to construct the black hole mass function. We bin in \logMBH\ and adjust the normalization so that $\log \Phi_\mathrm{BH}$ in bin $i$ is given by
		\begin{equation}
		\log \Phi_{\mathrm{BH}, i} = \log \left(\frac{n_i}{V(z_\mathrm{min}, z_\mathrm{max}, \Omega) \times \Delta \log M_\mathrm{BH}} \right).
		\end{equation}
		Here, $n_i$ corresponds to the number of simulated systems in bin $i$, $V(z_\mathrm{min}, z_\mathrm{max}, \Omega)$ is the volume that we have already used in equation \ref{eq:nr_draws} and $\Delta \log M_\mathrm{BH}$ represents the bin size.
	}
	\item{
		\textit{Construct the CDF of the ERDF and draw Eddington ratio values \loglambda}:
		We also assign randomly drawn Eddington ratio values to black hole mass value. In analogy to equation \ref{eq:cdf} we construct the CDF for the given ERDF in the range $\log \lambda_\mathrm{min}$ to $\log \lambda_\mathrm{max}$. The shape of the CDF is not affected by the normalization of the ERDF. However $\xi^{*}$ determines how many of the $N_{\rm draw}$ black hole mass values are assigned an Eddington ratio value. If the integral over $\xi(\lambda)$ from $\log \lambda_{\rm min}$ to $\log \lambda_{\rm max}$ is 1 we draw $N_{\rm draw}$ $\log \lambda$ values from the CDF. However, if the integral over the ERDF is $a$ with $a < 1$, then only $N_{\rm draw, AGN} = a \times N_{\rm draw}$ black holes are assigned a $\log \lambda$ value.
	}
	\item{
		\textit{Calculate bolometric luminosities to determine $\Phi_{\rm L}(L_{\rm bol})$}:
		We compute the bolometric luminosities corresponding to the $N_{\rm draw, AGN}$ black holes and their corresponding $\log \lambda$ values:
		\begin{equation}
		\log (L_\mathrm{bol}/\mathrm{erg\ s}^{-1})  =  \log \lambda + \log (M_\mathrm{BH}/M_\odot) + \rho.\\
		\end{equation} 
		Here, $\rho = \log \left(\frac{L_\mathrm{Edd}/\mathrm{erg\ s}^{-1}}{M_\mathrm{BH}/M_\odot} \right) = 38.2$.
		In analogy to the black hole mass function, we bin in \logLbol. $\log \Phi_L$ in bin $i$ is then given by:
		\begin{equation}\label{eq:lf_draw}
		\log \Phi_{L,i} = \log \left(\frac{n_i}{V(z_\mathrm{min}, z_\mathrm{max}, \Omega) \times \Delta \log L_\mathrm{bol}}\right).
		\end{equation}
	}
\end{itemize}

\subsection{The convolution method}\label{sec:convolution}
Our second method to predict the shape of the luminosity function is based on convolutions and follows the work of \citetalias{Caplar:2015aa}. We again use a stellar mass function and an ERDF as input and go through the following steps (see the bottom row of Figure \ref{fig:cartoon}):
\begin{itemize}
	\item {
		\textit{Convolve the stellar mass function with a normal distribution to predict the black hole mass function}:
		We convolve the observed stellar mass function with a normal distribution with mean $\mu$ and standard deviation $\sigma$: 
		\begin{equation}
		\begin{aligned}
		\Phi_\mathrm{BH}(\log M_\mathrm{BH})  = & (\Phi \ast N)(\log M_\mathrm{BH}) \\
		= & \int_{\log M_\mathrm{min}}^ {\log M_\mathrm{max}} [\Phi(\log M) \\
		& \times N(\log M_\mathrm{BH} - \log M, \mu, \sigma) ] d\log M.\\
		\end{aligned}
		\end{equation}
		Here, $N(x, \mu, \sigma)$ is the normal distribution and $\Phi(M)$ is the stellar mass function which is described by either a single or a double Schechter function (see equations \ref{eq:single_schechter} and \ref{eq:double_schechter}).
	}
	\item {
		\textit{Convolve the predicted black hole mass function with the given ERDF to determine the shape of the bolometric luminosity function}:
		First, we assume a constant Eddington ratio of $\log \lambda = 0$ and shift $\Phi_\mathrm{BH}(\log M_\mathrm{BH})$ from black hole mass to bolometric luminosity space
		\begin{equation}
		\Phi_{\mathrm{L}, \log \lambda = 0}(\log L) = \Phi_\mathrm{BH} (\log M_\mathrm{BH} + \rho). 
		\end{equation} 
		To take into account the fact that the Eddington ratio is not constant, we convolve $\Phi_{\mathrm{L}, \log \lambda = 0}$ with the ERDF, $\xi(\lambda)$:
		\begin{equation}
		\begin{aligned}
		\Phi_\mathrm{L}(\log L) = & (\Phi_{\mathrm{L}, \log \lambda = 0} \ast \xi)(\log L)\\
		=& \int_{\log \lambda_\mathrm{min}}^{\log \lambda_\mathrm{max}} [\Phi_{\mathrm{L}, \log \lambda =0}(\log L - \log \lambda)\\
		& \times \xi(\log \lambda)] d\log \lambda. \\
		\end{aligned}
		\end{equation}
	} 
\end{itemize}

Besides the ERDF parameters, both the random draw and the convolution method require choices for $\log M_{\rm min}$, $\log M_{\rm max}$, $\log \lambda_{\rm min}$ and $\log \lambda_{\rm max}$. We use $\log (M_{\rm min}/M_{\odot}) = 9$ and $\log (M_{\rm max}/M_{\odot}) = 12$ since this is the mass range over which the \citetalias{Weigel:2016aa} stellar mass functions are constrained. We discuss the effect of this choice on our results in Section \ref{sec:completeness}. We set $\log \lambda_{\rm min}$ and $\log \lambda_{\rm max}$ to \ERDFmin\ and \ERDFmax, respectively (also see Table \ref{tab:mcmc_para}). The values of $\log \lambda_{\rm min}$ and $\log \lambda_{\rm max}$ are degenerate with the normalization of the ERDF. $\xi^{*}$ determines what fraction of black holes are assigned an Eddington ratio between $\log \lambda_{\rm min}$ and $\log \lambda_{\rm max}$ and can therefore be considered as being `on' (see Section \ref{sec:flickering}). In the framework of the random draw technique assigning $\log \lambda$ values to all $N_{\rm draw}$ black hole mass values implies that all black holes in the sample can be considered AGN. The convolution method produces the same result if  the integral over $\xi(\lambda)$ from $\log \lambda_{\rm min}$ to  $\log \lambda_{\rm max}$ is 1. Without a priori knowledge of $\xi^{*}$ for the chosen $\log \lambda_{\rm min}$ and $\log \lambda_{\rm max}$ boundaries only the shape of the bolometric luminosity function, but not its normalization $\Phi^{*}_{\rm L}$ can be predicted. As we will discuss in more detail below, we determine $\xi^{*}$ by rescaling the predicted luminosity function so that the space density over the considered luminosity range matches the observed space density.

Once we have determined the bolometric luminosity function through either the random draw or the convolution method, we use the bolometric corrections $k_{\rm bol, X}$ and $k_{\rm bol, R}$ to predict the XLF and the RLF. In both cases, this results in a constant shift towards lower luminosities since our assumed bolometric corrections are constant with luminosity and Eddington ratio. We discuss the effect of a luminosity dependent bolometric correction in Section \ref{sec:lum_dep_kbol}.
\subsection{The predicted luminosity function}\label{sec:pred_lf}
Both methods allow us to predict the bolometric luminosity function once we have assumed the shape of $\xi(\lambda)$. As we discussed above, a broken power law shape is an appropriate first assumption for the ERDF since, in contrast to the stellar mass function, the observed AGN luminosity function is also power law shaped. \citetalias{Caplar:2015aa} derive and discuss the properties of the predicted luminosity function if a broken power law shaped ERDF is assumed. We summarize these characteristics of $\Phi_{\rm L}(L)$ and its dependence on $\Phi(M)$ and $\xi(\lambda)$ below and in Figure \ref{fig:cartoon_para}. 

\begin{itemize}
	
	\item{\textit{The bright end slope $\gamma_2$}:
		\citetalias{Caplar:2015aa} use a simplified ERDF with $\xi() = 0$ for $\log \lambda < \log \lambda^{*}$ to show how the $\xi(\lambda)$ shape affects the predicted luminosity function. They show analytically that the bright end of the luminosity function has the same slope as the high $\lambda$ end of the ERDF, that is:
		\begin{equation}\label{eq:delta2}
		\delta_2 = \gamma_2.
		\end{equation}
		As the stellar mass function falls off exponentially at high stellar masses, the shallower $\delta_2$ slope is necessary to reproduce the observed bright end of the luminosity function. To construct the black hole mass function, we convolve the stellar mass function with a normal distribution. In contrast to the stellar mass function, the black hole mass function does therefore no longer fall off exponentially at the high mass end. In the extreme case in which $\delta_2$ is steeper than the exponential cut off of $\Phi_\mathrm{BH}$, it is thus the high mass end of the black hole mass function and not $\delta_2$ that dominates $\gamma_2$.  
	}
	
	\item{\textit{The faint end slope $\gamma_1$}:
		The faint end of the luminosity function is determined by either the stellar mass function low mass end slope $\alpha$ or the ERDF low Eddington ratio end slope, $\delta_1$. Using their simplified model \citetalias{Caplar:2015aa} conclude that $\gamma_1 = - (\alpha_\mathrm{BH} + 1)$. Here, $\alpha_\mathrm{BH}$ is the black hole mass function slope. If $\delta_1$ is steeper than $\alpha_\mathrm{BH} + 1$, the ERDF slope determines $\gamma_1$. The linear relation which we assume between $M$ and $M_\mathrm{BH}$ ensures that the stellar and black hole mass functions have the same low mass end slopes. With $\alpha_\mathrm{BH} = \alpha$ we hence conclude\footnote{Our definition of $\alpha$ differs from the one used by \citetalias{Caplar:2015aa}. In our case, a flat Schechter function has a slope of $\alpha = -1$, whereas they use $\alpha = 0$. We are thus defining $\gamma_1$ in terms of $\alpha + 1$.} 
		\begin{equation}\label{eq:gamma1}
		\gamma_1 = \max [-(\alpha + 1), \delta_1].
		\end{equation}
		Equation \ref{eq:gamma1} shows that for luminosity functions with $\gamma_1 = -(\alpha + 1)$ we will not be able to constrain $\delta_1$ well, since it can take any value $\delta_1 < -(\alpha + 1)$.  
	}
	
	\item{\textit{The break $L^{*}$}:
		Under the assumption of $\delta_1 < \alpha + 1$, the position of the break of the luminosity function is given by:
		\begin{equation}\label{eq:lstar}
		\begin{aligned}
		\log L^{*} =& M^{*}_\mathrm{BH} + \log \lambda^{*} + \rho + \log \Delta_L(\delta_1, \gamma_2)\\
		=& M^{*} + \mu + \log \lambda^{*} + \rho + \log \Delta_L(\delta_1, \gamma_2).\\
		\end{aligned}
		\end{equation}
		Here, $\Delta_L(\delta_1, \gamma_2)$ is a small correction factor which is weakly dependent on the choice of $\delta_1$ and $\gamma_2$ and varies by less than 0.15 dex.
	}	
	
	\item{\textit{The normalization $\Phi^{*}_\mathrm{L}$}:
		At $L^{*}$ the normalization of the luminosity function can be predicted using 
		\begin{equation}\label{eq:phistar}
		\begin{aligned}
		\log \Phi^{*}_\mathrm{L} =& \log \Phi^{*}_\mathrm{BH} + \log \xi^{*} + \log \Delta_{\Phi}(\delta_1, \gamma_2)\\
		=& \log \Phi^{*} + \log \xi^{*} + \log \Delta_{\Phi}(\delta_1, \gamma_2).\\
		\end{aligned}
		\end{equation}
		$\xi^{*}$ is the normalization of the ERDF and we have used the fact that in our model $\Phi = \Phi_\mathrm{BH}$. To derive this relation, \citetalias{Caplar:2015aa} have again assumed $\delta_1 < \alpha + 1$ and introduced $\Delta_{\Phi}(\delta_1, \gamma_2)$, a small correction factor which is dependent on $\delta_1$ and $\gamma_2$ and varies by less than 0.15 dex. 
	}
\end{itemize}

\subsection{MCMC}\label{sec:mcmc}
\begin{figure*}
	\includegraphics[width=\textwidth]{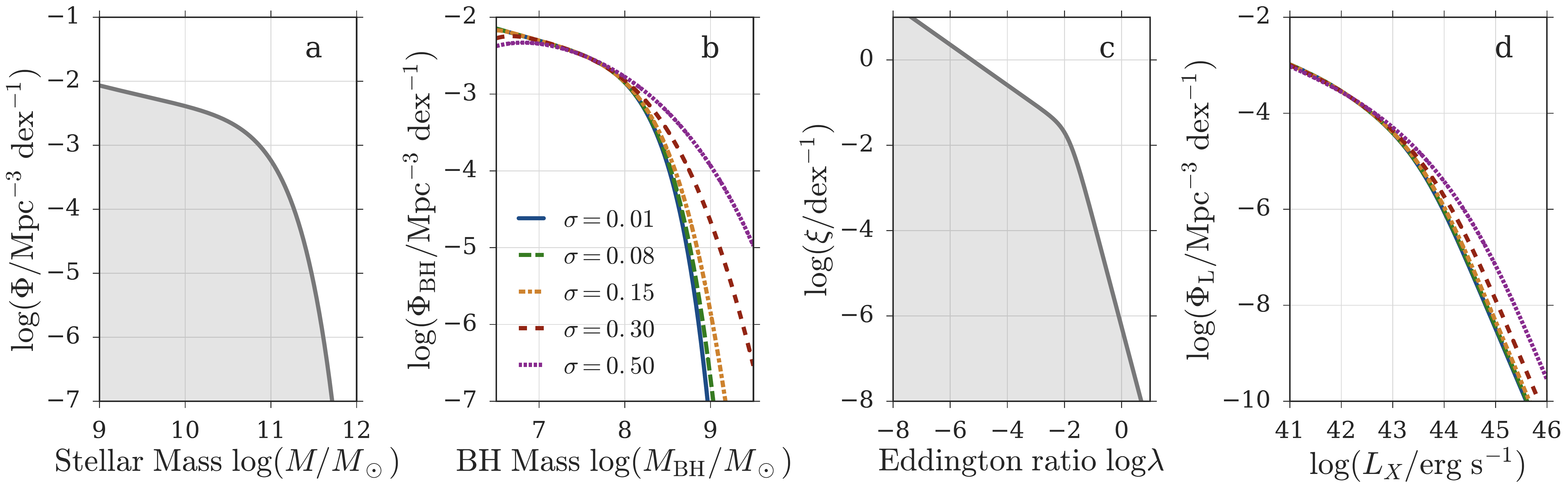}
	\caption{\label{fig:sigma} The effect of the $\log M$-$\log M_\mathrm{BH}$ ratio scatter and the resulting black hole mass function shape on the predicted luminosity function. Panel a shows the blue+green stellar mass function which we use as input. We convert $\Phi(\log M)$ to the black hole mass function, which is shown in panel b, by convolving it with a normal distribution. The normal distribution has a mean $\mu$ and a standard deviation $\sigma$ where we vary $\sigma$. We convolve all black hole mass functions with the same ERDF, shown in panel c, to predict the corresponding X-ray luminosity functions, which are shown in panel d. This figure illustrates that the bright end of the luminosity function, $\gamma_2$, is determined by either the high mass end of the black hole mass function or the high Eddington ratio end of the ERDF. For $\sigma \leq 0.15$, the high mass end of the black hole mass function is steeper than $\delta_2$. $\gamma_2$ of all black hole mass functions with $\sigma \leq 0.15$ is thus independent of $\sigma$ and solely determined by $\delta_2$. For $\sigma > 0.15$, the black hole mass function is shallower than the ERDF. $\gamma_2$ is therefore given by the high mass end of the black hole mass function, which depends on the assumed $\sigma$. The input stellar mass function and the assumed ERDF determine the $\sigma$ value above which $\gamma_2$ is no longer affected by $\delta_2$.}
\end{figure*}

The discussion in the previous section shows that we are likely to find ERDF parameters that allow us to reproduce the observed XLF and RLF. To constrain the best-fitting parameters for $\xi_{\rm X}(\lambda)$ and $\xi_{\rm R} (\lambda)$  and to quantify the corresponding uncertainties, we introduce a MCMC sampler. 

We use the broken power law ERDF defined in equation \ref{eq:broken_pl_erdf} and postulate $\delta_2 > \delta_1$. This ensures a predicted luminosity function shape similar to the observations and prevents the MCMC sampler from jumping between equivalent solutions during the sampling process. To incorporate this prior, we parametrize the broken power law ERDF in the following way:

\begin{equation}\label{eq:broken_pl_erdf_mcmc}
\begin{aligned}
\delta_2 =&  \delta_1 + \epsilon, \epsilon > 0\\
\xi(\lambda) =& \frac{dN}{d\log \lambda} = \xi^{*} \times \left[\left(\frac{\lambda}{\lambda^{*}}\right)^{\delta_1} + \left(\frac{\lambda}{\lambda^{}*}\right)^{\delta_1 + \epsilon} \right]^{-1}.\\
\end{aligned}
\end{equation}

We use the MCMC \textsc{python} package \textsc{cosmohammer}\footnote{\url{http://cosmohammer.readthedocs.org/}} \citep{Akeret:2013aa} to vary $\delta_1$, $\epsilon$ and $\log \lambda^{*}$. In each step, the MCMC proposes a new set of ERDF parameters. We use this prediction for the ERDF and the input stellar mass function to estimate the corresponding bolometric luminosity function $\Phi_\mathrm{L, pred}$ with the convolution technique. To shift the luminosity function to the hard X-ray or the 1.4 GHz radio regime we apply a constant bolometric correction. We then determine the predicted space densities ($\Phi_{\rm L, pred}$) in the luminosity bins of the observed luminosity function ($\log L_{\rm obs}$).

The normalization of the ERDF, $\xi^{*}$ is degenerate with $\log \lambda^{*}$, $\delta_1$ and $\delta_2$. To minimize the number of free parameters and degeneracies, we do not include $\xi^{*}$ in the MCMC. Instead we rescale $\Phi_{\rm L, pred}$ so that the space densities of the predicted and the observed luminosity functions match:
\begin{equation}\label{eq:normalization}
\tilde{\Phi}_\mathrm{L, pred} = \frac{n_\mathrm{obs}}{n_\mathrm{pred}} \times \Phi_{\mathrm{L, pred}}.
\end{equation} 
Here $n_{\rm obs}$ and $n_{\rm pred}$ represent the integrals over the predicted and the observed luminosity functions within the observed luminosity function's binning range.

To compute the log-likelihood for each set of new ERDF parameters, we use the observed $\log \Phi_{\rm L, obs}$ values and their errors and the predicted $\log \Phi_{\rm L, pred}$ values in the $\log L_{\rm obs}$ bins. We assume that the asymmetric observed errors on $\log \Phi_{\rm L, obs}$ follow a log-normal distribution and describe the details of the $\ln \mathcal{L}$ calculation in Section \ref{sec:likelihood}.

Once the MCMC has converged we use the MCMC chain to determine the median $\log \lambda^{*}$, $\delta_1$ and $\epsilon$ values and the corresponding 16 and 84 percentiles. Using the definition of $\epsilon$ in equation \ref{eq:broken_pl_erdf_mcmc}, we determine the sum of the $\delta_1$ and the $\epsilon$ chains to estimate the median $\delta_2$ value and its credible intervals.

It is important to acknowledge that the shape of the predicted luminosity function will not be affected by $\delta_1$ values that lie significantly below $-(\alpha + 1)$. According to equation \ref{eq:gamma1}, $\gamma_1$ will be given by $\alpha$, the low mass end of the stellar mass function, if $\delta_1 \ll -(\alpha + 1)$.  So, if $\delta_1 \sim -(\alpha + 1)$, we have to interpret $\delta_1$ as an upper limit. We can only fully constrain $\delta_1$, if $\gamma_1$ is significantly steeper than the stellar mass function ($\gamma_1 \gg -(\alpha + 1)$).

As we have pointed out in \ref{sec:pred_lf}, $\gamma_2$ does not solely rely on $\delta_2$, but is also affected by the steepness of the black hole mass function. We illustrate this effect in Figure \ref{fig:sigma}. We vary the scatter in the stellar to black hole mass conversion by convolving the blue+green stellar mass function (panel a) with normal distributions with different widths $\sigma$. The corresponding black hole mass functions (panel b) are then all convolved with the same ERDF (panel c) to estimate the corresponding X-ray luminosity functions (panel d). Figure \ref{fig:sigma} illustrates that for low $\sigma$ values the shape of the black hole mass function does not affect the shape of $\Phi_\mathrm{L}$; it is $\delta_2$ that determines $\gamma_2$. For high $\sigma$ values, $\delta_2$ is however irrelevant and $\gamma_2$ depends on the shallower high mass end of the black hole mass function.  

\subsection{Variability, Occupation fraction, duty cycle and active black hole fraction}\label{sec:flickering}
In this section, we discuss how in our model the normalizations of the stellar mass function, black hole mass function and the ERDF are linked to the black hole occupation fraction, the fraction of active black holes and the duty cycle. 

\begin{itemize}
	\item{\textit{Black hole occupation fraction}:
		We define the black hole occupation fraction as the fraction of galaxies that are hosting a black hole. In the local Universe most massive galaxies, including our own, host a black hole \citep{Genzel:1996aa,Ghez:2000aa,Ghez:2008aa,Magorrian:1998aa,Schodel:2003aa}. In our model, we thus assume an occupation fraction of $100$\%. At high redshift the occupation fraction might however be lower than $100$\% \citep{T13, Weigel:2015aa, Trakhtenbrot:2016aa}. Furthermore, depending on the black hole seed formation mechanism, the black hole occupation fraction in dwarf galaxies might be lower than in more massive galaxies \citep{Volonteri:2008aa, Natarajan:2011aa, Greene:2012aa, Reines:2013aa,  Moran:2014aa, Sartori:2015aa}. If this possible mass dependence is not taken into account, the black hole occupation fraction can be defined in the following way using the stellar mass and black hole mass function: 
			\begin{equation}
			\mathrm{occupation\ fraction} = \frac{\int_{-\infty}^{+\infty} \Phi_\mathrm{BH}(\log M_\mathrm{BH}) d\log M_\mathrm{BH} } {\int_{-\infty}^{+\infty} \Phi(\log M) d\log M}.
			\end{equation}
	}
\end{itemize}

\begin{itemize}
	\item {\textit{active black hole fraction}:
		In the MCMC we initially use an ERDF which is normalized so that the integral over $\log \lambda$ from $\log \lambda_{\rm min}$ to $\log \lambda_{\rm max}$ results in 1. In terms of the random draw technique this corresponds to assigning a $\log \lambda$ between $\log \lambda_{\rm min}$ and $\log \lambda_{\rm max}$ value to each black hole (or to each galaxy since the occupation fraction is 1) in the sample. Based on this ERDF we predict the luminosity function. We then rescale the predicted luminosity function so that the space density of the predicted and the observed luminosity functions match. This rescaling factor allows us to determine $\xi^{*}$. Furthermore we can compute the actual fraction of all black holes that have to be assigned a $\log \lambda$ value. We refer to this as the active black hole fraction and define it in the following way:
		\begin{equation}
		\mathrm{active\ black\ hole\ fraction}(\lambda_\mathrm{min}, \lambda_\mathrm{max}) = \frac{\xi^{*}}{\xi^{*}_\mathrm{norm}}.
		\end{equation}
		$\xi^{*}_\mathrm{norm}$ corresponds to the normalization of the normalized ERDF, $\xi^{*}$ represents the normalization of the ERDF after the rescaling factor has been applied. Note that the active black hole fraction is always linked to the definition of $\lambda_\mathrm{min}$ and $\lambda_\mathrm{max}$.
	}
\end{itemize}

During its lifetime an AGN is expected to change its Eddington ratio and therefore its luminosity \citep{de-Vries:2003aa, MacLeod:2010aa, Novak:2011aa, Kasliwal:2015aa, Schawinski:2015aa}. The ERDF represents the distribution of Eddington ratios, for all black hole masses at one moment in time. For instance, the ERDF shape implies that at the time of observation only few galaxies have high Eddington ratios. By postulating that the ERDF does not change significantly over a certain time range, we can further conclude that black holes evolve along the ERDF, spending a small fraction of their lifetime at high Eddington ratios. The ERDF alone does not contain time scale information. We are unable to constrain how quickly black holes change their Eddington ratios and move along the ERDF. Yet by assuming a lifetime model, the ERDF can be constrained from the luminosity function \citep{Hopkins:2009aa}. This leads us to the definition of the AGN duty cycle:

\begin{itemize}
	\item{\textit{duty cycle}:
	The AGN duty cycle is a unitless quantity. It describes the fraction of black holes that, at a given moment in time, have an Eddington ratio above a certain value $\log \lambda_{\rm lim}$. Alternatively, the duty cycle corresponds to the fraction of a black hole's lifetime that it is likely to spend at $\log \lambda > \log \lambda_{\rm lim}$. In the framework of the random draw method this corresponds to the fraction of black holes (or galaxies since the occupation fraction is 1) that are assigned a $\log \lambda$ value and for which $\log \lambda > \log \lambda_{\rm lim}$. We thus define the duty cycle in the following way:
		\begin{equation}
		\begin{aligned}
		\rm{duty\ cycle}(\log \lambda_{\rm lim}) = & \rm{active\ black\ hole\ fraction}\\
		 & \times\frac{\int_{\log \lambda_{\rm lim}}^{\log \lambda_{\rm max}} \xi(\lambda, \xi^{*} = 1) d\log \lambda}{\int_{\log \lambda_{\rm min}}^{\log \lambda_{\rm max}} \xi(\lambda, \xi^{*}=1) d\log \lambda}\\
		 =& \frac{\xi^{*}}{\xi^{*}_{\rm norm}} \times \xi^{*}_{\rm norm}\\
		  &\times \int_{\log \lambda_{\rm lim}}^{\log \lambda_{\rm max}} \xi(\lambda, \xi^{*} = 1) d\log\lambda\\
		  =&\int_{\log \lambda_{\rm lim}}^{\log \lambda_{\rm max}} \frac{\xi^{*}}{\left(\frac{\lambda}{\lambda^{*}}\right)^{\delta_1} + \left(\frac{\lambda}{\lambda^{*}}\right)^{\delta_2}} d\log \lambda
		\end{aligned}
		\end{equation}
		The duty cycle depends on the definition of $\log \lambda_{\rm lim}$, but not the chosen $\log \lambda_{\rm min}$ value.
	}
\end{itemize}
\section{Application and results}\label{sec:app_results}
After introducing both our model and our method, we now discuss the application to observations. We already argued that our model produces a broken power law shaped AGN luminosity function. We are thus likely to find ERDF parameters that allow us to reproduce the observed XLF and RLF. To find the best-fitting ERDF parameters and to quantify the corresponding uncertainties, we now use the MCMC which we introduced in Section \ref{sec:mcmc}. 

First, we introduce the stellar mass functions which we use as input for our model and the observed XLF and RLF with which we compare our predictions (see Section \ref{sec:data}). In Section \ref{sec:mcmc_results} we discuss our MCMC results for radiatively efficient and inefficient AGN. This is followed by Section \ref{sec:erdf_mass} in which we examine a possible ERDF mass dependence. 
\subsection{Input galaxy stellar mass and AGN luminosity functions}\label{sec:data}
\begin{deluxetable*}{ll|lllll}
	\tablecolumns{7}
	\tablewidth{0pt}
	\tablecaption{\label{tab:mass_lum_fcts} Input stellar mass and luminosity functions}
	\tablehead{
		\colhead{Stellar mass functions} & 
		\colhead{reference} & 
		\colhead{$\log (M^{*}/M_\odot)$} & 
		\colhead{$\log (\Phi^{*}_1/\rm Mpc^{-3})$} & 
		\colhead{$\alpha_1$} & 
		\colhead{$\log (\Phi^{*}_2/\rm Mpc^{-3})$} & 
		\colhead{$\alpha_2$}
		}
		\startdata
		{blue + green mass function} & {\citetalias{Weigel:2016aa}}  & {$10.67\pm0.02$} & {$-3.10\pm0.10$} & {$-1.38\pm0.05$} & {$-2.91\pm0.18$} & {$-0.70\pm0.11$}\\
		{red mass function} & {\citetalias{Weigel:2016aa}}  & {$10.77\pm0.01$} & {$-7.12\pm0.77$} & {$-3.08\pm0.50$} & {$-2.67\pm1.09$} & {$-0.46\pm0.02$}\\
		{}\\
		\hline
		\multicolumn{1}{c}{Luminosity functions} & \multicolumn{1}{c}{reference} & \multicolumn{1}{c}{$\log L^{*}$} & \multicolumn{1}{c}{$A$/$\mathrm{Mpc}^{-3}$} & \multicolumn{1}{c}{$\gamma_1$} & \multicolumn{1}{c}{$\gamma_2$} & {}\\
		\hline
		{$Swift$/BAT 15-55 keV XLF} & {\citetalias{Ajello:2012aa}}  & {$43.71 \pm 0.12$ [$\mathrm{erg\ s}^{-1}$]} & {$113.1 \pm 6.0 \times 10^{-7}$} & {$0.79 \pm 0.08$} & {$2.39 \pm 0.12$} & {}\\
		{$Swift$/BAT 14-195 keV XLF} & {\cite{Tueller:2008aa}}  & {$43.85\pm0.26$ [$\mathrm{erg\ s}^{-1}$]} & {$1.80^{+2.7}_{-1.1}\times 10^{-5}$} & {$0.84^{+0.16}_{-0.22}$} & {$2.55^{+0.43}_{-0.30}$} & {}\\
		{INTEGRAL 17-60 keV XLF} & {\cite{Sazonov:2007aa}}  & {$43.40^{+0.28}_{-0.28}$ [$\mathrm{erg\ s}^{-1}$]} & {$3.55\times 10^{-5}$} & {$0.76^{+0.18}_{-0.20}$} & {$2.28^{+0.28}_{-0.22}$} & {}\\
		{}\\
		{FIRST/NVSS 1.4 GHz RLF} & {\cite{Pracy:2016aa}}  & \multicolumn{5}{l}{data points for radio AGN, see Table 2 in \cite{Pracy:2016aa}}\\
		{NVSS 1.4 GHz RLF} & {\citetalias{Mauch:2007aa}}  & {$24.59\pm0.30$ [$\mathrm{W\ Hz}^{-1}$]} & {$7.91 \pm 4.55 \times 10^{-6}$} & {$0.49\pm0.04$} & {$1.27\pm0.18$} & {}
		\enddata
		\tablecomments{Overview of the stellar mass functions and luminosity functions that we use as input for our model. To construct the stellar mass function of green and blue galaxies we use the method and the sample presented in \citetalias{Weigel:2016aa}. $\Phi^{*}$ values are given in units of $\mathrm{Mpc}^{-3}$ and not in units of  $\mathrm{h}^{3} \mathrm{Mpc}^{-3}$ as in \citetalias{Weigel:2016aa}. For the XLF by \protect\citetalias{Ajello:2012aa} we use their non-evolving model fit. For the RLF by \protect\citetalias{Mauch:2007aa} we have converted the normalization $A$ from $\mathrm{Mpc}^{-3}\ \mathrm{mag}^{-1}$ to $\mathrm{Mpc}^{-3}\mathrm{dex}^{-1}$. \cite{Pracy:2016aa} do not report functional fits to their radio luminosity functions. In Figure \ref{fig:rlf_xlf_comp}, for instance, we thus only show their data points.}
\end{deluxetable*}
\subsubsection{Galaxy stellar mass functions}\label{sec:mass_fcts}
Our model is based on the stellar mass functions of red, blue and green galaxies in the local Universe which we determine using the method and sample by \citetalias{Weigel:2016aa}. Stellar mass functions are constructed using SDSS DR7 \citep{York:2000aa, Abazajian:2009aa}  data with apparent magnitudes from the New York Value-Added Galaxy Cataloge (NYU VAGC, \citealt{Blanton:2005aa,Padmanabhan:2008aa}) and stellar mass values from the Max Planck Institute for Astrophysics John Hopkins University catalog (MPA JHU, \citealt{Brinchmann:2004aa,Kauffmann:2003ab}). The sample is restricted to the redshift range $0.02 < z < 0.06$. In the $u-r$ dust \citep{Calzetti:2000aa, Oh:2011aa} and $k$-corrected \citep{Blanton:2007aa} color-mass diagram galaxies lying above 
\begin{equation}\label{eq:red}
u - r = 0.6 + 0.15 \times \log M
\end{equation}
are referred to as being red and galaxies lying below 
\begin{equation}\label{eq:blue}
u-r = 0.15 + 0.15 \times \log M
\end{equation}
are classified as being blue \citepalias{Weigel:2016aa}. Galaxies lying between equations \ref{eq:red} and \ref{eq:blue} are part of the green valley  and referred to as being green. For more detailed information on the sample see \citetalias{Weigel:2016aa}.

\citetalias{Weigel:2016aa} combine the following three independent methods to generate the stellar mass functions of various subsamples: the classical 1/\Vmax\ approach \citep{Schmidt:1968aa}, the non-parametric maximum likelihood method by \citeauthor{Efstathiou:1988aa} (\citeyear{Efstathiou:1988aa}, SWML) and the parametric maximum likelihood technique by \citeauthor{Sandage:1979aa} (\citeyear{Sandage:1979aa}, STY). To estimate the stellar mass completeness \citetalias{Weigel:2016aa} use the method by\cite{Pozzetti:2010aa}. In contrast to previous work, \citetalias{Weigel:2016aa} do not make any a priori assumptions on which subsamples should be fit with a single or a double Schechter function. Instead, a likelihood ratio test is used to determine the better fitting model. 

We use the method and the sample presented in \citepalias{Weigel:2016aa} to compute the stellar mass functions of the combination of optically blue and green and of optically red galaxies. Due to some randomness in the MCMC, the best-fitting STY parameters for the red stellar mass function are not equivalent to the values reported in \citetalias{Weigel:2016aa}. They do however lie well within the errors. Both, the stellar mass functions of blue+green and of red galaxies, are well described by doubled Schechter functions (see equation \ref{eq:double_schechter}). The best-fitting Schechter function parameters are given in Table \ref{tab:mass_lum_fcts}.
\subsubsection{AGN luminosity functions}\label{sec:lum_fcts}
In our model, we use the AGN luminosity functions by \citetalias{Mauch:2007aa} for 1.4 GHz and \citetalias{Ajello:2012aa} for hard X-rays to compare our predictions to observations. 

The luminosity function by \citetalias{Ajello:2012aa} is based on the 60-month \textit{Swift}/Burst Alert Telescope (BAT, \citealt{Gehrels:2004aa, Barthelmy:2005aa}) catalog \citep{Ajello:2008aa,Ajello:2008ab}. To identify optical counterparts and determine redshifts, \citetalias{Ajello:2012aa} use the work by \cite{Masetti:2008aa,Masetti:2009aa,Masetti:2010aa}. The BAT survey is an all-sky survey and the sample used here contains 428 AGN, with a median redshift of 0.029, which were detected in the 15-55 keV energy range. 

Compared to, for instance, an optical selection, the detection of AGN in the X-rays is less biased \citep{Mushotzky:2004aa}, especially against low luminosity AGN. The hard X-ray selection in particular  represents the least biased method to select AGN at the moment (e.g. \citealt{Alexander:2012aa}). Nonetheless, it might be incomplete and could be missing a population of heavily obscured Compton-thick AGN which are too faint to be detected with current facilities \citep{Ricci:2015aa}. These sources would have to be taken into account in future work once the data is available. The X-ray selection of AGN has been found to often select galaxies that are bluer in color than mass matched inactive galaxies and optically selected Seyferts \citep{Koss:2011aa}.

To constrain the hard X-ray luminosity function, \citetalias{Ajello:2012aa} use a maximum likelihood method \citep{Ajello:2009ab}. In analogy to \citetalias{Ajello:2012aa} we ignore their first data point at $\log (L_X/\rm erg\ s^{-1}) = 41.2$ as it is affected by incompleteness and might be contaminated by X-ray binary emission. In the MCMC we thus consider luminosities between $\log (L_{X, \rm min}/\rm erg\ s^{-1})=41.5$ and $\log (L_{X, \rm max}/\rm erg\ s^{-1})=45.6$.

\citetalias{Mauch:2007aa} construct their 1.4 GHz luminosity function using data from the NRAO VLA Sky Survey (NVSS, \citealt{Condon:1998aa}). To determine redshifts and to distinguish between star-forming galaxies and radio AGN they cross match their sample with the 6 degree Field Galaxy Survey (6dFGS, \citealt{Jones:2004aa}). This results in a sample of $\sim8000$ galaxies with a median redshift of 0.043. Galaxies are classified as either star-forming or as AGN using their optical spectra and the method detailed in \cite{Sadler:2002aa}. To construct the luminosity function, \citetalias{Mauch:2007aa} use the classical 1/$V_\mathrm{max}$ approach by \cite{Schmidt:1968aa}. In contrast to the more recent 1.4 GHz AGN luminosity function by \cite{Pracy:2016aa} ($\log (P_{1.4\ \rm GHz, min}/\rm W\ Hz^{-1}) = 21.8$, $\log (P_{1.4\ \rm GHz, max}/\rm W\ Hz^{-1}) = 26.2$, 4.4 orders of magnitude), the \citetalias{Mauch:2007aa} RLF covers 6 orders of magnitude in terms of luminosity. We thus use the \citetalias{Mauch:2007aa} rather than the \cite{Pracy:2016aa} results and set $\log (P_{1.4\ \rm GHz, min}/\rm W\ Hz^{-1}) = 20.4$ and $\log (P_{1.4\ \rm GHz, max}/\rm W\ Hz^{-1}) = 26.4$. We note that the 1.4 GHz luminosity function is not directly coupled to radio jet power \citep{Godfrey:2016aa} and that it is one of the main observables from current and future radio continuum surveys. 

As a reference, we summarize the best-fitting XLF and RLF parameters by \citetalias{Ajello:2012aa},  \citetalias{Mauch:2007aa} and by additional studies in Table \ref{tab:mass_lum_fcts}.

\subsection{MCMC results}\label{sec:mcmc_results}
\begin{figure*}
	\centering
	\begin{minipage}{\textwidth}
		\includegraphics[width=\textwidth]{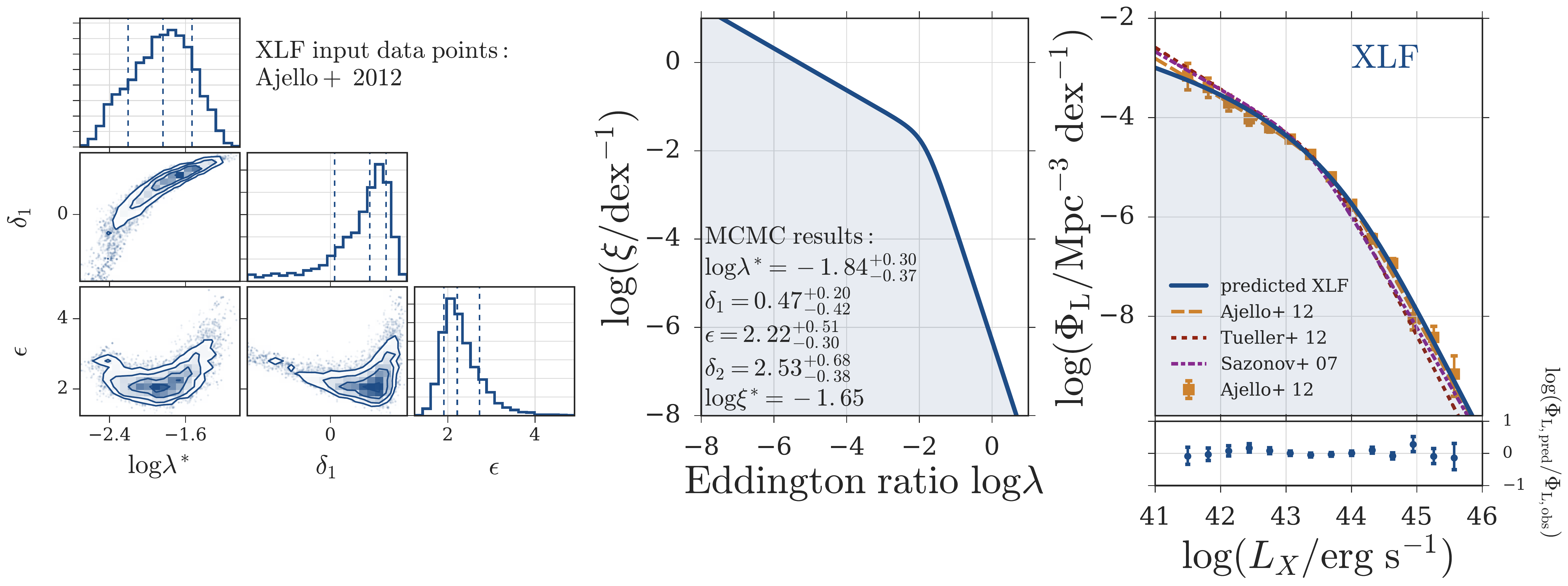}
	\end{minipage}
	\begin{minipage}{\textwidth}
		\includegraphics[width=\textwidth]{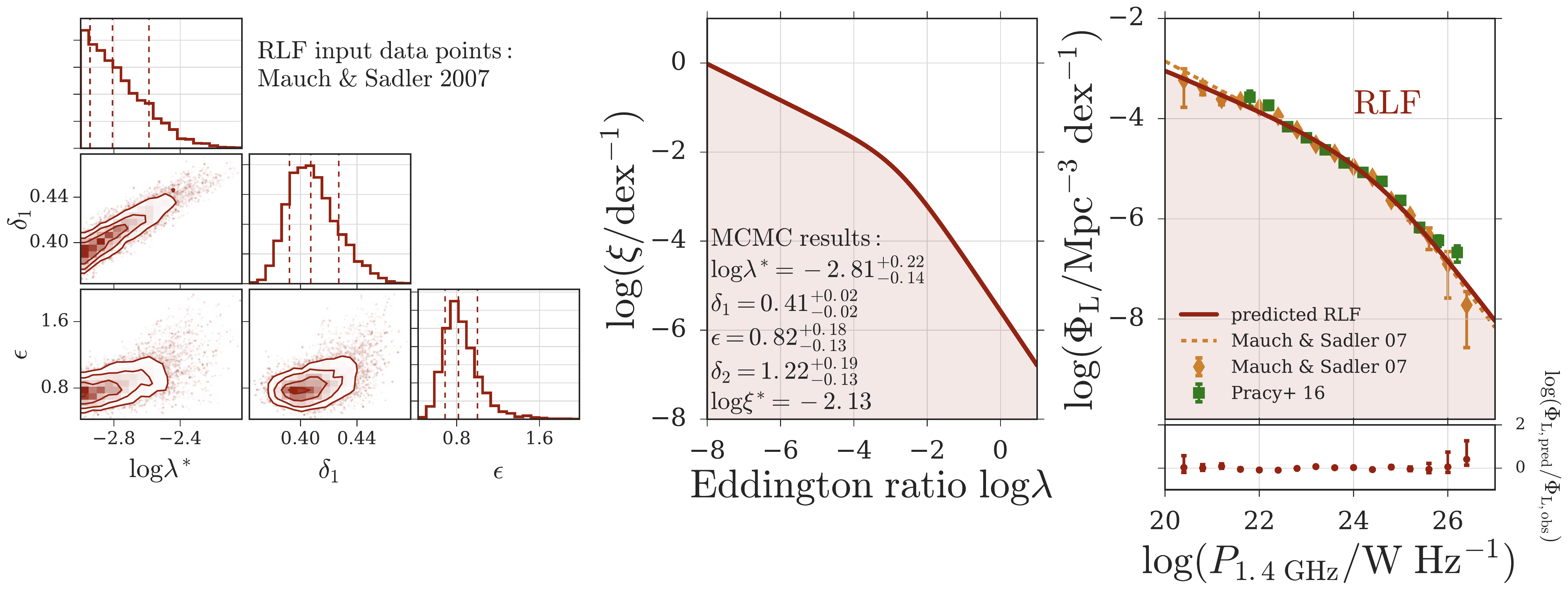}
	\end{minipage}
	\caption{\label{fig:mcmc_results}MCMC results for radiatively efficient (top row) and inefficient (bottom row) AGN. To predict the shapes of the underlying ERDFs for X-ray and radio AGN we use a convolution method based MCMC (see Section \ref{sec:mcmc}) which we run twice: to predict $\xi_{\rm X}$ we compare to the XLF by \protect\citetalias{Ajello:2012aa} and to constrain $\xi_{\rm R}$  we use the RLF by \protect\citetalias{Mauch:2007aa}. In the MCMC we vary $\log \lambda^{*}$, $\delta_1$ and $\epsilon$ ($\delta_1 + \epsilon = \delta_2$). The normalization of the ERDF is not a free parameter. In the left-hand panels we show the three dimensional probability distribution and the marginalized distributions for $\log \lambda^{*}$, $\delta_1$ and $\epsilon$. The central panels summarize the best-fitting ERDFs. The right-hand panels show the predicted luminosity functions compared to observed XLFs and RLFs. Below the right-hand panels we show the residuals of the predicted luminosity functions relative to the observed XLF by \protect\citetalias{Ajello:2012aa} (top row) and the observed RLF by \protect\citetalias{Mauch:2007aa} (bottom row). The left part of this figure was created using the \textsc{corner}\footnote{\url{http://corner.readthedocs.io}} \textsc{python} package \protect\citep{Foreman-Mackey:2013aa}.}
\end{figure*}

We run the MCMC twice. First, we use the stellar mass function of blue and green galaxies, a logarithmic bolometric correction of $k_{\rm bol, X} =\kbolX$ and the XLF by \citetalias{Ajello:2012aa} to find the best-fitting parameters for $\xi_{\rm X}(\lambda)$, the ERDF of radiatively efficient AGN. To constrain $\xi_{\rm X}(\lambda)$ we consider luminosities between $\log (L_{X, \rm min}/\rm erg\ s^{-1})=41.5$ and $\log (L_{X, \rm max}/\rm erg\ s^{-1})=45.6$. Second, for the ERDF of radiatively inefficient AGN, $\xi_{\rm R}(\lambda)$,  we use the red stellar mass function, a logarithmic bolometric correction of $k_{\rm bol, R} =\kbolR$ and the RLF by \citetalias{Mauch:2007aa}. We consider luminosities between $\log (P_{1.4\ \rm GHz, min}/\rm W\ Hz^{-1}) = 20.4$ and $\log (P_{1.4\ \rm GHz, max}/\rm W\ Hz^{-1}) = 26.4$. For both, $\xi_{\rm X}(\lambda)$ and $\xi_{\rm R}(\lambda)$, we use $\log \lambda_{\rm min} = -8$ and $\log \lambda_{\rm max} = 1$. We consider stellar masses between \Mmin\ and \Mmax. This is the stellar mass range that was considered in \citetalias{Weigel:2016aa} and we refrain from extrapolating the Schechter function fits to lower stellar masses. In Section \ref{sec:completeness} we discuss what effect constraining our analysis to this mass range might have on our results. 

When running the MCMC we include an initial guess for $\lambda^{*}$, $\delta_1$ and $\delta_2$ (through $\epsilon$) and constrain the parameters to lie within certain ranges. These initial values and priors are summarized in Table \ref{tab:mcmc_para}. To derive an initial guess for $\lambda^{*}$ we use equation \ref{eq:lstar}, neglecting the correction factor $\log \Delta_\mathrm{L}(\delta_1, \gamma_2)$. For $\lambda^{*}_{\rm X}$ and $\lambda^{*}_{\rm R}$ we use $L^{*}$ from \citetalias{Ajello:2012aa} and \citetalias{Mauch:2007aa} (see Table \ref{tab:mass_lum_fcts}), respectively:
\begin{equation}\label{eq:pred_break_X}
\begin{aligned}
\log \lambda^{*}_\mathrm{X} =& \log (L^{*}_{\mathrm{X}}/\mathrm{erg\ s}^{-1}) - k_\mathrm{bol, X} \\
&- \log (M^{*}_\mathrm{blue+green}/M_\odot) - \mu - \rho \\
=& \breakAjello - (\kbolX) -\logMstarbluegreen -(\mmmu) - \mmrho\\
=& -1.41.\\
\end{aligned}
\end{equation}
\begin{equation}\label{eq:pred_break_R}
\begin{aligned}
\log \lambda^{*}_\mathrm{R} =& \log (L^{*}_{\mathrm{R}}/\mathrm{erg\ s}^{-1}) - k_\mathrm{bol, R}\\
 &- \log (M^{*}_\mathrm{red}/M_\odot) - \mu - \rho \\
=& \breakMaucherg - (\kbolR) -\logMstarred - (\mmmu) - \mmrho\\
=&-2.48.
\end{aligned}
\end{equation}
Constraining the allowed parameter ranges when running the MCMC is especially important for the RLF. Equation \ref{eq:lstar}  shows that the break of the luminosity function $\log L^{*}$ is dependent on the $\log \lambda^{*}$ value. The RLF is shallow and so without a constraint on $\log \lambda^{*}_{\rm R}$, the MCMC tries to fit the entire RLF with a single power law: by pushing $\log \lambda^{*}_{\rm R}$ to low values the entire RLF is fit with the high Eddington ratio end of the ERDF ($\delta_2$) and the break $L^{*}_{\rm R}$ is ignored. To ensure that instead $\delta_1$ and $\delta_2$ are determined by $\gamma_1$ and $\gamma_2$, respectively, we include a stringent constraint for $\lambda^{*}_{\rm R}$ in the MCMC by restricting it to values between -3.0 and-2.0. In contrast to the RLF, the XLF is steeper. The MCMC thus automatically projects $\log \lambda^{*}_{\rm X}$ onto $\log L^{*}_{\rm X}$. We hence use $-3.0 \leq \log \lambda^{*}_{\rm X}\leq 0.0$, testing a wider range of $\log \lambda^{*}$ values than for the RLF.

We show the MCMC results in Figure \ref{fig:mcmc_results}. The upper and lower rows show the results for the ERDF of radiatively efficient and inefficient AGN, respectively.  The left-hand panels show the three dimensional probability distributions and the marginalized distributions for $\lambda^{*}$, $\delta_1$ and $\epsilon$. In the central panels we illustrate the best-fitting ERDFs. The best-fitting ERDF parameters and the corresponding errors are given within the panels. The right-hand panels show a comparison between the best-fitting predicted and the observed AGN luminosity functions. Also included in the right-hand panels are the residuals of the predicted luminosity functions relative to the observed XLF by \citetalias{Ajello:2012aa} (upper panel) and the observed RLF by \citetalias{Mauch:2007aa} (lower panel).

The best-fitting broken power law parameters for the ERDF of radiatively efficient AGN are:
\begin{equation}
\begin{aligned}
\log \lambda^{*}_{\rm X} &= -1.84^{+ 0.30}_{- 0.37}\\
\delta_1 &= 0.47^{+ 0.20}_{- 0.42}\\
\delta_2 &= 2.53^{+ 0.68}_{- 0.38}\\
\log \xi^{*} &= -1.65\\
(\epsilon &= 2.22^{+ 0.51}_{- 0.30})\\
\end{aligned}
\end{equation}
For radiatively inefficient AGN we find:
\begin{equation}
\begin{aligned}
\log \lambda^{*}_{\rm R} &= -2.81^{+ 0.22}_{- 0.14}\\
\delta_1 &= 0.41^{+ 0.02}_{- 0.02}\\
\delta_2 &= 1.22^{+ 0.19}_{- 0.13}\\
\log \xi^{*} &= -2.13\\
(\epsilon &= 0.82^{+ 0.18}_{- 0.13})\\
\end{aligned}
\end{equation}

As we mentioned in Section \ref{sec:mcmc}, the normalization of the ERDF $\xi^{*}$ is not a free parameter in the MCMC. Instead we predict the luminosity functions based on a $\xi(\lambda)$ that is normalized to have an integral of 1. We then rescale the predicted $\Phi_{\rm L}(L)$ so that the integral over the predicted matches the integral over the observed luminosity function. This rescaling factor then allows us to constrain $\xi^{*}$.
	
Given our $\log \lambda_{\rm min}$ and $\log \lambda_{\rm max}$ values, $\xi^{*}_{\rm X}$ and $\xi^{*}_{\rm R}$ imply active black fractions of (see Section \ref{sec:flickering}) $\sim 17$ and $\sim 1$, respectively. For radiatively inefficient AGN this implies that every black hole is assigned an Eddington ratio. The unintuitive value of $\sim 17$ for radiatively efficient AGN is an artifact of our $\log \lambda_{\rm min}$ choice. We chose to use the same $\log \lambda_{\rm min}$ values for radiatively efficient and inefficient AGN to allow for a direct comparison between the resulting ERDFs. However, this shows that to give a physical meaning to the active black hole fraction of X-ray AGN we would have to choose a higher $\log \lambda_{\rm min}$ value. Equivalently we could introduce a cut-off in $\xi_{\rm X}(\lambda)$ at the low $\lambda$ end.

Figure \ref{fig:mcmc_results} shows that the shapes of the XLF and the RLF are reflected in the corresponding ERDFs. $\xi_{\rm X}(\lambda)$ and $\xi_{\rm R}(\lambda)$ have similar $\delta_1$ values, yet $\delta_2$ of  $\xi_{\rm X}(\lambda)$ is significantly steeper than $\delta_2$ of $\xi_{\rm R}(\lambda)$.

For the XLF $\delta_1$ = \deltaXa\ and hence $\delta_1 \gtrsim -(\alpha_1 + 1)$. Here $\alpha_1$ is the slope of the stellar mass function which dominates $\Phi(M)$ of blue and green galaxies. The ERDF does thus determine the faint end of the XLF (see equation \ref{eq:gamma1}). The marginalized distribution for $\delta_1$ has a tail towards lower $\delta_1$ values. This shows that if $\delta < -(\alpha + 1)$, $\delta_1$ can no longer be constrained since the low mass end of the stellar mass function determines $\gamma_1$. Similarly, the marginalized distribution of $\epsilon$ has a tail towards higher values. As we discussed in Section \ref{sec:mcmc}, the bright end of the luminosity function does not only depend on $\delta_2$, i.e. $\delta_1 + \epsilon$, but also  on $\sigma$, the assumed scatter in the stellar to black hole mass conversion. For steep $\delta_2$ values $\gamma_2$ is determined by the high mass end of the black hole mass function (see Figure \ref{fig:sigma}) and $\epsilon$ can no longer be constrained.

A similar trend can be seen in the marginalized $\epsilon$ distribution for $\xi_{\rm R}(\lambda)$. The stellar mass function of red galaxies is dominated by $\alpha_2 = \alphabred$. Since $\delta_1 >> -(\alpha_2 + 1)$, the $\xi_{\rm R}(\lambda)$ $\delta_1$ is better constrained than the $\xi_{\rm X}(\lambda)$ $\delta_1$. The marginalized $\delta_1$ distribution for $\xi_{\rm R}(\lambda)$ lacks the tail towards lower values. Due to the stringent constraints, the marginalized $\log \lambda^{*}_{\rm R}$ distribution is cut off at its maximum.

In our model the error on the input stellar mass function is not taken into account. Table \ref{tab:mass_lum_fcts} summarizes the errors on $M^{*}$, $\Phi^{*}$ and $\alpha$. $M^{*}$ which is degenerate with $\lambda^{*}$ is well constrained for both the blue+green and the red population. As we are not constraining $\xi^{*}$ in the MCMC, the errors on $\Phi^{*}_1$ and $\Phi^{*}_2$ are not taken into account. Furthermore, as we discussed, the faint ends of the XLF and the RLF are dominated by $\delta_1$ and not the slopes of the stellar mass function. The errors on $\alpha_1$ for the blue and green stellar mass function and on $\alpha_2$ for the red stellar mass function thus do not affect our results significantly.

Equations \ref{eq:pred_break_X} and \ref{eq:pred_break_R} show that $\lambda^{*}_{\rm X}$ and $\lambda^{*}_{\rm R}$ are degenerate with the bolometric corrections $k_\mathrm{bol, X}$ and $k_\mathrm{bol, R} $. As we have discussed in Section \ref{sec:bol_corr_radio}, our choice of $k_{\rm bol, R} = \kbolR$ is subject to significant uncertainties. Due to it being constant, changing the assumed $k_{\rm bol, R}$ value causes a shift of radio ERDF, but not a change in the $\xi_{\rm R}(\lambda)$ shape. We have thus shown that the observed RLF is consistent with a mass independent ERDF and a constant bolometric correction. We have determined the shape of $\xi_{\rm R}(\lambda)$, but we have not constrained the absolute value of $\log \lambda^{*}_{\rm R}$. 

\subsection{The ERDF mass dependence}\label{sec:erdf_mass}
Figure \ref{fig:mcmc_results} shows that both the observed XLF and RLF are consistent with the simplest form of an Eddington ratio distribution, a mass independent one. However, mass dependent ERDF models have been used by, for example, \cite{Schulze:2015aa} and \cite{Bongiorno:2016aa}. We now investigate the effect of a mass dependent ERDF and quantify the allowed $\xi$ variation with $\log M$. We  use the random draw method (see Section \ref{sec:random_draw}) since it allows us to change the ERDF CDF as a function of stellar mass. We separately vary $\log \lambda^{*}$, $\delta_1$ and $\delta_2$, while keeping the other two parameters constant. Our tests assume a linear $\log M$ dependence, for instance $\log \lambda^{*} = a \times \log M + b$. Section \ref{sec:erdf_mass_app} contains the details of the test. Figures  \ref{fig:mass_dep_break_xray} to \ref{fig:mass_dep_slopes_radio} show the results.

In summary, our analysis in Section \ref{sec:erdf_mass_app} shows that a mild mass dependence of either $\log \lambda^{*}$ or $\delta_2$ is consistent with the observed AGN luminosity functions. For the RLF and the XLF, $\lambda^{*}$ can be increased by an order of magnitude per order of magnitude in stellar mass. 
 
For $\xi_{\rm X}(\lambda)$ $\delta_2$ can also be varied by up to $\pm 1$ per magnitude in stellar mass. $\xi_{\rm R}(\lambda)$ only allows $\delta_2$ values that  decrease by up to 1 per magnitude in stellar mass. For both ERDFs $\delta_1(\log M)$ models lead to luminosity functions that are no longer power law shaped and do not resemble the observed $\Phi_{\rm L}(L)$. In Section \ref{sec:completeness} we show that for our chosen stellar mass range, the RLF and the XLF are dominated by galaxies with $M\sim M^{*}$. This also affects the mass dependent models which we consider here. A mass dependence of $\log \lambda^{*}$ or $\delta_2$ only leads to agreement with the observations if at $M^{*}$ the mass dependent ERDF parameter is either equal to its best-fitting mass independent value. Galaxies with $M \sim M^{*}$ are hence still convolved with the best-fitting mass independent ERDFs which we determined in Section \ref{sec:mcmc_results}.

We conclude that our model predicts X-ray and radio luminosity functions that are consistent with the observations. Making the most straightforward assumptions possible and, for instance, ignoring all complexities that might be affecting $k_{\rm bol, R}$, we are able to derive these AGN luminosity functions from the galaxy population. The shapes of the XLF and the RLF can be traced back to characteristic, broken power law shaped, mass independent ERDFs. We explored a first order perturbation to the model and quantified the mild mass dependence of the ERDF that is still consistent with the data. We showed that only certain $\xi(\lambda, \log M)$ models are allowed. Specifically, the ERDFs have to resemble our best-fitting mass independent ERDFs for $M \sim M^{*}$ galaxies. A more extreme dependence of $\xi(\lambda)$ on $\log M$ leads to a deviation from the broken power law shape of the observed XLF and RLF. Choosing the simplest model possible and not making any assumptions about the mass dependence of $\xi(\lambda)$, we carry on to discuss the implications of mass independent ERDFs. 

\section{Implications}\label{sec:implications}
\begin{figure*}
	\includegraphics[width=\textwidth]{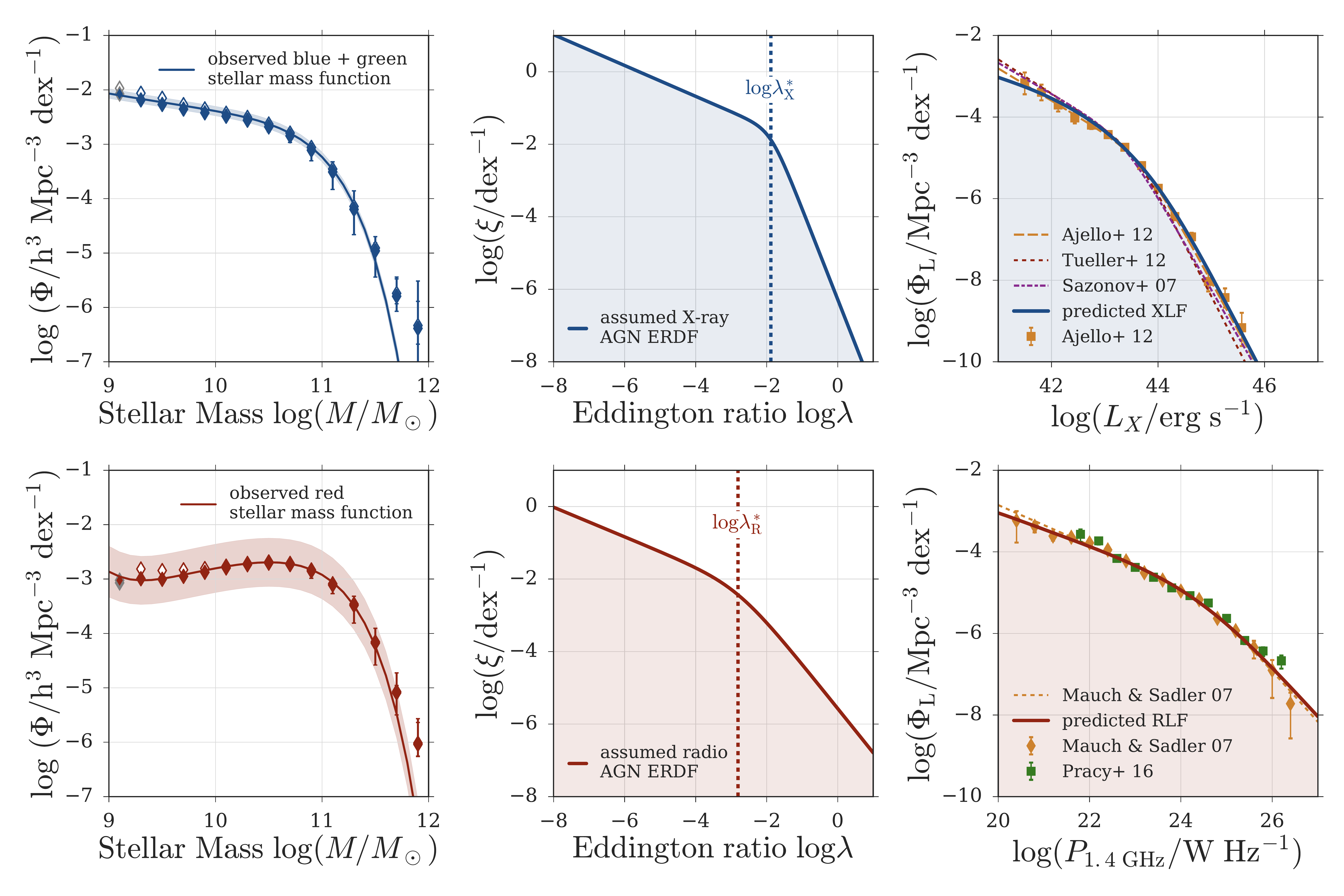}
	\caption{\label{fig:summary}Summary of our results. We use a simple model to predict the shape of the Eddington ratio distribution functions (ERDFs, central panel) for radiatively efficient and inefficient AGN. We assume that radiatively efficient and inefficient AGN are predominantly hosted by blue+green and red galaxies, respectively and thus use these mass function as input for our model. Additionally, we assume that the ERDFs are mass independent and broken power law shaped and use constant bolometric corrections. The right-hand panels show that based on these simple assumptions we are able to predict X-ray and radio luminosity functions which are consistent with observations.}
\end{figure*}

After showing that our simple model is capable of reproducing the observed AGN luminosity functions, we now discuss the implications of our results. We return to the previously postulated need for two ERDFs and examine the impact of the input stellar mass functions. We discuss the AGN fraction (Section \ref{sec:agn_frac}), AGN feedback and quenching (Section \ref{sec:feedback}). Furthermore, we provide a possible physical interpretation for our results (Section \ref{sec:picture}) and discuss the effect of black hole populations with different ERDFs (Section \ref{sec:sec_pop}). 
\subsection{The need for two ERDFs and the impact of the input stellar mass functions}\label{sec:two_erdfs}
When we introduced our assumption and our model in Section \ref{sec:model} we postulated the need for two ERDFs: one for radiatively efficient and one for radiatively inefficient AGN. After having discussed our results and how the predicted luminosity function depends on the assumed ERDF (see Section \ref{sec:pred_lf}), we now revisit this fundamental assumption.

The fact that there is no global ERDF which describes the Eddington ratio distribution of both, radiatively efficient and inefficient AGN, is primarily due to the significantly different shapes of the XLF and the RLF. With a large difference between $\gamma_1$ and $\gamma_2$ the XLF is steep with a clear break at $L^{*}$. Compared to the XLF, the RLF is shallow and only has a weak break. 

Figures \ref{fig:mcmc_results} and \ref{fig:summary} show that $\xi_{\rm X}(\lambda)$ and $\xi_{\rm R}(\lambda)$ differ in their $\lambda^{*}$ values. However, $\lambda^{*}$ is degenerate with the bolometric correction (see equation \ref{eq:lstar}). As we have discussed above, due to the large uncertainties that affect $k_{\rm bol, R}$ we do not claim to have constrained $\lambda^{*}_{\rm R}$. We thus do not use the difference in $\lambda^{*}$ to argue for the need of two ERDFs. Besides the difference in $\lambda^{*}$, $\xi_{\rm X}(\lambda)$ and $\xi_{\rm R}(\lambda)$ have different $\delta_2$ slopes. $\delta_2$ determines $\gamma_2$ and is thus steeper for the XLF than for the RLF. Both the XLF and the RLF have a faint end that is steeper than the respective input stellar mass functions. $\gamma_1$ is hence determined by $\delta_1$. Unlike $\delta_2$, the $\delta_1$ values that we determine for $\xi_{\rm X}(\lambda)$ and $\xi_{\rm R}(\lambda)$ are consistent with each other. We conclude that we are unable to reproduce both the steep XLF and the shallow RLF with a single, global ERDF.

As both the XLF and the RLF are steeper than their respective input stellar mass functions, the blue+green and the red stellar mass functions do not significantly affect the luminosity function shapes. Their $M^{*}$ values have an effect on $\lambda^{*}_{\rm X}$ and $\lambda^{*}_{\rm R}$, their $\Phi^{*}$ values impact $\xi^{*}_{\rm X}$ and $\xi^{*}_{\rm R}$. Nonetheless $\gamma_1$ and $\gamma_2$ are unaffected by $\alpha$. 

Our results thus show that the different shapes of the XLF and the RLF can be accounted for by using different ERDFs. The fact that $\xi_{\rm X}(\lambda)$ and $\xi_{\rm R}(\lambda)$ differ in shape is however not due to us using different input stellar mass functions for the X-ray and the radio AGN populations. Using the same stellar mass function, for instance the mass function of the entire galaxy sample, would still result in a steep $\xi_{\rm X}(\lambda)$ and a shallow $\xi_{\rm R}(\lambda)$. This is a result of our analysis and not an assumption that we could have made a priori. We hence do not claim to have shown that AGN in blue+green and red galaxies have different ERDFs. Instead our results imply that X-ray and radio selected AGN must have different ERDF shapes. 

\subsection{AGN fraction}\label{sec:agn_frac}
A mass independent ERDF implies that the fraction of galaxies that host AGN, the AGN fraction, is mass independent. Galaxies which host AGN are randomly drawn from the galaxy population. Due to a flux or luminosity limit the AGN fraction can however be observed to be stellar mass or black hole mass dependent: At low black hole masses only AGN with high Eddington ratios will be bright enough to lie above the flux or luminosity limit. The AGN fraction will thus be low. At high black hole masses, we will be able to observe AGN with a range of Eddington ratios since all of them are bright enough to be detected or included in the sample. Compared to low black hole masses, the AGN fraction at high black hole masses will hence be higher. So, even though the AGN fraction is intrinsically mass independent, a luminosity or flux limit will make it appear mass dependent. Previously, this has for instance been discussed by \cite{Aird:2012aa}.

We have shown that to zeroth order the observed XLF and RLF are consistent with mass independent ERDFs which implies mass independent AGN fractions. In Section \ref{sec:erdf_mass} we quantified the allowed mass dependence of $\xi_{\rm X}(\lambda)$ and $\xi_{\rm R}(\lambda)$. A mild mass dependence of the ERDFs is consistent with the data and would manifest itself in a mass dependent AGN fraction. Previous studies have reported AGN fractions that vary as a function of stellar mass \citep{Best:2005aa, Kauffmann:2008aa, Hickox:2009aa, Silverman:2009aa, Xue:2010aa,  Haggard:2010aa, Tasse:2011aa, Janssen:2012aa, Hernan-Caballero:2014aa, Williams:2015aa}. However, when interpreting these results it is important to ensure that a possible selection effect has been accounted for.

\subsection{AGN feedback and quenching}\label{sec:feedback}
AGN with radiatively efficient and inefficient accretion are thought to play different roles in the quenching of star formation. AGN with radiatively efficient accretion are considered to be the cause of quenching \citep{Sanders:1988aa, Di-Matteo:2005aa, Cattaneo:2009ab,  Fabian:2012aa}. AGN with radiatively inefficient accretion may be keeping their host galaxies quenched \citep{Croton:2006aa, Bower:2006aa,  Springel:2006aa,  Somerville:2008aa}. 

In their phenomenological model, \cite{Peng:2010aa, Peng:2012aa} showed that the stellar mass function of red and quiescent galaxies can be reproduced by splitting the quenching mechanism into two distinct processes. `Mass quenching' is a mass dependent, but environment independent mechanism. `Environment quenching' summarizes mass independent, but environment dependent processes. While environment quenching could, for example, be associated with ram pressure stripping \citep{Gunn:1972aa} or strangulation \citep{Larson:1980aa, Balogh:2000aa}, AGN feedback could be the physical origin of mass quenching. 

One could imagine a simplistic model in which mass quenching is caused by more AGN activity and thus more AGN feedback at high stellar masses. Massive galaxies could for example be more likely to host AGN with particularly high Eddington ratios. We have shown that the observations are consistent with a mass independent $\xi_{\rm X}(\lambda)$ model. This is inconsistent with this simplest form of mass quenching. We also discussed models which allow for a mild mass dependence of the ERDF. The linear increase of $\log \lambda^{*}$ with $\log M$ of up to 1 magnitude per magnitude in stellar mass is however inconsistent with the exponential increase of the mass quenching probability which the \cite{Peng:2010aa, Peng:2012aa} model requires. Nonetheless, more sophisticated models could link AGN feedback and mass quenching. For instance, regardless of a mass independent AGN fraction, AGN feedback in low mass galaxies could be more effective than in high mass galaxies due to a shallower potential well. Furthermore, AGN feedback might still be a necessary but not sufficient condition for the quenching of star formation. 

For AGN with radiatively inefficient accretion, the mass independent ERDF implies that galaxies of all masses are equally likely to host radio AGN of high and low Eddington ratios. Massive ellipticals contain massive black holes and are thus radio  bright, even for low Eddington ratios. Flux limited radio surveys bias us towards these massive galaxies. Our results imply that radio activity is possible at all stellar and black hole mass scales and not just in massive ellipticals. Furthermore, maintenance mode can occur on all mass scales, if reaching a certain  Eddington ratio is a sufficient condition. 

\subsection{The physical interpretation of two mass independent ERDFs}\label{sec:picture}
After having discussed the need for two ERDFs, the implications of mass independent ERDFs for the AGN fraction and for the quenching of star formation, we now consider what might cause X-ray and radio selected AGN to have different $\xi(\lambda)$ shapes.

\begin{itemize}
	\item {\textit{host galaxies:}
		As we discussed above, X-ray selected AGN tend to be found in optically blue and green galaxies, whereas radio selected AGN are often hosted by optically red galaxies \citep{Silverman:2009aa, Smolcic:2009aa, Treister:2009ab, Smolcic:2009ab,  Koss:2011aa,Schawinski:2009ab, Hickox:2009aa, Janssen:2012aa, Rosario:2013ab, Goulding:2014aa, Ching:2017aa}. Star forming and quiescent galaxies are thought to be accreting in the cold and hot phases \citep{Shabala:2008aa}. A first assumption could thus be that AGN in red galaxies have to have a different ERDF, since they contain less gas than blue and green galaxies. Black hole growth is however not limited by the amount of available gas, but by the loss of angular momentum \citep{Jogee:2006aa}\footnote{For example, a $M^{*}$ galaxy of $10^{10.8} M_\odot$ requires a SFR of $\sim 5 M_\odot/\rm yr$ to be on the main sequence. At the corresponding black hole mass of $10^{8} M_\odot$, a black hole reaches the break of the X-ray luminosity function for $\lambda \sim 0.03$ and an accretion rate of only $0.1 M_\odot/\rm yr$.}.}
	\item {\textit{stellar mass: }
		We assumed that both, $\xi_{\rm X}(\lambda)$ and $\xi_{\rm R}(\lambda)$, are mass independent and showed that this choice of ERDFs is consistent with the observations. Stellar or black hole mass is hence unlikely to be the physical property that determines the ERDF shape. }
	\item {\textit{large scale properties: }
		Large scale properties, such as merger events, environment or halo mass, might be promoting higher or lower Eddington ratios. However it remains unclear how these physical conditions would be transmitted over multiple orders of magnitude in distance, from the galaxy outskirts to the central black hole accretion disc (see e.g. \citealt{Alexander:2012aa}). The ERDF shape is thus unlikely to be set by large scale galaxy properties.}
	\item {\textit{accretion process:}
		On small scales, it might be the accretion process itself that explains the difference between the two galaxy populations and determines the Eddington ratio distribution. For accretion rates $L/L_{\rm Edd} \lesssim 0.3$, we would expect accretion via a classical \cite{Shakura:1973aa} disc. At $\lambda \lesssim 0.01$  accretion via advection-dominated accretion flows \citep{Narayan:1994aa, Narayan:2008aa} becomes dominant. If the shape of the ERDF, and especially $\lambda^{*}$, is purely set by this change between accretion modes, we would expect $\xi_{\rm X}(\lambda)$ and $\xi_{\rm R}(\lambda)$ to be similar in shape.}
\end{itemize}

None of the aforementioned processes is likely to solely determine the ERDF shape. We hence conclude that the difference between $\xi_{\rm X}(\lambda)$ and $\xi_{\rm R}(\lambda)$ could be caused by processes on intermediate scales: the shape of the ERDFs could be set by how efficiently the gas is driven from the inner parsecs to the accretion disc, by the properties of the gas itself or by both.

According to \cite{Hardcastle:2007aa}, cold and hot gas lead to radiatively efficient and inefficient accretion, respectively.  The host galaxies of X-ray selected AGN are thus likely to contain cold gas, which, due to its temperature, has a clumpy structure. Radio AGN host galaxies are likely to have reservoirs of smooth, hot gas. Radio AGN could therefore be fuelled by a continuous Bondi-like accretion flow. For X-ray selected AGN the clumpy structure of the gas could cause short, episodic bursts of high accretion rate.

Compared to $\xi_{\rm R}(\lambda)$, $\xi_{\rm X}(\lambda)$ has a steeper slope $\delta_2$. As $\lambda$ increases we expect the radiative efficiency to decrease once we approach the Eddington limit. This is caused by the increasing significance of the radiation pressure and the transition from a thin to a slim accretion disc (e.g.  \citealt{Abramowicz:1988aa, Laor:1989aa, Netzer:2014aa, Sc-adowski:2016aa}). This change in radiative efficiency could be connected to the steep high Eddington ratio slope that we observe for the ERDF of blue+green galaxies. A physical limit to the clump size of the cold gas could provide an alternative explanation.  In contrast, the shallow $\delta_2$ slope of the radio AGN ERDF could show that the continuous hot gas accretion flow is not affected by such limits.

Even though we cannot constrain $\log \lambda^{*}_{\rm R}$ well, it is clear that the two ERDFs overlap at least to some degree. Within the overlapping $\lambda$ range, a galaxy can thus be part of both, the radio and the X-ray population. At high Eddington ratios, such a galaxy would be bright in the radio and the X-rays and we would identify it as a radio-loud quasar. At lower $\lambda$ values we would classify such a galaxy as a HERG. The fraction of AGN that are detected both in the X-rays and the radio does not only depend on the overlapping $\lambda$ range, but also $\xi^{*}_{\rm X}$ and $\xi^{*}_{\rm R}$.

\subsection{Black hole populations with different ERDFs}\label{sec:sec_pop}
\begin{figure*}
	\includegraphics[width=\textwidth]{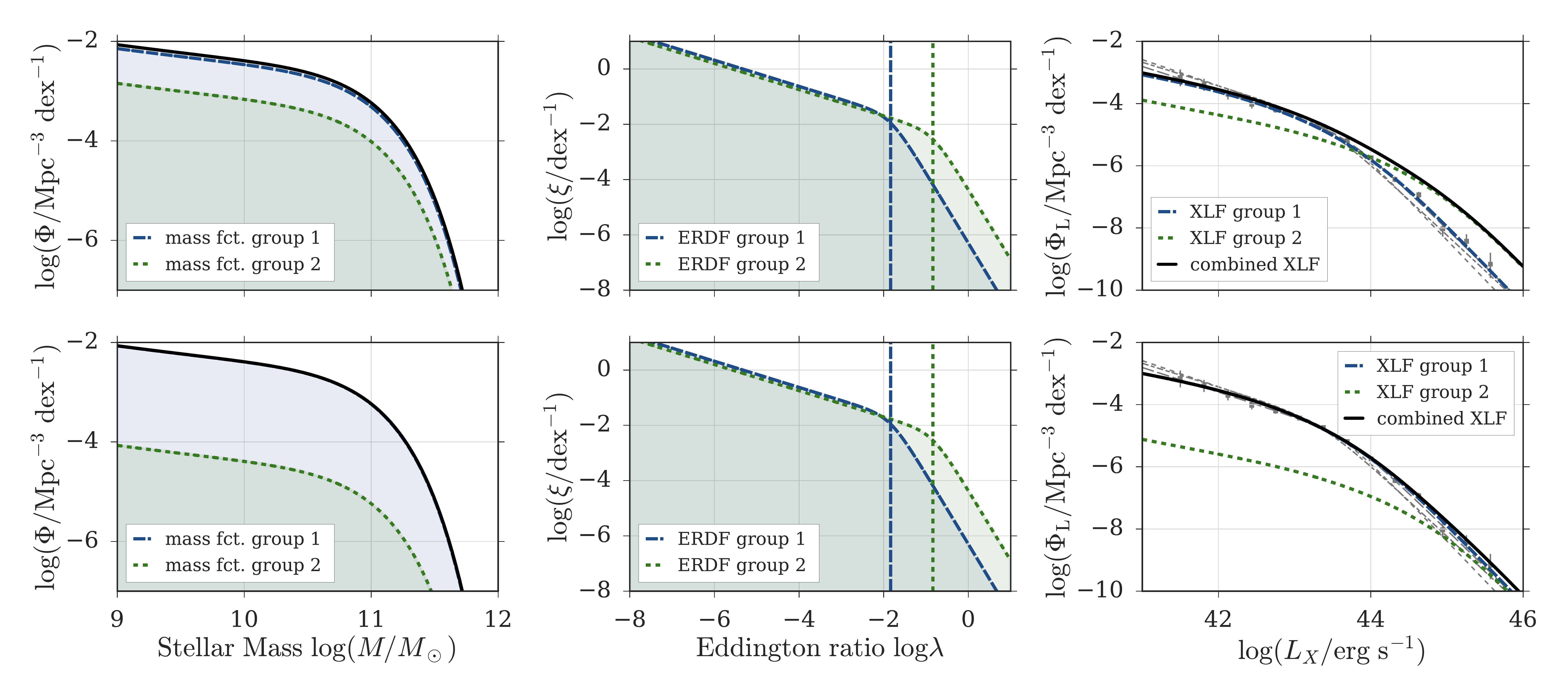}
	\caption{\label{fig:sec_pop}Example for two black hole populations with different ERDFs. We assume that the population of X-ray AGN is made up of two black hole groups, group 1 and group 2. In the top row their number densities are comparable, in the bottom row group 2 contains significantly fewer galaxies than group 1. Both groups have the same stellar mass function shape (left-hand panels), but their ERDFs have significantly different $\lambda^{*}$ values (central panels, vertical dashed lines). The ERDF of group 1 corresponds to $\xi_{\rm X}(\lambda)$, the ERDF that best reproduces the observed XLF (see Section \ref{sec:mcmc_results}). The predicted luminosity functions for group 1 and 2 are shown in the right-hand panels. The combined XLF is shown as a solid black line. As a reference the observed XLFs are shown in grey. The figure illustrates that if all X-ray AGN consist of groups with different ERDFs, we have to recover  $\xi_{\rm X}(\lambda)$ when summing over all ERDFs and weighing them by the space density of AGN that they apply. In the first case, shown in the top row, this is not the case as the groups have comparable space densities and group 1 was given our best-fitting X-ray ERDF. The weighted sum of these two ERDFs does not resemble $\xi_{\rm X}(\lambda)$ and we are thus unable to recover the observed XLF. In the second case, shown in the bottom row, the space density of group 2 is low and so the observed XLF can still be recovered as the weighted sum of the ERDFs of group 1 and group 2 does not deviate significantly from $\xi_{\rm X}(\lambda)$.}
\end{figure*}

In Section \ref{sec:mcmc_results} we determined the ERDF of all X-ray and all radio AGN. However, within these two populations there might be groups of AGN that follow different ERDF shapes. For instance, the ERDF of AGN in galaxies that are undergoing a galaxy merger might differ from the overall $\xi_{\rm X}(\lambda)$. While such subpopulations are possible, $\xi_{\rm X}(\lambda)$ and $\xi_{\rm R}(\lambda)$, which we determined in Section \ref{sec:mcmc_results}, have to be recovered when we consider the sum of all X-ray AGN and radio AGN. For simplicity we focus our discussion on the ERDF of radiatively efficient AGN. The same arguments can be applied to $\xi_{\rm R}(\lambda)$ though.

We consider two possibilities for how the ERDFs of such subpopulations might differ from the overall, general ERDF. First, $\xi(\lambda)$ of a subpopulation might still be power law shaped, but might have $\lambda^{*}$, $\xi^{*}$, $\delta_1$ and $\delta_2$ values that differ from $\xi_{\rm X}(\lambda)$. Second, $\xi_{\rm X}(\lambda)$ could be split into two subpopulations: one group of AGN that is assigned $\lambda$ values from the high Eddington ratio end of $\xi_{\rm X}(\lambda)$ and one group that covers the low $\lambda$ range of the best-fitting X-ray ERDF. 

In the simplest case $\xi_{\rm X}(\lambda)$ could be made up of groups of AGN that have the same $\lambda^{*}$, $\delta_1$ and $\delta_2$ values as $\xi_{\rm X}(\lambda)$, but differ in their normalization $\xi^{*}$. Physically this would imply that the active black hole fraction varies among certain groups of galaxies.

Alternatively, $\xi_{\rm X}(\lambda)$ could consist of groups of AGN that have broken power law shaped ERDFs with different $\lambda^{*}$, $\delta_1$ and $\delta_2$ values. Figure \ref{fig:sec_pop} shows a generic example: We assume that the population of X-ray selected AGN consist of two groups. The first group has an ERDF which has the same $\lambda^{*}$, $\delta_1$ and $\delta_2$ values as $\xi_{\rm X}(\lambda)$. The ERDF of group 2 has the same slopes as the ERDF of group 1, but a significantly higher $\lambda^{*}$ value. We consider two cases. In the top row of Figure \ref{fig:sec_pop} we show the stellar mass functions, the ERDFs and the resulting individual and the total XLF for the assumption that group 1 and group 2 have comparable space densities. The panels in the bottom row show the results for the case in which group 2 has a significantly lower space density than group 1. In the right-hand panels we show the observed XLFs in grey.

Figure \ref{fig:sec_pop} shows that in the case where group 1 and group 2 have comparable space densities, the significantly higher $\lambda^{*}$ value of group 2 leads to the combined XLF no longer being consistent with the observations. This is not the case in the bottom row, where the space density of group 2 is significantly lower than the space density of group 1. The very few AGN in group 2 having a higher $\lambda^{*}$ value does not alter the shape of the combined XLF. 

For Figure \ref{fig:sec_pop} we assumed that the AGN in groups 1 and 2 are drawn from host galaxy populations with similar mass function shapes. Furthermore, we assumed that the active black hole fraction among both groups is the same. If for group 2 we would have assumed a stellar mass function with a significantly lower $M^{*}$, this would have accounted for the higher $\lambda^{*}$ value. Similarly, assuming that group 2 has a lower active black hole fraction that group 1, would have diminished the deviation from the observations which we highlight in the upper panels.

Figure \ref{fig:sec_pop} shows that within the population of X-ray AGN groups of AGN with ERDFs that differ from $\xi_{\rm X}(\lambda)$ can exist. By summing over the ERDFs of all subpopulations and weighing them by the space density of AGN that they apply to, we have to recover $\xi_{\rm X}(\lambda)$, as it represents the average ERDF of all X-ray AGN. Introducing subpopulations with different ERDFs still leads to consistency with the observations, if the space density of these AGN is low. However, if we model the population of all X-ray AGN as consisting of, for instance, two groups of comparable space densities, it is crucial that the sum of their weighted ERDFs reproduces $\xi_{\rm X}(\lambda)$. For the example shown in the first row of Figure \ref{fig:sec_pop} this would require changing the ERDF of group 1. 

Instead of having subpopulations with power law shaped ERDFs, $\xi_{\rm X}(\lambda)$ could also be split into different parts at a given $\lambda$ value. In the most extreme case, one group of X-ray AGN could have an ERDF which covers $\lambda > \lambda^{*}$. With a cut-off at $\lambda^{*}$, the second group could dominate the low $\lambda$ end. With similar stellar mass functions, the sum of these two AGN populations would reproduce $\xi_{\rm X}(\lambda)$ and the observed XLF.

Such a scenario is possible mathematically, would not break the mass independence of the ERDF and cannot be excluded by our model. A split of the ERDF close to $\lambda^{*}$ implies that some AGN are constantly growing, whereas others remain at low accretion rates. If AGN are unlikely to move between the two groups, we expect their to be a population of non-flickering AGN \citep{Schawinski:2015aa} that constantly exhibit high accretion rates. Furthermore, runaway growth for the high $\lambda$ group of AGN and a total stalling of growth among the low $\lambda$ group is likely to break local scaling relations.  If AGN move among the two populations, for instance because high accretion rates are caused by temporary events such as starbursts or galaxy mergers, we return to our previous discussion of what shapes the ERDF. While determining if certain conditions promote high accretion rates is an interesting question, there is no need to think of the ERDF as being made up of strictly separate parts. Furthermore, the amount of time that AGN spend at high and low $\lambda$ values is directly reflected by the ERDF shape.

Our model does not allow us to constrain the ERDF of such subpopulations. However, by constraining $\xi_{\rm X}(\lambda)$ and $\xi_{\rm R}(\lambda)$, we have determined the average ERDF of all X-ray and radio AGN. We stress that recovering the XLF and the RLF does not require the introduction of subpopulations with different ERDFs and the necessary additional assumptions. Instead, the multifaceted populations of X-ray and radio AGN can be described by two separate, simple and global ERDFs. Dissecting these populations into their subpopulations and determining how they contribute to their respective ERDFs, represents the first order perturbation to our zeroth order model and requires additional observational constraints.

\section{Discussion}\label{sec:discussion}
Below we discuss alternative ERDF shapes (Section \ref{sec:erdf_shapes}) and summarize the caveats of our model (Section \ref{sec:caveats}). We also compare our results to previous work  (Section \ref{sec:compare}) and end with an outlook regarding future work and surveys (Section \ref{sec:future}). 

\subsection{Alternative ERDF shapes}\label{sec:erdf_shapes}
A broken power law shaped ERDF resolves the discrepancy between the exponential decline at the high mass end of the stellar mass function and the power law shaped bright end of the AGN luminosity function. Furthermore, a broken power law shaped ERDF allows us to control both, $\delta_1$ and $\delta_2$, the high and low Eddington ratio ends of the distribution. Other studies have used alternative functional forms \citep{Kollmeier:2006aa, Kauffmann:2009aa, Hopkins:2009aa, Cao:2010aa, Aird:2012aa, Bongiorno:2012aa, Nobuta:2012aa, Conroy:2013aa, Aird:2013ab, Veale:2014aa, Hickox:2014aa, Schulze:2015aa, Trump:2015aa, Jones:2016aa, Bongiorno:2016aa}.  With an exponential cut-off at high $\lambda$, a Schechter function ERDF provides a natural limit to the number of high and super-Eddington ratio sources. However it causes a steep decline at the bright end of the luminosity function.  Besides Schechter function shaped ERDFs, log normal shaped distributions have been used in the literature. We adjust the MCMC method and fit the observed XLF and RLF with Schechter and log normal shaped ERDFs to investigate the challenges that arise from these $\lambda$ distributions. The details of how we adjust the MCMC are given in Section \ref{sec:erdf_shapes_app} and the results are shown in Figures \ref{fig:schechter_normal_xray} and \ref{fig:schechter_normal_radio}. 

While Schechter function and log normal shaped ERDFs fail to reproduce the broken power law shaped fits to the observed luminosity functions, they lead to broad consistency with the measured $\Phi_{\rm L}(L)$ values. On one hand, a log normal shaped ERDF succeeds in reproducing the $\Phi_{\rm L}(L)$ values measured by \citetalias{Ajello:2012aa}. The \citetalias{Mauch:2007aa} RLF values can be reproduced by a Schechter function shaped ERDF. On the other hand, a Schechter function shaped $\xi_{\rm X}(\lambda)$ and a log normal shaped $\xi_{\rm R}(\lambda)$ lead to inconsistency between the predictions and the respective observations in the brightest luminosity bins.

Our analysis in Section \ref{sec:erdf_shapes_app} shows that Schechter and log normal shaped ERDFs are likely to lead to significant disagreement between predictions and observations if a wider luminosity range is considered. If AGN luminosity functions are truly broken power law shaped, Schechter and log normal shaped ERDFs are likely to fail at the bright and faint end, respectively. To be able to determine if a broken power law, Schechter function or log normal shaped ERDF leads to better agreement between predictions and observations, we thus need luminosity functions that probe a wider luminosity range than the ones that we considered here.
\subsection{Caveats to our model}\label{sec:caveats}
Our approach and our assumptions make our model subject to caveats. 

Observationally, the local black hole mass function is often fit by a Schechter, a modified Schechter, a double power law or a log normal function (\citealt{Aller:2002aa, Shankar:2004aa, Marconi:2004aa, Greene:2007aa}). In our model, the black hole mass function high mass end depends on the $\sigma$ of the normal distribution that we assume to convert from stellar to black hole mass. The steepness of this slope affects our MCMC results, especially if $\sigma$ is large and the black hole mass function is thus shallow at high masses.  Our black hole mass function high mass end might therefore not reflect the observations and might have affected the MCMC results for $\delta_2$. 

Making the simplest assumptions possible, we chose constant bolometric corrections for the hard X-rays and 1.4 GHz and did not take luminosity or Eddington ratio dependencies into account (see Sections \ref{sec:bol_corr_xray} and \ref{sec:bol_corr_radio}). Due to many underlying complexities affecting especially $k_\mathrm{bol, R}$ we were only able to determine shape of $\xi_{\rm R}(\lambda)$. As we discuss in Section \ref{sec:lum_dep_kbol}, a luminosity dependent hard X-ray bolometric correction results in change of $\xi_{\rm X}(\lambda)$'s $\delta_1$ and $\delta_2$. The shallower $\delta_1$ value lies within the uncertainties of the MCMC results where we assumed $k_{\rm bol, X} = \rm const.$. For $\delta_2$ this is not the case. Yet even after adjusting $\delta_2$ to compensate for the luminosity dependent bolometric correction, $\delta_2$ is still steeper than the $\delta_2$ value of the RLF that we determined for a constant bolometric correction.

We chose not to extrapolate the input stellar mass functions beyond the mass range considered in \citetalias{Weigel:2016aa}. While stellar mass bins beyond $\log(M/M_\odot) = 12$ do not affect our results, we are neglecting galaxies with $M < \log(M/M_\odot) = 9$. We quantify this effect in Section \ref{sec:completeness}.  

Furthermore, we were unable to constrain the ERDF normalization $\xi^{*}$ with our approach. $\xi^{*}$ is not a free parameter in the MCMC, since every combination of $\lambda^{*}$, $\delta_1$ and $\delta_2$ requires its own $\xi^{*}$ value. $\xi^{*}$ is thus degenerate with the ERDF parameters. Instead of varying $\xi^{*}$, we adjusted the ERDF normalization so that the predicted space density $n_\mathrm{pred}$ matched the observed one $n_\mathrm{obs}$ (see equation \ref{eq:normalization}). 
\subsection{Comparison to previous work}\label{sec:compare}
\subsubsection{Phenomenological models}
Our method is based on the work by \citetalias{Caplar:2015aa}. For different AGN luminosity bins, \citetalias{Caplar:2015aa} use the convolution method (see Section \ref{sec:convolution}) to make testable predictions for the black hole mass function and the stellar mass function of AGN hosts. They fit the stellar mass functions by \cite{Ilbert:2013aa} with a smoothly varying, redshift dependent model. Furthermore, they use data by \cite{Hopkins:2007aa} to calibrate a similar redshift evolution model for the quasar luminosity function. They find that within $0 < z < 3$, the normalization of the star-forming stellar mass function and the normalization of the quasar luminosity function evolve in the same way. This implies a redshift independent AGN duty cycle. \citetalias{Caplar:2015aa} also show that while $M^{*}$ stays constant with redshift, the break of the quasar luminosity function, $L^{*}$, evolves $\propto (1 + z)^3$ out to $z \sim 2$. They attribute this evolution, to a change in both, the knee of the ERDF $\lambda^{*}$ and the black hole mass to stellar mass conversion, $\mu$.   

Significantly different from the approach by \citetalias{Caplar:2015aa} and our method, is, for instance, the work by \cite{Veale:2014aa}. They base their model on halo mass functions which are converted to stellar mass functions using the empirical stellar mass - halo mass conversion by \cite{Behroozi:2013aa}. The black hole growth rate is then linked to the galaxy growth rate. \cite{Veale:2014aa} predict the `intrinsic' quasar luminosity function at a certain redshift, by making the quasar luminosity dependent on the black hole mass or Eddington ratio at that $z$. To determine the shape of the observed quasar luminosity function, the `intrinsic' luminosity function is convolved with a distribution of instantaneous luminosities. This instantaneous luminosity distribution has a log normal (`scattered light bulb') or a power law (`luminosity-dependent lifetime') shape and entails the distribution of Eddington ratios, efficiencies and duty cycles. Considering $1 < z < 6$, \cite{Veale:2014aa} conclude that both instantaneous luminosity distributions fit the observed luminosity function well. The difference between the models is only apparent at the faint end, where the quasar luminosity function is poorly constrained. 

\cite{Kelly:2012aa} provide an overview of theoretical models for the derivation of the black hole mass function across cosmic time. These models include assumptions for the distribution of Eddington ratios. Most models consider X-ray AGN, taking radiatively inefficient accretion or radio mode AGN into account is less common. Recently, \cite{Saxena:2017aa} used the log normal shaped ERDF by \cite{Shankar:2013ab} and the black hole mass function by \cite{Shankar:2009aa} to model the RLF and the linear size distribution of radio AGN at $z\sim6$. For the ERDF, \cite{Shankar:2013ab} assume two functional forms: a standard log normal and a log normal with a power law at low $\lambda$. To reproduce the mass dependent AGN fraction and the observed ERDFs, they introduce redshift dependent modifications to these ERDF shapes. In contrast to our log normal shaped ERDF for radiatively inefficient AGN (see Figure \ref{fig:schechter_normal_radio}, $\log \lambda^{*}$ = \lambdastarlognormalR, $\tilde{\sigma}$=\sigmalognormalR), the $z\sim2$ ERDF by \cite{Shankar:2013ab} is significantly narrower ($\tilde{\sigma} =0.5$) and has a significantly higher break ($\log \lambda^{*}=-0.8$). To predict both the luminosity and size evolution of radio AGN, \cite{Saxena:2017aa}  model the radio jet power by taking different energy loss phases into account and by assuming different spin parameter distributions.

As we have highlighted above in Section \ref{sec:flickering}, the ERDF can be directly determined from the luminosity function, if an AGN lifetime model is assumed. For example, \cite{Hopkins:2009aa} use this approach to test different lifetime models. Within $0 < z < 1$, they find that predictions based on self-regulated black hole growth models agree well with the observed ERDFs. In accordance with these results, \cite{Shen:2009aa} ($0.5 < z < 4.5$) and \cite{Cao:2010aa} ($z \sim 0$) focus on such self-regulated black hole lifetime models. \cite{Aversa:2015aa} ($0 < z < 6$) also consider changes in $\lambda$ and the radiative efficiency connected to different accretion stages. \cite{Conroy:2013aa} use a `scattered light bulb' model, assuming that during an accretion episode all AGN take on a certain Eddington ratio. They reproduce the observed quasar luminosity functions  within $0.5 < z <2.5$ by assuming a redshift dependent, but mass independent duty cycle. \cite{Merloni:2008aa} ($0 < z < 5$) do not assume a black hole lifetime mode or an ERDF, but link the observed mass and luminosity functions. In analogy to X-ray binaries, they assume different accretion modes: a low $\lambda$ kinetic mode, a purely radiative mode and a high $\lambda$ kinetic mode. \cite{Tucci:2016aa} ($0 < z < 4$) aim to study both, the active and the inactive, black hole population. In contrast to our model, they assume different ERDFs for type-1 and type-2 AGN.

\subsubsection{Observational ERDF constraints}\label{sec:compare_obs}
Observationally, the ERDF of AGN in the local Universe has, for example, been studied by \cite{Aird:2012aa}. They use Prism multi-object survey (PRIMUS; \citealt{Coil:2011aa, Cool:2013aa}) data in combination with X-ray data from $Chandra$ and $XMM-Newton$ to measure Eddington ratios for $\sim 240$ AGN in the ranges $0.2 < z < 1$ and $42 < \log (L_\mathrm{2- 10\ keV}) < 44$. They fit the ERDF with a single power law (slope = -0.65) and show that the data is consistent with a mass independent ERDF. Similar to our discussion in Section \ref{sec:agn_frac}, they point out that observations of a mass dependent AGN fraction are due to a selection effect. By splitting the sample into redshift bins, \cite{Aird:2012aa} show that the normalization of the ERDF, and therefore also the overall AGN fraction, decreases between $z = 1$ and $z = 0$. They find the same ERDF shape for X-ray selected AGN in blue, green and red galaxies. The normalization of the ERDF is higher for AGN in blue and green galaxies. Based on this result, \cite{Aird:2012aa} conclude that AGN feedback is unlikely to be the main reason for star formation quenching. 

\cite{Aird:2017aa} expand the work by \cite{Aird:2012aa} to $0.1 < z < 4$ using near-infrared data from CANDELS \citep{Grogin:2011aa,Koekemoer:2011aa} and \textit{Chandra} X-ray imaging for these fields \citep{Alexander:2003aa, Xue:2011aa, Nandra:2015aa, Civano:2016aa}. Using a non-parametric approach they find that the specific black hole accretion rate distribution of star-forming and quiescent galaxies is well described by a power law with a steep cut-off at high $\lambda$ and a flattening at low $\lambda$ values. For quiescent galaxies, \cite{Aird:2017aa} find that the specific accretion rate distribution is generally consistent with being mass independent. For $z<1.5$ they find an indication of a possible shift towards lower $\lambda$ values in the $\log (M/M_\odot) > 11$ bins. In a given $z$ bin ($0.5 \lesssim z \lesssim 2$), star-forming galaxies show a similar cut-off at high accretion rates but a mass dependent normalization at the low $\lambda$ end. A similar, but less clear mass dependence is found for the combination of AGN in star-forming and quiescent hosts.

In contrast to our work, \cite{Aird:2012aa} and \cite{Aird:2017aa} only use X-ray selected AGN, measure the accretion rate distribution instead of inferring it and consider galaxies at higher redshift. To constrain the ERDF of X-ray selected AGN in blue/green/red or star-forming/quiescent galaxies using our model, we would have to split the observed XLF by host galaxy properties. If the observed XLF has a similar shape after we split it by colour or sSFR, we would expect to find similarly shaped ERDFs for blue, green and red or star-forming and quiescent hosts as long as the host galaxy mass functions are shallower than the observed XLF. In accordance with \cite{Aird:2012aa} we would expect the normalization of the ERDF of X-ray AGN in red galaxies to be lower than in blue and green hosts, reflecting the lower active black hole fraction of X-ray selected AGN in red galaxies (see Section \ref{sec:blue_Xray_AGN}).

We note that \cite{Aird:2017aa} consider galaxies at $z > 0.1$, while we focus on $z < 0.1$. Compared to higher redshifts, the mass dependence that \cite{Aird:2017aa} find for the ERDF of all and of star-forming galaxies in their lowest redshift bin ($0.1 < z < 0.5$) is less significant (see for example their Figure 6). The mass dependence that they find for the low $\lambda$ end of the ERDF could be interpreted as a mass dependent slope $\delta_1$. In Sections \ref{sec:erdf_mass} and \ref{sec:erdf_mass_app} we discussed the effect of $\delta_1$ being dependent on $\log M$. We concluded that a $\delta_1(\log M)$ model is unlikely as it produces a XLF that is no longer broken power law shaped. However, by assuming a linear dependence and by fixing $\log \lambda^{*}$ and $\delta_2$ to our best-fitting values, we only investigated the simplest $\delta_1(\log M)$ model. To estimate the change in $\delta_1$ that would be required by the observations by \cite{Aird:2017aa}, we consider the ERDF  of all galaxies in the $0.1 < z < 0.5$ bin (see their Figure 5). We assume that only $\delta_1$ is mass dependent, that $\log \lambda^{*}\sim-1$ and that $\delta_1$ changes by $\sim1.2$ over the 3 orders of magnitude in stellar mass that are considered. Compared to the $\delta_1(\log M)$ models that we tested in Section \ref{sec:erdf_mass_app}, this corresponds to a relatively mild mass dependence. By, for instance, deviating from our best-fitting values for $\log \lambda^{*}$ and $\delta_2$ or not assuming a linear dependence of $\delta_1$ on $\log M$ we are thus likely to find a $\delta_1(\log M)$ model that allows us to reproduce the local XLF.  \cite{Aird:2017aa} show that towards higher $z$ the mass dependence of the ERDF becomes more prominent. Testing if such a mass dependence is still consistent with our model, requires a more careful analysis. 

Observationally, the ERDF has also been constrained by \cite{Kollmeier:2006aa} ($0.3 < z < 4$). They find that the ERDF is well described by a log normal shaped ERDF with $\log \lambda* = -0.6$ and $\tilde{\sigma} = 0.3$. Compared to our results (see Figure \ref{fig:schechter_normal_radio}, $\log \lambda^{*}$ = \lambdastarlognormalR, $\tilde{\sigma}$=\sigmalognormalR), the ERDF by \cite{Kollmeier:2006aa} hence has a significantly higher break and is significantly narrower. \cite{Shen:2008aa} ($0.1 < z < 4.5$) find a log normal shaped ERDF with a similar $\tilde{\sigma}$ of 0.4. However, they use a black hole mass dependent $\log \lambda^{*}$. Using optically selected AGN from the SDSS at low $z$, \cite{Kauffmann:2009aa} find a mass independent, log normal shaped ERDF for star-forming galaxies (`feast mode') and a bulge mass dependent, power law shaped ERDF for galaxies with old central stellar populations (`famine mode'). \cite{Schulze:2010aa} ($z < 0.1$) select type-1 AGN from the Hamburg/ESO Survey (HES; \citealt{Wisotzki:2000aa}) and, after accounting for selection effects, fit the ERDF with a Schechter and a log normal function. When simultaneously fitting the black hole mass function with a modified Schechter function and the ERDF with a normal Schechter function, they find $\log \lambda^{*} = -0.55$ and $\tilde{\alpha} = -1.95$. In comparison to our Schechter function fits for radiatively efficient AGN (see Figure  \ref{fig:schechter_normal_xray}, $\log \lambda^{*} = $\lambdastarSchechterX, $\tilde{\alpha} = $\alphaSchechterX), they thus find a steeper low $\lambda$ slope and an ERDF break at significantly higher Eddington ratios. At $z \sim 1.4$, \cite{Nobuta:2012aa} fit Schechter functions to the black hole mass function and the ERDF.  \cite{Schulze:2015aa} ($1 < z < 2$) allow for a mass dependence in the ERDF and fit a Schechter function and a log normal ERDF to the data. Covering $z \sim 0 - 4$, \cite{Padovani:2015aa} find $\lambda$ distributions that resemble a log normal shape for both radio quiet ($\log \lambda^{*} \sim -1.3$) and radio loud ($\log \lambda^{*} \sim -3$) AGN (defined in terms of the $24\mu\rm{m}$ to 1.4 GHz flux ratio). For the \textit{XMM}-COSMOS point source catalog \citep{Hasinger:2007aa, Cappelluti:2009aa}, \cite{Bongiorno:2012aa} ($0 < z < 4$) and \cite{Bongiorno:2016aa} ($0.3 < z < 2.5$) find power law (slope $\sim 1$) and double power law (mass dependent break) shaped ERDFs , respectively. \cite{Jones:2016aa} ($0 < z < 0.33$) show that for young galaxies, an intrinsic Schechter shaped ERDF can be reduced to the log normal ERDF by \cite{Kauffmann:2009aa}, if the optical selection effects are taken into account. For the Schechter shaped ERDF for star-forming galaxies, \cite{Jones:2016aa} fix $\log \lambda^{*} = 0$ and find $\tilde{\alpha} = -1.4$ \footnote{Note that in this section, we adjust all $\alpha$ values to be conform with our Schechter function definition in equation \ref{eq:single_schechter}.}. When using a Schechter function ERDF, we find a steeper slope for radiatively efficient AGN, as mentioned above. \cite{Jones:2017aa} use the same prescription for the ERDF as \cite{Jones:2016aa}. By coupling their accretion rate distribution to the dark matter halo formation model by \cite{Mutch:2013aa} they investigate the effect of a luminosity and $\lambda$ limit on key observables such as the AGN halo occupation or sSFR-M distribution out to $z \sim 3.5$.
\subsection{Future work and surveys}\label{sec:future}
So far we treated radiatively efficient and inefficient AGN as separate populations and did not   distinguish between AGN within these two groups. In future work we will investigate these straightforward assumptions further and explore the first order perturbations to our model. For instance, as we discussed in Section \ref{sec:acc_modes}, HERGs represent the overlap between the populations of X-ray and radio selected AGN. Constraining the HERG ERDF will allow us to test our model further and to expand our discussion of which physical process might be shaping the ERDF. Furthermore, among the population of X-ray selected AGN, the ERDF of AGN in major galaxy mergers is of great interest. In the context of the classical \cite{Sanders:1988aa} model and the results by \cite{Treister:2012ab} it will allow us to investigate the role of AGN and mergers in the quenching of star formation further.

For AGN, and especially X-ray selected AGN, we observe the effect of cosmic downsizing: the space density of more luminous AGN seems to peak at higher redshift than the space density of fainter AGN \citep{Cowie:1996aa, Ueda:2003aa, Hasinger:2005aa} . Within the local volume, the measurement of the X-ray luminosity function is thus limited by the small volume and the low number of luminous AGN. While the errors on the local X-ray luminosity function are large, the broken power law shape is supported by luminosity functions of the same shape at higher redshift \citep{Vito:2014aa, Miyaji:2015aa, Georgakakis:2015aa, Fotopoulou:2016aa}. Repeating our analysis for the X-ray luminosity function at higher $z$ is hence possible. However, we need to consider that at higher redshift the stellar mass range within which we can constrain the stellar mass function decreases and the errors on the mass function itself increase \citep{Ilbert:2013aa, Muzzin:2013aa}. 

Our results can be applied to hydrodynamical simulations and semi-analytical models. Without having to resolve or describe the inner few parsecs and the details of the accretion process, the gas temperature could indicate whether the black hole is accreting via efficient or inefficient accretion. Our results thus not only predict the ERDF shape, but also imply whether quasar or radio mode feedback should be used. 

Overall, our analysis would benefit from a wider fitting range and smaller errors on the observed luminosity functions for which large surveys are necessary. In the hard X-rays, the upcoming 105-month Swift-BAT data release (flux limit $\sim 10^{-11}\ \mathrm{erg\ s}^{-1}\ \mathrm{cm^{-2}}$; Oh et al. in prep.) will allow for smaller errors on $\Phi_\mathrm{L}$. At lower energies, the \textit{Chandra} and \textit{XMM-Newton} coverage of SDSS Stripe 82 (Stripe 82X; flux limit $\sim 10^{-15}\ \mathrm{erg\ s}^{-1}\ \mathrm{cm^{-2}}$; \citealt{LaMassa:2013aa,LaMassa:2013ab}) will be probing fainter luminosities than Swift-BAT. In the near future, eROSITA \citep{Merloni:2012aa}, and specifically the eROSITA all-sky survey (eRASS; flux limit $\sim 10^{-14}\ \mathrm{erg\ s}^{-1}\ \mathrm{cm^{-2}}$), will be able to significantly increase the number of soft X-ray detected AGN \citep{Kolodzig:2013aa}. Compared to eRASS, ATHENA will be probing even fainter luminosities over a wider redshift range and area, in the far future (flux limit $\sim 10^{-17}\ \mathrm{erg\ s}^{-1}\ \mathrm{cm^{-2}}$; \citealt{Aird:2013aa, Georgakakis:2013aa}). 

In the radio regime, upcoming surveys and facilities such as, VLASS, MeerKAT, ASKAP and SKA, will not only allow us to constrain the local radio luminosity function better, but will also enable us to probe it at much higher redshifts. The VLA Sky Survey \footnote{\url{https://science.nrao.edu/science/surveys/vlass}} (VLASS) will cover $34000 \mathrm{deg}^2$ at a flux density limit of $100 \mu \mathrm{Jy}$ in the 2 - 4 GHz range. VLASS will thus reach a similar depth to FIRST with a significantly better resolution of 2.5 arcsec. The Meer Karoo Array Telescope (MeerKAT; \citealt{Jonas:2009aa}) and  the Australian SKA pathfinder (ASKAP; \citealt{Johnston:2007aa, DeBoer:2009aa}),  are the South African and the Australian SKA pathfinder telescopes, respectively. Planned surveys include the 1.4 GHz MeerKAT international GigaHertz Tiered Extragalactic Exploration survey (MIGHTEE; 35 $\mathrm{deg}^2$ at rms $1\mu\mathrm{Jy}$/beam, \citealt{Jarvis:2012aa}) and the Evolutionary Map of the Universe (EMU; entire southern sky, rms $10 \mu\mathrm{Jy}/\mathrm{beam}$, \citealt{Norris:2011aa}). Photometric and spectroscopic redshifts for EMU will be provided by SkyMapper \citep{Keller:2007aa}, WALLABY \citep{Koribalski:2012aa} and TAIPAN  \citep{Beutler:2011aa}. \cite{Norris:2013aa} summarize all upcoming radio continuum surveys with the SKA pathfinders. During its first stage of completion, the Square Kilometre Array (SKA; \citealt{Dewdney:2009aa}) will include two instruments: SKA1-LOW (50 - 350 MHz) and SKA1-MID (350 - 14 GHz) (see e.g. \citealt{Kapinska:2015aa}). Wide and deep radio continuum surveys with the SKA will increase the number of low redshift radio AGN by orders of magnitude and will allow us to detect radio quiet AGN of luminosities as low as $\log L_{1\ \mathrm{GHz}} \sim 23$ out to redshifts of $z \sim 6$ \citep{Smolcic:2015aa}. 
\section{Summary}\label{sec:summary}
Our aim was to test if black hole growth in the local Universe can be described with a phenomenological model, based on the following simple, straightforward assumptions:
\begin{itemize}
	\item black holes can be classified into two independent categories: radiatively efficient and inefficient AGN (see Section \ref{sec:acc_modes}), 
	\item radiatively efficient AGN are predominantly hosted by optically blue and green galaxies and are detected in hard X-rays (see Section \ref{sec:blue_Xray_AGN}),
	\item radiatively inefficient AGN are mostly hosted by red galaxies and are detected at 1.4 GHz (see Section \ref{sec:red_radio_AGN}),
	\item black hole mass and stellar mass are proportional to each other (see Section \ref{sec:MM}),
	\item the Eddington ratio distributions of radiatively efficient and inefficient AGN are broken power law shaped and mass independent.
\end{itemize}
To convert between bolometric and 1.4 GHz luminosities we furthermore assumed that: 
\begin{itemize}
	\item bolometric luminosities can be converted to hard X-rays and 1.4 GHz by using constant logarithmic bolometric corrections of $k_\mathrm{bol, X} = -1$ and $k_\mathrm{bol, R} = -3$, respectively (see Sections \ref{sec:bol_corr_xray} and \ref{sec:bol_corr_radio}),
\end{itemize}
We showed that, based on these assumptions, we can predict AGN luminosity functions that are consistent with the observed hard X-ray \citepalias{Ajello:2012aa} and 1.4 GHz \citepalias{Mauch:2007aa} luminosity functions. Two simple ERDFs thus describe the X-ray and radio AGN population. 

We used a MCMC to constrain $\xi_{\rm X}(\lambda)$ and $\xi_{\rm R}(\lambda)$, the best-fitting, mass independent ERDFs for radiatively efficient and inefficient AGN. The results are summarized in Figure \ref{fig:summary}. They are consistent with:
\begin{itemize}
	\item AGN hosts being randomly drawn from the galaxy population,
	\item a mass independent AGN fraction, 
	\item massive galaxies not being more likely to host high Eddington ratio AGN, compared to low mass galaxies (see Section \ref{sec:agn_frac}), 
	\item all red galaxies being equally likely to host radio AGN,
	\item maintenance mode occurring in low mass galaxies, not just in massive ones, if a certain Eddington ratio is a sufficient condition for this form of feedback. 
\end{itemize}
The results are inconsistent with:
\begin{itemize}
	\item the simplest form of mass quenching where AGN with higher accretion rates in massive galaxies lead to the quenching of star formation (see Section \ref{sec:feedback}).
\end{itemize}

In addition to our mass independent, broken power law shaped ERDF model, we quantified the effects of an ERDF for which either $\log \lambda^{*}$, $\delta_1$ or $\delta_2$ are mass dependent (see Sections \ref{sec:erdf_mass} and \ref{sec:erdf_mass_app}). We showed that while a mild level of mass dependence in the ERDF parameters still leads to consistency with the observations, at $M \sim M^{*}$ the ERDF has to resemble our best-fitting mass independent solution. This ensures that galaxies at $M^{*}$, which dominate both the XLF and RLF, are convolved with either $\xi_{\rm X}(\lambda)$ or $\xi_{\rm R}(\lambda)$ and so on average, the predicted luminosity function is consistent with the observations. The observations might allow for a mass dependence of $\xi_{\rm X}(\lambda)$ and $\xi_{\rm R}(\lambda)$, yet the mass independent ERDFs represent the simplest form of the model that is consistent with the observations and they require the least assumptions.

We also showed that introducing a luminosity dependent hard X-ray bolometric correction does still allow us to predict a X-ray luminosity function that is consistent with the observations (see Section \ref{sec:lum_dep_kbol}).

In Section \ref{sec:picture} we discussed what might be shaping the ERDF and what might cause radiatively efficient and inefficient AGN to have different ERDFs. We concluded that the shape of the ERDF might be determined by how efficiently gas can be driven from the inner few parsecs to the accretion disc or by the properties of the gas itself. 

Besides a broken power law, we also considered a Schechter function and log normal distribution as the functional form of the ERDF (see Sections \ref{sec:erdf_shapes} and \ref{sec:erdf_shapes_app}). With fewer free parameters, reproducing the observed XLF and RLF with these alternative ERDF shapes is more challenging. Luminosity functions that cover a wider luminosity range are necessary to show the disagreement between observations and predictions for log normal and Schechter function shaped ERDFs which are likely to become apparent at the faint and bright end, respectively.

We presented a phenomenological model for black hole growth in the local Universe which provides an intuitive framework to interpret observations and estimate their impact. Having constrained the zeroth order properties of X-ray and radio AGN, it is the first order perturbations and small deviations from our model that we will investigate in the future. Rather than seeing these mismatches as failures of the model, it is these discrepancies that will help us further understand the underlying physics.
\bibliographystyle{apj}
\bibliography{Bib_Lib.bib}
\acknowledgments
We thank Anna Kapi\'{n}ska, Meg Urry, Mike Koss, Franz Bauer and Adam Amara for helpful discussions. We also thank the anonymous referee for helpful comments. AKW and KS gratefully acknowledge support from Swiss National Science Foundation Grants PP00P2\_138979 and PP00P2\_166159. ET  acknowledges support from  CONICYT-Chile grants Basal-CATA PFB-06/2007 and FONDECYT Regular 1160999. This research made use of NASA's ADS Service. This publication made use of Astropy, a community-developed core \textsc{Python} package for Astronomy (Astropy Collaboration, 2013). This publication made extensive use of the Tool for OPerations on Catalogues And Tables (\textsc{TOPCAT}), which can be found at \url{http://www.starlink.ac.uk/topcat/}. This research used the NVSS and FIRST radio  surveys, carried out using the National Radio Astronomy Observatory (NRAO) Very Large Array.  The NRAO is a facility of the National Science Foundation operated under cooperative agreement by Associated Universities, Inc. This publication made use of the \textit{Swift}/Burst Alert Telescope (BAT) 60-month all sky data release. Funding for the SDSS has been provided by the Alfred P. Sloan Foundation, the Participating Institutions, the National Aeronautics and Space Administration, the National Science Foundation, the US Department of Energy, the Japanese Monbukagakusho, and the Max Planck Society. The SDSS website is http://www.sdss.org/. The SDSS is managed by the Astrophysical Research Consortium (ARC) for the Participating Institutions. The Participating Institutions are The University of Chicago, Fermilab, the Institute for Advanced Study, the Japan Participation Group, The Johns Hopkins University, Los Alamos National Laboratory, the Max-Planck-Institute for Astronomy (MPIA), the Max-Planck-Institute for Astrophysics (MPA), New Mexico State University, University of Pittsburgh, Princeton University, the United States Naval Observatory and the University of Washington.
\appendix

\section{Extended method}
\subsection{MCMC likelihood computation}\label{sec:likelihood}
When running the MCMC to find the best-fitting ERDF parameters we compute the log-likelihood of each set of tested parameters. Given $\log \lambda^{*}$, $\delta_1$ and $\epsilon$, we determine $\log \Phi_{\rm L, pred}$ in the bins of the observed luminosity function ($\log L_{\rm obs}$) using the convolution method.  After rescaling $\log \Phi_{\rm L, pred}$ to match the space density of the observed luminosity function, we then use $\log \Phi_{\rm L, obs}$ and its corresponding errors to compute $\ln \mathcal{L}$.

The errors on $\log \Phi_{\rm L, obs}$ are asymmetric. It is thus inappropriate to determine $\ln \mathcal{L}$ based on a normal distribution and reduced $\chi^2$ values. Instead we assume that $\log \Phi_{\rm L}$ is distributed log-normally,  parametrize $\log \Phi_{\rm L}$ as $x$ and express the probability density function of $x$ in the following way:

\begin{equation}
p(x, \mu, \sigma) = \frac{1}{x \sigma \sqrt{2\pi}} \times \exp \left( -\frac{(\ln x - \mu)^2}{2 \sigma^2} \right).
\end{equation}

In each $\log L_{\rm obs}$ bin, we use the observed $\log \Phi_{\rm L, obs}$ values and their errors to determine the medians of the log-normal distributions and the $\log \Phi_{\rm L}$ values that correspond to the 16 and 84 percentiles:
\begin{equation}\label{eq:median}
\begin{aligned}
\bar{x} &= \log \Phi_{\rm L, obs} + a\\
\bar{x}_{16} &= \bar{x} - \sigma_{\Phi, \rm low}\\
\bar{x}_{84} &= \bar{x} + \sigma_{\Phi, \rm up}.
\end{aligned}
\end{equation}
$a$ represents a shift that ensures that all $x$ values are $> 0$. It is the same in each MCMC step. $\sigma_{\Phi, \rm low}$ and $\sigma_{\Phi, \rm up}$ correspond to the lower and the upper $1 \sigma$ errors on $\log \Phi_{\rm L, obs}$, respectively.

To be able to compute $\ln \mathcal{L}$ we determine $\mu$ and $\sigma$ of the log-normal distributions in each of the $\log L_{\rm obs}$ bins:
\begin{equation}
\begin{aligned}
\mu &= \ln(\bar{x})\\
\sigma_{16} &= \frac{\ln(\bar{x}_{16}) - \mu}{\rm{PPF}(0.16)}\\
\sigma_{84} &= \frac{\ln(\bar{x}_{84}) - \mu}{\rm{PPF}(0.84)}.\\
\end{aligned}
\end{equation}
$\rm{PPF}$ represents the percent point function, the inverse of the cumulative density function. $\rm{PPF}(0.16)$ ($\rm{PPF}(0.84)$) thus corresponds to the value at which the integral over a normal distribution with mean 0 and standard deviation 1 reaches 16 (84) percent.

$\sigma_{16} = \sigma_{84}$ if $x$ is truly log-normally distributed. However, the log-normal distribution is only an approximation for the $\log \Phi_{\rm L, obs}$ distribution. We thus add $\sigma_{16}$ and $\sigma_{84}$ in quadrature to compute $\sigma$: 
\begin{equation}
\sigma = \sqrt{\sigma_{16}^2 + \sigma_{84}^2}.
\end{equation}

Once $\mu$ and $\sigma$ have been determined in each of the $N$ $\log L_{\rm obs}$ bins, we compute $\ln \mathcal{L}$ using $x = \log \Phi_{\rm L, pred}$: 
\begin{equation}
\begin{aligned}
\mathcal{L} (x) &= \prod_{i}^{N} p(x_i, \mu_i, \sigma_i)\\
\ln \mathcal{L} (x) &\propto - \sum_{i}^{N} \ln (x_i + a) - \sum_{i}^{N} \frac{(\ln (x_i+a) - \mu_i)^2}{2\sigma_i^2}.\\
\end{aligned}
\end{equation}

\section{Extended analysis}

\begin{deluxetable*}{lll}
	\tablecolumns{3}
	\tablecaption{\label{tab:mcmc_para} Summary of MCMC settings}
	\tablehead{
		\colhead{MCMC parameters} & 
		\colhead{radiatively efficient AGN} & 
		\colhead{radiatively inefficient AGN} 
	}
	\startdata
	{comparison LF} & {\citetalias{Ajello:2012aa}}  & {\citetalias{Mauch:2007aa}}\\
	{$\log M_{\rm min}$, $\log M_{\rm max}$} & {9, 12} & {9, 12}\\
	{log. bolometric correction $k_\mathrm{bol}$} & {\kbolX} & {\kbolR}\\
		{fitting $\log L_{\rm min}$, $\log L_{\rm max}$} & {\fitLminX, \fitLmaxX\ [$\rm erg\ s^{-1}$]} & {\fitLminR, \fitLmaxR\ [$\rm W\ Hz^{-1}$]}\\
	{}\\
	{broken power law} & {} & {}\\
	{see Section \ref{sec:mcmc_results}} & {Fig. \ref{fig:mcmc_results}} & {Fig. \ref{fig:mcmc_results}}\\
	{initial $\log \lambda^{*}$} & {-1.4} & {-2.5}\\
	{initial $\delta_1$} & {0.5} & {0.5}\\
	{initial $\delta_2$} & {2.4} & {1.3}\\
	{$\log \lambda^{*}_{\rm min}$, $\log \lambda^{*}_{\rm max}$} & {-3.0, 0.0} & {-3.0, -2.0}\\
	{$\delta_{1, \rm{min}}$, $\delta_{1, \rm{max}}$} & {-1.0, 1.0} & {0.0, 2.0}\\
	{$\epsilon_{\rm{min}}$, $\epsilon_{\rm{max}}$} & {0.0, 5.0} & {0.0, 2.0}\\
	{}\\
	{Schechter function} & {} & {}\\
	{see Section \ref{sec:erdf_shapes_app}} & {Fig. \ref{fig:schechter_normal_xray}} & {Fig. \ref{fig:schechter_normal_radio}}\\
	{initial $\log \lambda^{*}$} & {-1.3} & {-2.7}\\
	{initial $\tilde{\alpha}$} & {-1.2} & {-1.5}\\
	{$\log \lambda^{*}_\mathrm{min}$, $\log \lambda^{*}_\mathrm{max}$} & {-2.0, -1.0} & {-3.0, 0.0}\\
	{$\tilde{\alpha}_{\mathrm{min}}$, $\tilde{\alpha}_{\mathrm{max}}$} & {-2.0, 1.0} & {-3.0, 0.0}\\
	{}\\
	{log-normal} & {} & {}\\
	{see Section \ref{sec:erdf_shapes_app}} & {Fig. \ref{fig:schechter_normal_xray}} & {Fig. \ref{fig:schechter_normal_radio}}\\
	{initial $\log \lambda^{*}$} & {-2.8} & {-2.7}\\
	{initial $\tilde{\sigma}$} & {0.2} & {0.2}\\
	{$\log \lambda^{*}_\mathrm{min}$, $\log \lambda^{*}_\mathrm{max}$} & {-7.0, -2.0}& {-10.0, -2.0}\\
	{$\tilde{\sigma}_{\mathrm{min}}$, $\tilde{\sigma}_{\mathrm{max}}$} & {0.1, 1.5} & {0.0, 2.0}\\
	\enddata
	\tablecomments{Initial guess and priors for the MCMC runs. In Section \ref{sec:mcmc} we discuss the details of the MCMC. We assume a broken power law ERDF and vary $\log \lambda^{*}$, $\delta_1$ and $\epsilon = \delta_2 - \delta_1$. In Section \ref{sec:erdf_shapes_app} we extend this discussion to log-normal and Schechter function ERDFs.}
\end{deluxetable*}

\subsection{The effect of $M_{\rm min}$, $M_{\rm max}$ on the predicted luminosity function}\label{sec:completeness}
\begin{figure*}
	\centering
	\begin{minipage}{\textwidth}
		\includegraphics[width=\textwidth]{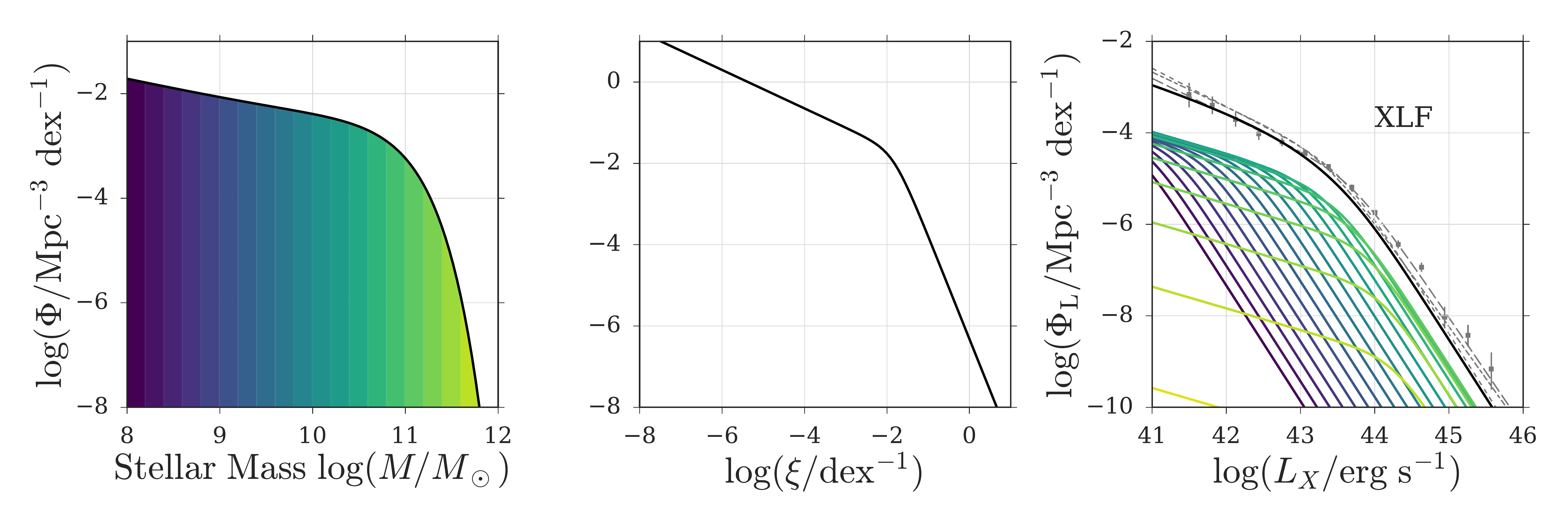}
	\end{minipage}
	\begin{minipage}{\textwidth}
		\includegraphics[width=\textwidth]{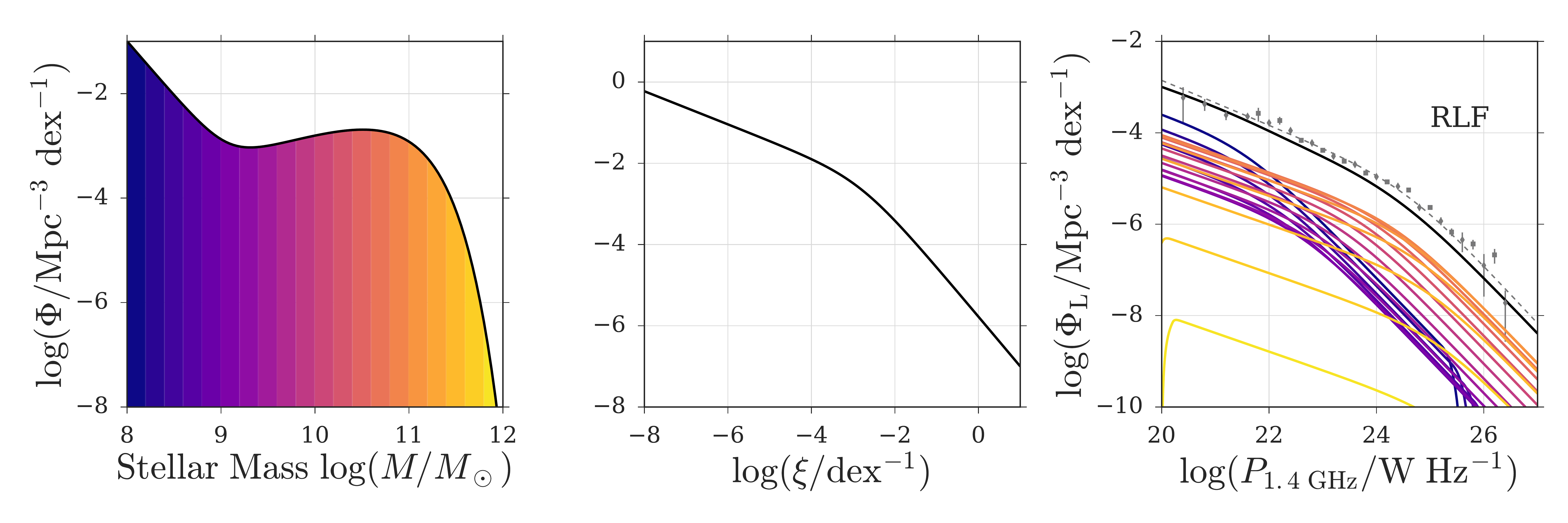}
	\end{minipage}
	\caption{\label{fig:completeness}The contribution of individual stellar mass bins to the predicted XLF (top row) and the RLF (bottom row). When constraining the best-fitting ERDF for radiatively efficient and inefficient AGN, we only consider stellar masses above \Mmin\ and below \Mmax. This figure shows the effect that this cut-off has on the predicted luminosity function. The right-hand panels show the observed blue + green (top row) and red (bottom row) stellar mass functions, which we determined by using the sample and the method by \protect\citetalias{Weigel:2016aa}, extrapolated to $\log(M/M_\odot) = 8$. The central panels illustrate the assumed ERDFs, $\xi_{\rm X}(\lambda)$ and $\xi_{\rm R}(\lambda)$. The colored lines in the right-hand panels show the contribution of individual stellar mass bins to the predicted luminosity functions. The black solid line shows the combined luminosity functions. We illustrate the observed XLFs and RLFs in grey. Due to the exponential cut-off of the stellar mass function the space density of stellar mass bins beyond $\log (M/M_\odot) = 12$ is too low to contribute significantly to the predicted luminosity function. Stellar mass bins below $\log (M/M_\odot) = 9$ do however affect the faint end of the luminosity function. For the XLF masses in the $8\leq \log (M/M_\odot) < 9$ range contribute $\sim 5\%$ to the overall space density within $\fitLminX < \log (L_{\rm X}/\rm{erg\ s^{-1}}) < \fitLmaxX$. Due to the steep double Schechter function shape of the red mass function, this effect is more extreme for the RLF.  Masses in the $8\leq \log (M/M_\odot) < 9$ range account for $\sim 38 \%$ of the total space density within $\fitLminR < \log(P_{1.4\ \rm GHz}/\rm{W\ Hz^{-1}}) < \fitLmaxR$. The steep increase of the red stellar mass function that we predict by extrapolating the \protect\citetalias{Weigel:2016aa} results is however not in accordance with observations, as we discuss in more detail in the text. }
\end{figure*}

When applying the MCMC to the observed luminosity functions by \citetalias{Ajello:2012aa} and \citetalias{Mauch:2007aa}, we only consider stellar masses between \Mmin\ and \Mmax. This is the stellar mass range that was considered for the mass function determination in \citetalias{Weigel:2016aa}. By not extrapolating the stellar mass functions to lower and higher masses we are introducing a bias in the ERDF determination since the contribution of mass bins beyond $9 \leq \log (M/M_\odot) \leq 12$ is neglected. Here we quantify this effect for the red and the blue+green stellar mass function.

Figure \ref{fig:completeness} shows the contribution of individual stellar mass bins to the XLF in the top row and to the RLF in the bottom row. The left-hand panels show the input stellar mas functions, the central panels illustrate the assumed ERDF and the right-hand panels show the predicted luminosity function. For this figure we extrapolated the STY \citep{Sandage:1979aa} results for the blue+green (top row) and the red (bottom row) stellar mass function to $\log(M/M_\odot) = 8$. These additional stellar mass bins were  also considered in the convolution with $\xi_{\rm X}(\lambda)$ and $\xi_{\rm R}(\lambda)$ for which we assumed our results from Section \ref{sec:mcmc_results}. In the right-hand panels we show the contribution of individual mass bins with colored lines and the combined luminosity functions with solid black lines. To guide the eye the observed XLF and RLF are shown in grey.

The top row of Figure \ref{fig:completeness} shows that the XLF is dominated by stellar mass bins at $\sim M^{*}$.  The space density of stellar mass bins beyond $\log(M/M_\odot) \sim 11.5$ is too low to have a significant effect on the XLF. Mass bins below $\log(M/M_\odot) = 9$ only contribute to the faint end of the XLF. Within our fitting range of $\fitLminX < \log (L_{\rm X}/\rm{erg\ s^{-1}}) < \fitLmaxX$ masses in the $8 \leq \log(M/M_\odot) < 9$ range contribute $\sim 5\%$ to the overall space density of the XLF. 

While the blue+green stellar mass function is fit with a mild double Schechter function, the red stellar mass function is described by a strong double Schechter function. As the lower panels of Figure \ref{fig:completeness} show, the steep $\alpha_1$ of the red stellar mass function causes a significant increase in the space density of low mass objects if we extrapolate to masses below $\log(M/M_\odot) = 9$. The faint end of the RLF is hence dominated by the contribution of these low stellar mass bins. Within the $\fitLminR < \log(P_{1.4\ \rm GHz}/\rm{W\ Hz^{-1}}) < \fitLmaxR$ range, $8 \leq \log(M/M_\odot) < 9$ masses account for $\sim 38\%$ of the total space density. Similar to the blue+green stellar mass function, the exponential cut-off of the red stellar mass function results in high mass bins not affecting the predicted luminosity function. 

Extrapolating the red stellar mass function to low stellar masses thus significantly changes the predicted luminosity function and as a consequence, would affect our ERDF fitting results. The strong double Schechter function shape and the steep $\alpha_1$ value of the red stellar mass function is primarily due to galaxies at the edge of the $\log(M/M_\odot)\sim 9$ completeness limit of the \citetalias{Weigel:2016aa} sample. An increase in $\Phi(M)$ of more than a dex between $\log(M/M_\odot) =8$  and $\log(M/M_\odot) = 9$ is unphysical as previous work by, for instance, \cite{Baldry:2012aa} shows. The effect on the RLF is thus likely to be less than $\sim 38\%$ and likely to be comparable to the result that lower stellar mass bins have on the XLF. 

We acknowledge that by choosing not to extrapolate the input stellar mass functions we are introducing a bias in our analysis. The cut-off at \Mmax\ does not affect our results. However, the lower limit of \Mmin\ does result in us missing lower mass galaxies which contribute to the faint end of the predicted luminosity function. Including these lower mass bins in our MCMC analysis (see Section \ref{sec:mcmc_results}) would result in a change in the best-fitting $\delta_1$ value. These lower mass bins would however not change the fact that based on mass independent ERDFs, our simple model is capable of predicting luminosity functions that are consistent with the observations.

\subsection{The effect of a luminosity dependent bolometric correction}\label{sec:lum_dep_kbol} 
\begin{figure*}
	\includegraphics[width=\textwidth]{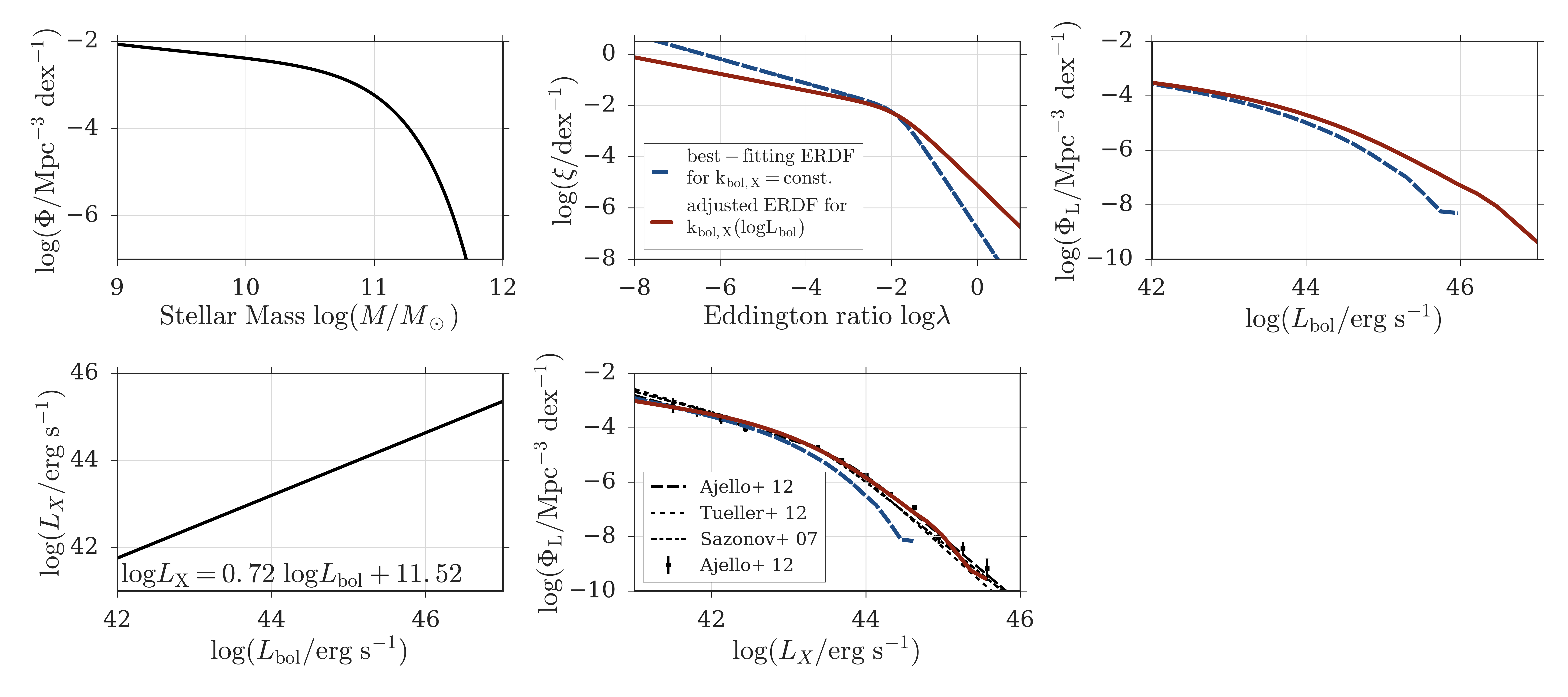}
	\caption{\label{fig:lum_dep_kbol}The effect of a luminosity dependent bolometric correction on the XLF. We introduce a linear dependence of $\log L_{\rm X}$ on $\log L_{\rm bol}$ (bottom left-hand panel) and use the random draw method to investigate the effect on the predicted XLF. We illustrate the predicted XLFs for two different ERDFs. The blue dashed lines show the results for our best-fitting X-ray ERDF from Section \ref{sec:mcmc_results} which we determined using $k_{\rm bol, X} = \rm const.$. The central bottom panel shows that this ERDF does no longer lead to consistency with the observed XLFs (shown in black) if a luminosity dependent bolometric correction is introduced. The effect can however be compensated by adjusting the slopes of the ERDF, as the red solid lines illustrate. For this second model we have decreased $\delta_1$ by 0.15 and $\delta_2$ by 0.9 relative to our best-fitting X-ray ERDF. The central bottom panel illustrates that for a luminosity dependent bolometric correction, this modified ERDF leads to consistency with the data. Introducing a luminosity dependent bolometric correction does hence result in a change of the slopes $\delta_1$ and $\delta_2$. Nonetheless a $k_{\rm bol, X}(L_{\rm bol})$ model does not change our main result that the observed XLF can be reproduced with a mass independent ERDF.}
\end{figure*}

For our model we assume a constant bolometric correction for the conversion between bolometric luminosities and hard X-rays. Based on the work by \cite{Rigby:2009aa} and \cite{Vasudevan:2009aa} we use $k_{\rm bol, X} =  \log(L_{\rm X}/L_{\rm bol}) = \kbolX$. Here we illustrate the effect that a luminosity dependent bolometric correction has on our results.

To include $k_{\rm bol, X}(L_{\rm bol})$ in our model we use the random draw method (see Section \ref{sec:random_draw}). We introduce a linear dependence of $k_{\rm bol, X}$ on $\log L_{\rm bol}$:

\begin{equation}\label{eq:lum_dep_kbol}
\log L_{\rm X} = 0.72 \times \log L_{\rm bol} + 11.52
\end{equation}

The slope of 0.72 is motivated by work by \cite{Steffen:2006aa} who determine 0.72 to be the best-fitting slope for the conversion between the 2500 \AA\ and the 2 keV luminosity. To adapt this relation for our model, we renormalize the relation so that $\log L^{*}_{\rm X} = \log L^{*}_{\rm bol} \kbolX$. For $\log L^{*}_{\rm X}$ we assume the value by \citetalias{Ajello:2012aa} (see Table \ref{tab:mass_lum_fcts}). 

Figure \ref{fig:lum_dep_kbol} illustrates our results. The blue dashed and red solid lines show the random draw results including a luminosity dependent bolometric correction for two different ERDFs. For the blue model $\xi_{\rm X}(\lambda)$ corresponds to our best-fitting solution for the X-ray ERDF with $k_{\rm bol, X} = \rm const.$ (see Section \ref{sec:mcmc_results}). The red solid lines show the results of an ERDF with shallower slopes $\delta_1$ and $\delta_2$. The lower panels show the predicted XLFs for these ERDFs and the assumed $k_{\rm bol, X}(L_{\rm bol})$ dependence. Shown in black are the observed XLFs by \cite{Sazonov:2007aa}, \cite{Tueller:2008aa} and \citetalias{Ajello:2012aa}. 

Instead of a constant shift, a luminosity dependent bolometric correction causes the XLF to be compressed when we convert from bolometric to hard X-ray luminosities. The lower middle panel of Figure \ref{fig:lum_dep_kbol} shows that in this case our best-fitting X-ray ERDF from Section \ref{sec:mcmc_results} is no longer consistent with the data. The effect can be compensated by changing the ERDF slopes $\delta_1$ and $\delta_2$. In the upper central panel the red solid line illustrates a possible modification of the ERDF which leads to consistency with the data given a luminosity dependent bolometric correction. Compared to our best-fitting ERDF with a constant bolometric correction, we have decreased $\delta_1$ by 0.15 and $\delta_2$ by 0.9. For $\delta_1$ this change lies within the uncertainties that we determined for the best-fitting $k_{\rm bol, X} = const.$ model. The change in $\delta_2$ exceeds these uncertainties. However, the shallower $\delta_2$ value that is necessary to compensate for a luminosity dependent bolometric correction is not consistent with the $\delta_2$ value that we determined for the radio ERDF. Note that a change in the normalization of equation \ref{eq:lum_dep_kbol} would require and additional change of $\lambda^{*}$. 

We conclude that introducing a luminosity dependent bolometric correction in our model requires a change of the slopes $\delta_1$ and $\delta_2$ to ensure consistency with the data. A luminosity dependent bolometric correction does however  not change our main conclusion that a mass independent ERDF is consistent with the observed XLF. 

\subsection{The effect of a mass dependent ERDF}\label{sec:erdf_mass_app} 
For our approach we assume that the broken power law ERDF is constant and does neither depend on stellar nor on black hole mass. In Section \ref{sec:mcmc_results} we showed that, based on this simple assumption, we are able to reproduce the observed XLF and RLF. \cite{Aird:2012aa} argue for a mass independent ERDF, but \cite{Schulze:2015aa} and \cite{Bongiorno:2016aa} use mass dependent ERDFs. An ERDF for which, for example, $\log \lambda^{*}$ increases with $\log M$ could thus be plausible. We now test if the observations can also be reproduced with such a $\xi(\log \lambda, \log M)$ and quantify the allowed variation in $\xi$. In general, we cannot claim that the ERDF is mass independent or that it has to follow a certain mass dependence, simply because we are unable to test all possible ways in which $\log \lambda^{*}, \delta_1$ and $\delta_2$ could depend on $\log M$. We can only test if a certain $\xi$ model is consistent with the observations. For simplicity, we limit our analysis to broken power law ERDFs for which only one of the three parameters is mass dependent. 

In the following we will not be examining $\xi^{*}(\log M)$, a mass dependent ERDF normalization, which implies a mass dependent active black hole fraction. We consider the simplest case and assume that $\xi^{*}$ is mass independent.

To incorporate a mass dependent ERDF into our model, we use the random draw technique (see Section \ref{sec:random_draw}). To make the method computationally less expensive, we pre-determine the ERDF CDF in stellar mass bins ($\Delta \log M = \massdeperdfmbinsize$). We then calculate $N_\mathrm{draw}$, draw $\log M$ values and assign $\log M_\mathrm{BH}$. We determine the stellar mass bin for each $\log M$ value and use the corresponding ERDF CDF to draw a $\log \lambda$ value. The $\log L$ values and the corresponding luminosity function are calculated and constructed in the same way as discussed in Section \ref{sec:random_draw}. 

To allow for an easy comparison, we renormalize the luminosity functions that are predicted by mass dependent ERDFs when plotting our results in the following way: 
\begin{equation}
\tilde{\Phi}_{\mathrm{L, pred}} = \frac{\int_{\log L_\mathrm{min}}^{\log L_\mathrm{max, val}} \Phi_\mathrm{L,obs} d\log L}{\int_{\log L_\mathrm{min}}^{\log L_\mathrm{max, val}} \Phi_\mathrm{L, pred} d\log L} \times \Phi_\mathrm{L, pred}.
\end{equation}
Here, $\log L_\mathrm{max, val}$ is the maximum luminosity value that is reached with $\xi(\log \lambda, \log M)$ and $\log (L_\mathrm{min}/\mathrm{erg\ s}^{-1}) = \fitLminX$ for the XLF and $\log L_\mathrm{min}/\mathrm{W\ Hz}^{-1} = \fitLminR$ for radiatively inefficient AGN.

\subsubsection{A mass dependent break}
\begin{figure*}
	\includegraphics[width=\textwidth]{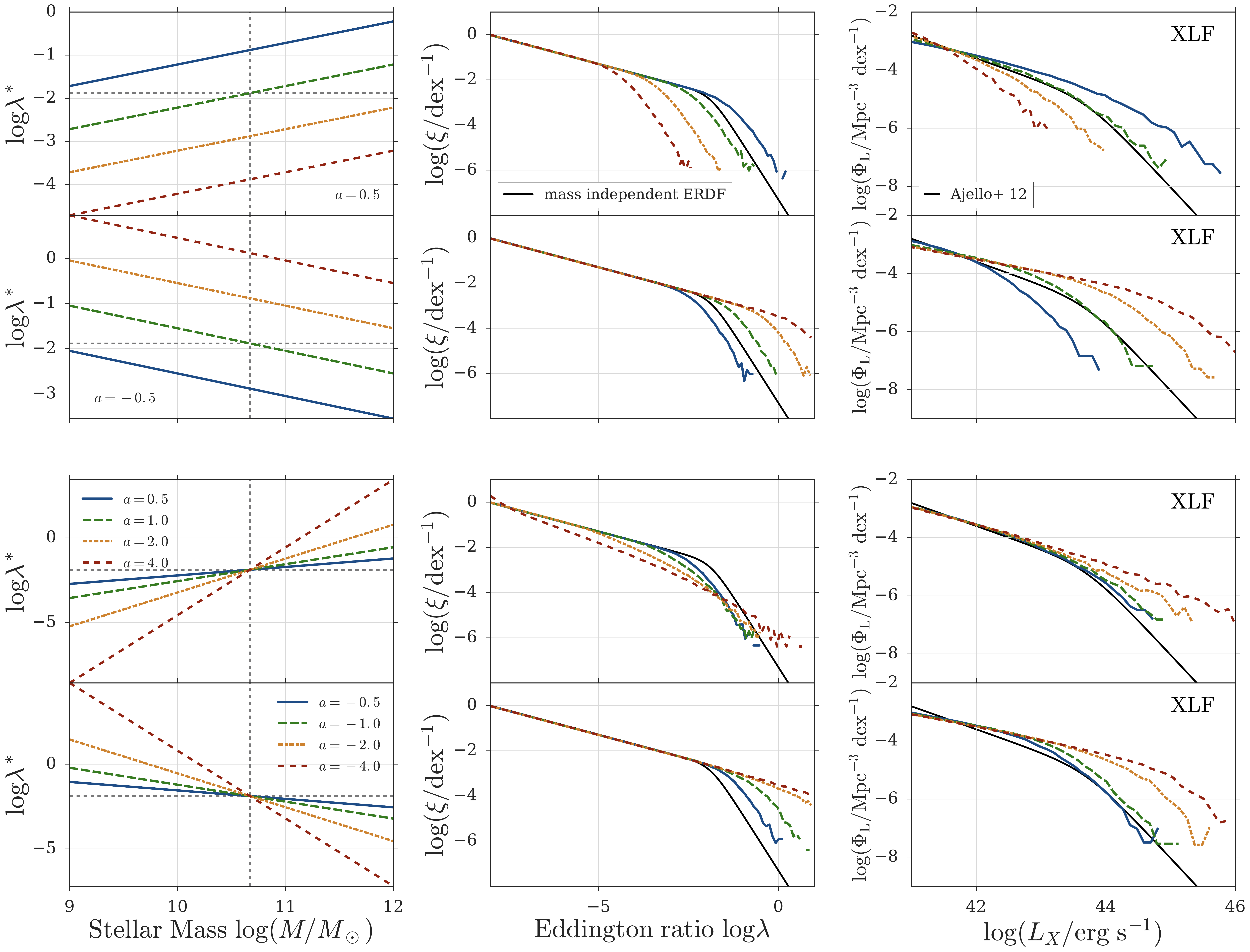}
	\caption{\label{fig:mass_dep_break_xray} Random draw results for \textit{radiatively efficient AGN} with a broken power law shaped ERDF with a mass dependent break $\lambda^{*}$. We parametrize $\log \lambda^{*}$ as $\log \lambda^{*}(\log M) = a \times \log M + b$ and use the random draw method to predict the corresponding luminosity function (see Sec. \ref{sec:random_draw}). We keep $\delta_1$ and $\delta_2$ fixed at the values we determined for the mass independent ERDF (see Section \ref{sec:mcmc_results}). In the left-hand panels we show different models which we assume for the $\log \lambda^{*}$ mass dependence. The vertical dashed lines highlight $M^{*}$ for blue+green galaxies. The horizontal dashed lines indicate the $\log \lambda^{*}$ value we determined under the assumption of a mass independent ERDF ($\log \lambda^{*}_{M\ \mathrm{indep}}$). Taking stellar mass values between $10^9$ and $10^{12} M_\odot$ into account, we show the resulting ERDFs in the central panels. The black solid line illustrates our best-fitting solution for a mass independent ERDF. The right-hand panels illustrate the predicted luminosity functions for the assumed $\log \lambda^{*}(\log M)$ models. The black solid line highlights the observed luminosity function by  \protect\citetalias{Ajello:2012aa}. When interpreting these results it is important to remember that, due to the shape of the blue+green stellar mass function, a sample of randomly drawn stellar mass values will always contain more low mass than high mass objects. Furthermore, the simple relations we derived for the predicted luminosity function in Section \ref{sec:pred_lf} are no longer applicable. This figure illustrates  that most of the assumed $\log \lambda^{*}_{M\ \mathrm{indep}}$ models are unable to reproduce the observed luminosity function. Only for $-1 < a < 1$ with $\log \lambda^{*}(\log M^{*}) = \log \lambda^{*}_{M\  \mathrm{indep}}$ does the predicted luminosity function resemble the observed one.}
\end{figure*}

\begin{figure*}
	\includegraphics[width=\textwidth]{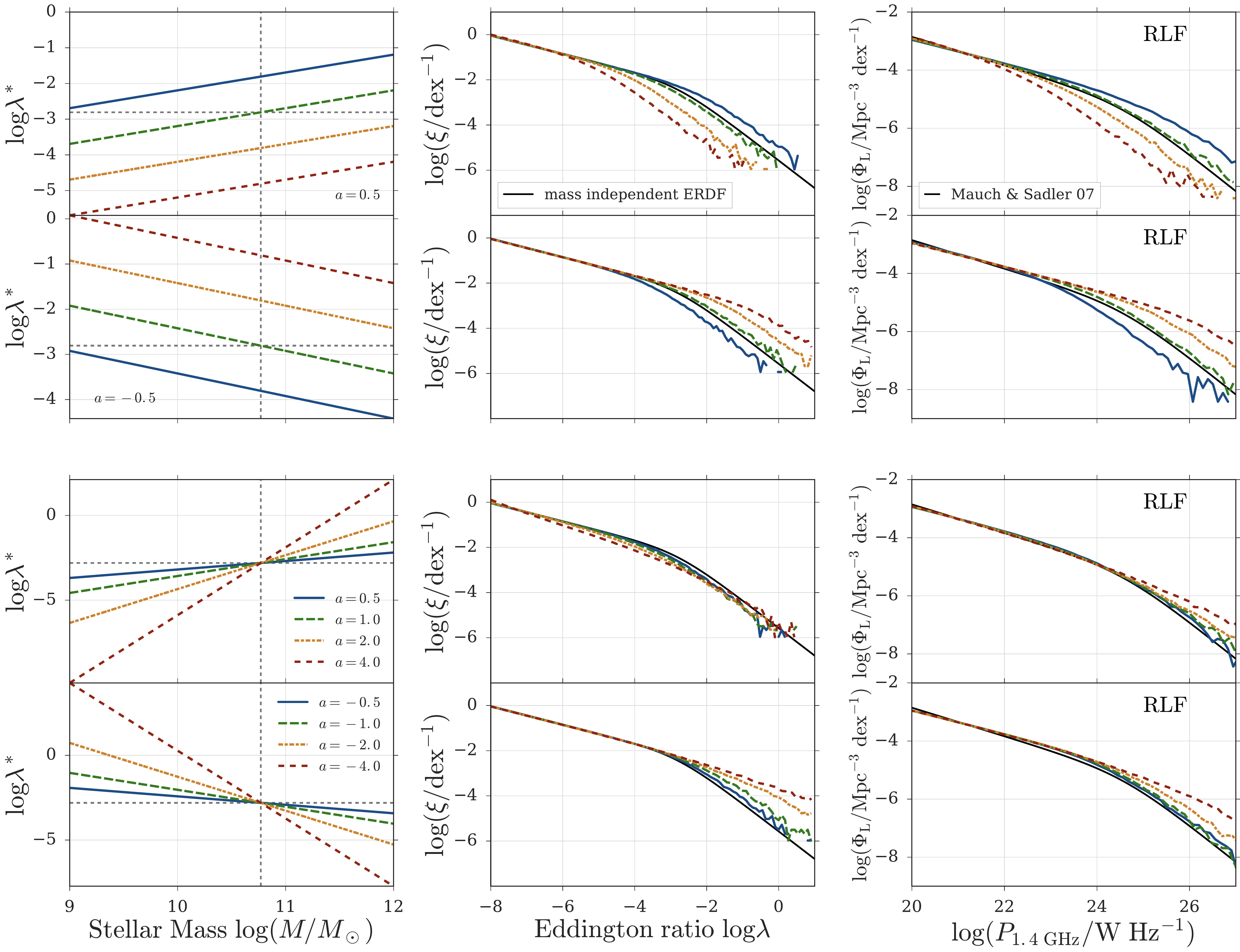}
	\caption{\label{fig:mass_dep_break_radio}Random draw results for \textit{radiatively inefficient AGN} with a broken power law shaped ERDF with a mass dependent break $\lambda^{*}$, in analogy to Figure \ref{fig:mass_dep_break_radio}. We assume different $\log \lambda^{*}(\log M)$ models (left-hand panels) and show the resulting ERDFs for all masses between $10^9$ and $10^{12} M_\odot$ (central panels) and the predicted 1.4 GHz luminosity functions (right-hand panels). In the left-hand panels the vertical dashed lines mark $M^{*}$ for red galaxies. The horizontal lines highlight our best-fitting solution for $\log \lambda^{*}$ under the assumption that the ERDF is mass independent ($\log \lambda^{*}_{M\ \mathrm{indep}}$, see Section \ref{sec:mcmc_results}). In the central panels we show this best-fitting, mass independent ERDF which a black solid line. In the right-hand panels the observed luminosity function by \protect\citetalias{Mauch:2007aa} is shown in the same way. Compared to the results for a mass dependent ERDF for radiatively efficient AGN, the observed radio luminosity function allows for a similar amount of variation in $\log lambda^{*}$. The figure illustrates that for $-1 \leq a = \leq 1$ and $\log \lambda^{*}(\log M^{*}) = \log \lambda^{*}_{M\  \mathrm{indep}}$ the predicted luminosity function matches the observations.}
\end{figure*}

Figure \ref{fig:mass_dep_break_xray} and \ref{fig:mass_dep_break_radio} show the random draw results for ERDFs with mass dependent breaks for radiatively efficient and inefficient AGN, respectively. We show $\log \lambda^{*}(\log M)$,  the resulting $\xi(\log \lambda, \log M)$ for all masses and the predicted luminosity functions in the left-hand, central and right-hand columns, respectively. In the left-hand columns the vertical dashed lines highlight $\log M^{*}$ of the blue+green mass function (Figure \ref{fig:mass_dep_break_xray}) and $\log M^{*}$ of the red mass function (Figure \ref{fig:mass_dep_break_radio}). The horizontal dashed lines mark $\log \lambda^{*}_{M\ \mathrm{indep}}$, which we determined with the MCMC, assuming mass independent ERDFs (see Section \ref{sec:mcmc_results}). For comparison, we illustrate the best-fitting, mass independent ERDFs as black solid lines in the central columns. In the right-hand column of Figure \ref{fig:mass_dep_break_xray}, the black solid lines show the observed X-ray luminosity function by \citetalias{Ajello:2012aa}. In the right-hand column of Figure \ref{fig:mass_dep_break_radio}, the black solid lines illustrate the 1.4 GHz luminosity function by \citetalias{Mauch:2007aa}. 

We assume $\log \lambda^{*} = a \log M + b$ and keep $\delta_1$ and $\delta_2$ constant, setting them to the values we determined with the MCMC for a mass independent ERDF. In the top two rows, we vary the y-intercept $b$, but keep the slope $a$ constant. While in the bottom two rows, we change $a$, but enforce $\log \lambda^{*}(\log M^{*}) = \log \lambda^{*}_{M\ \mathrm{indep}}$. 

When interpreting these mass dependent ERDFs, it is important to keep the stellar mass function in mind. Our randomly drawn sample contains more $9 < \log (M/M_\odot) < 10$ galaxies than $11 < \log (M/M_\odot) < 12$ galaxies. The $\log \lambda^{*}$ values at low stellar masses thus affect the predicted luminosity function more strongly, than at higher stellar masses. Furthermore, the simple relations that we derived for the predicted luminosity function in Section \ref{sec:pred_lf} are no longer valid due to the mass dependence. 

For the XLF Figure \ref{fig:mass_dep_break_xray} shows that for mass dependent break $\log \lambda^{*}$ the predicted and the observed XLF are in agreement if $-1 \leq a \leq 1$ and $\log \lambda^{*}(\log M^{*}) = \log \lambda^{*}_{M\ \mathrm{indep}}$. More extreme $a$ values are less likely since they result in $L^{*}$ values that are significantly higher or lower than the observed break. 

For the RLF Figure \ref{fig:mass_dep_break_radio} shows a similar trend. We find an agreement between our predictions and the observations if $-1 \leq a \leq 1$ and $\log \lambda^{*}(\log M^{*}) \sim \log \lambda^{*}_{M\ \mathrm{indep}}$.
\subsubsection{A mass dependent slope}
\begin{figure*}
	\includegraphics[width=\textwidth]{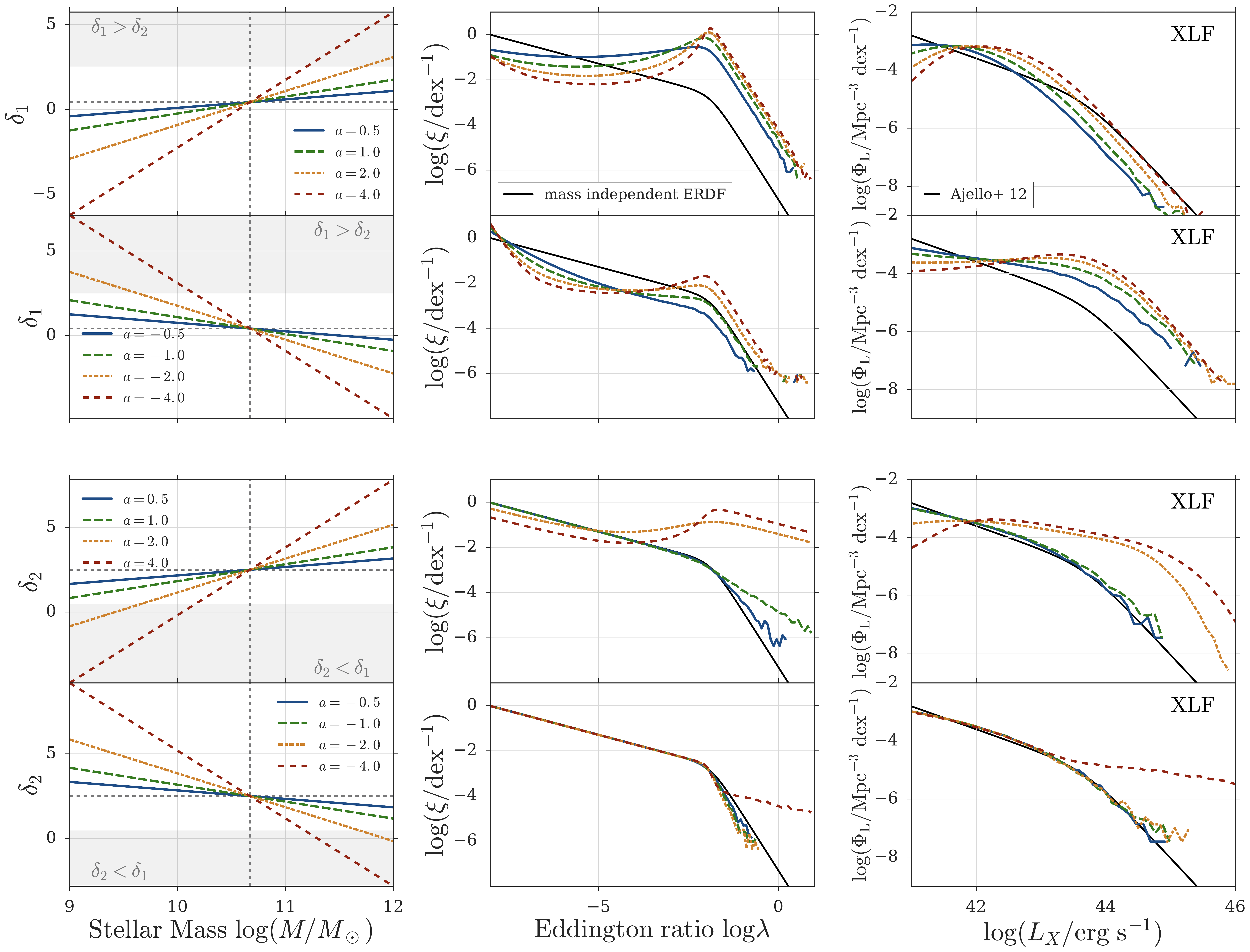}
	\caption{\label{fig:mass_dep_slopes_xrays} Random draw results for \textit{radiatively efficient AGN} with a broken power law shaped ERDF with a mass dependent slopes $\delta_1$ (top two rows) and $\delta_2$ (bottom two rows). In the top two rows, we parametrize $\delta_1$ as $\delta_1(\log M) = a \times \log M + b$ and keep $\delta_2$ and $\lambda^{*}$ fixed at the best-fitting values which we determined for a mass independent ERDF (see Section \ref{sec:mcmc_results}). The vertical dashed line in the left-hand panels marks $M^{*}$ for blue+green galaxies. In the top (bottom) two rows, the horizontal dashed lines highlight our best-fitting $\delta_1$ ($\delta_2$) values for a mass independent ERDF. Additionally, we shade the region in which $\delta_1 > \delta_2$ (top two rows)/$\delta_2 < \delta_1$ (bottom two rows). We refer to the low Eddington ratio end of $\xi(\lambda)$ as $\delta_1$ and the high $\log \lambda$ end as $\delta_2$. Once $\delta_1 > \delta_2$($\delta_2 < \delta_1$), we are thus changing the high (low) Eddington ratio end instead of the intended slope at low (high) $\log \lambda$ values. Our best-fitting mass independent ERDF is shown with a black solid line in the central panels. In the right-hand panels, the observed luminosity function by  \protect\citetalias{Ajello:2012aa} is highlighted in the same way. The figure shows that for radiatively efficient AGN a mass dependent low Eddington ratio slope $\delta_1$ can be excluded. A mass dependent $\delta_2$ is only permitted for $-1 \leq a \leq 1$ with $\delta_2{\log M^{*}} = \delta_{2, M\ \rm{indep}}$.}
\end{figure*}

\begin{figure*}
	\includegraphics[width=\textwidth]{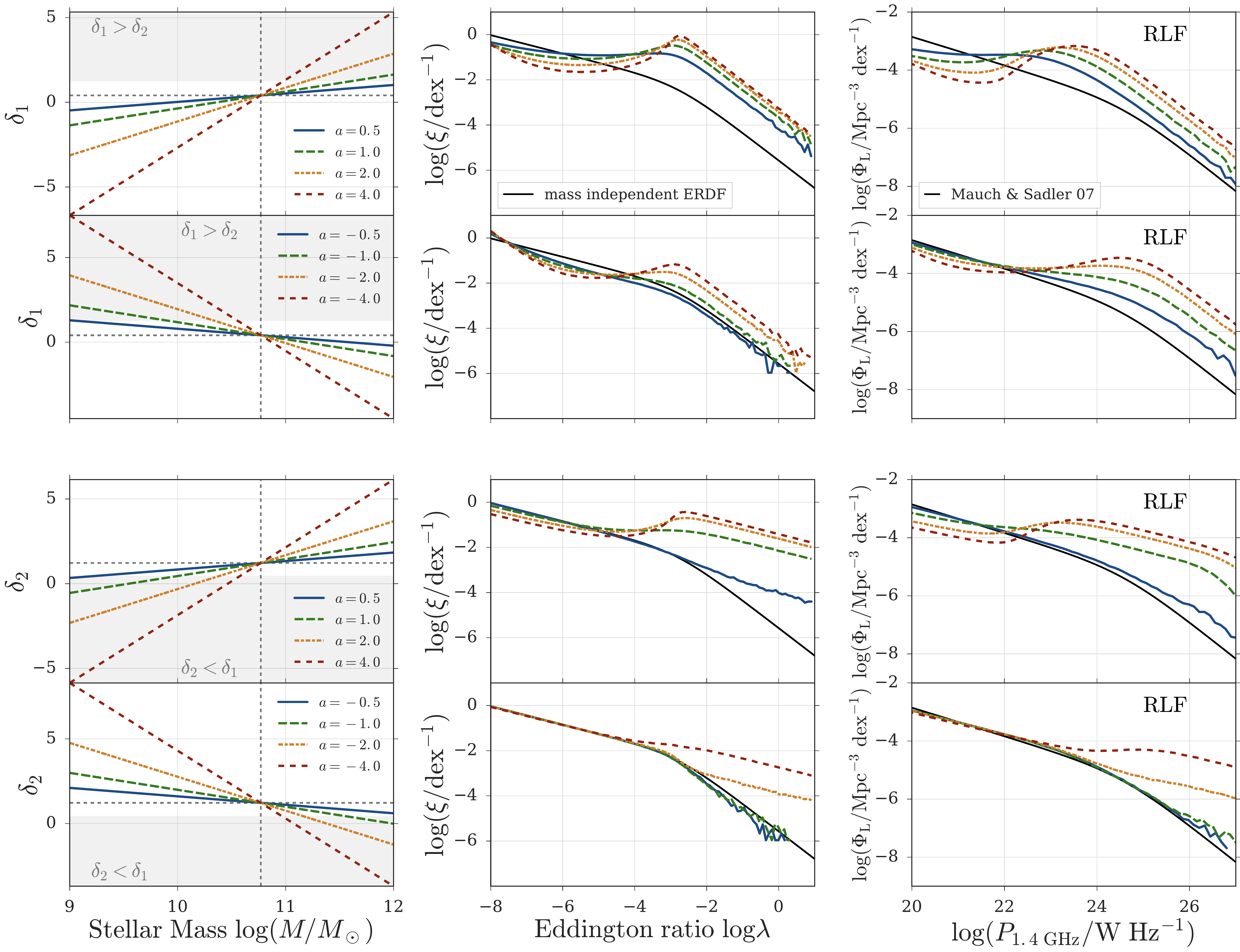}
	\caption{\label{fig:mass_dep_slopes_radio}Random draw results for \textit{radiatively inefficient AGN} with a broken power law shaped ERDF with a mass dependent slopes $\delta_1$ (top two rows) and $\delta_2$ (bottom two rows). In analogy to Figure \ref{fig:mass_dep_slopes_xrays}, we vary the low $\lambda$ (top two rows) and the high $\lambda$ (bottom two rows) end of the ERDF as a function of stellar mass. In the left-hand panels the vertical dashed lines mark $M^{*}$ for red galaxies. The horizontal dashed lines highlight our best-fitting $\delta_1$/$\delta_2$ value for a mass independent ERDF (see Section \ref{sec:mcmc_results}). In the left-hand panels, we also shade the regions where $\delta_1 > \delta_2$ (top two rows) and $\delta_2 < \delta_1$ (bottom two rows). In the central panels the black solid line highlights our best-fitting, mass independent $\xi$ model. The right-hand panels show the resulting luminosity functions and the observed luminosity function by \protect\citetalias{Mauch:2007aa}.  The figure illustrates that $\delta_1(\log M)$ models can be excluded. To find agreement between the predicted and the observed RLF, $a$ has to be within $-1$ and $0$ if $\delta_2$ is mass dependent. For $\delta_2(\log M)$ is is also important that at $M^{*}$ $\delta_2$ corresponds to the value that we determined for the mass independent case.}
\end{figure*}

Our results for mass dependent ERDF slopes for radiatively efficient and inefficient AGN are shown in Figure \ref{fig:mass_dep_slopes_xrays} and \ref{fig:mass_dep_slopes_radio}, respectively. In the upper two rows we vary the low Eddington ratio slope $\delta_1$. The bottom two rows show $\delta_2(\log M)$. $\log \lambda^{*}$ is always constant and set to $\log \lambda^{*}_{M\ \mathrm{indep}}$. Similar to Figure \ref{fig:mass_dep_break_xray} and \ref{fig:mass_dep_break_radio}, we show $\log M^{*}$ and $\delta_{1, M \mathrm{indep}}$/$\delta_{2, M\ \mathrm{indep}}$ as vertical and horizontal dashed lines in the left-hand columns, respectively. The mass independent ERDFs that we determined with the MCMC are shown as black solid lines in the central columns. In Figure \ref{fig:mass_dep_slopes_xrays}, the luminosity function by \citetalias{Ajello:2012aa} is shown with black solid lines in the right-hand column. In Figure \ref{fig:mass_dep_slopes_radio}, we show the luminosity function by \citetalias{Mauch:2007aa} in the same way.

In these figures, we additionally highlight the regions in which $\delta_1 > \delta_{2, M\ \mathrm{indep}}$ or $\delta_2 < \delta_{1, M\ \mathrm{indep}}$ in the left-hand column. We previously defined $\delta_1$ and $\delta_2$ to be the ERDFs low $\log \lambda$ and high $\log \lambda$ ratio slopes, respectively and postulated $\delta_2 > \delta_1$. In the grey shaded regions in the left-hand columns, this definition is no longer valid and by varying $\delta_1$($\delta_2$), we are changing the high (low) Eddington ratio slope.  

Figure \ref{fig:mass_dep_slopes_xrays} shows that a mass dependent slope $\delta_1$ can be excluded for the XLF. It leads to XLFs which are no longer broken power law shaped (first and second row). A $\delta_2(\log M)$ leads to agreement between the predicted and the observed XLF if $-1 \leq a \leq 1$ and $\delta_2(\log M^{*}) = \delta_{2, M\ \rm{indep}}$ (third and fourth row).

For the RLF we examine mass dependent ERDF slopes in Figure \ref{fig:mass_dep_slopes_radio}. The first two rows show that a mass dependent $\delta_1$ does not lead to agreement between the predictions and observations. For $\delta_2(\log M)$ only $-1 \leq a \leq 0$ with $\delta_2(\log M^{*}) = \delta_{2, M\ \rm{indep}}$ leads to a RLF that resembles the observations (fourth row). 

We have not quantified the goodness of fit for these mass dependent models and have not taken the errors on the observed luminosity functions into account. Furthermore, we fixed two of the three free parameters to the best-fitting values for the mass independent case which we determined in Section \ref{sec:mcmc_results}. Rather than fitting the constant parameters to achieve consistency with the observations for the mass dependent ERDF model, we thus simply added the mass dependence on top of our best-fitting ERDF results.  Nonetheless, our analysis shows that mild mass dependencies of the ERDF still result in luminosity functions that resemble the observations. For the XLF and the RLF, mass dependencies of the order of up to 1 magnitude in $ \lambda^{*}$ per magnitude in $M$ are consistent with the observations. Changing $\delta_1$ and $\delta_2$ as a function of stellar mass leads more extreme variations of the predicted luminosity function. For both, $\xi_{\rm X}(\lambda)$ and $\xi_{\rm R}(\lambda)$, the simple $\delta_1(\log M)$ models that we proposed here can be excluded. For $\xi_{\rm X}(\lambda)$ $\delta_2$ can be varied by up to $\pm 1$ per magnitude in stellar mass. For the ERDF of radiatively inefficient AGN a $\delta_2$ that decreases by up to 1 per magnitude in $M$ still leads to consistency with the observations. For all models which we considered here it is important that at $M^{*}$ the mass dependent ERDF parameter corresponds to its respective best-fitting mass independent value. $M^{*}$ galaxies, which within our mass range dominate the XLF and the RLF (see Section \ref{sec:completeness}), are thus still convolved with the best-fitting, mass independent ERDFs which we determined in Section \ref{sec:mcmc_results}. 

\section{Extended discussion}
\subsection{Alternative ERDF shapes}\label{sec:erdf_shapes_app}
\begin{figure*}
	\centering
	\begin{minipage}{\textwidth}
		\includegraphics[width=\textwidth]{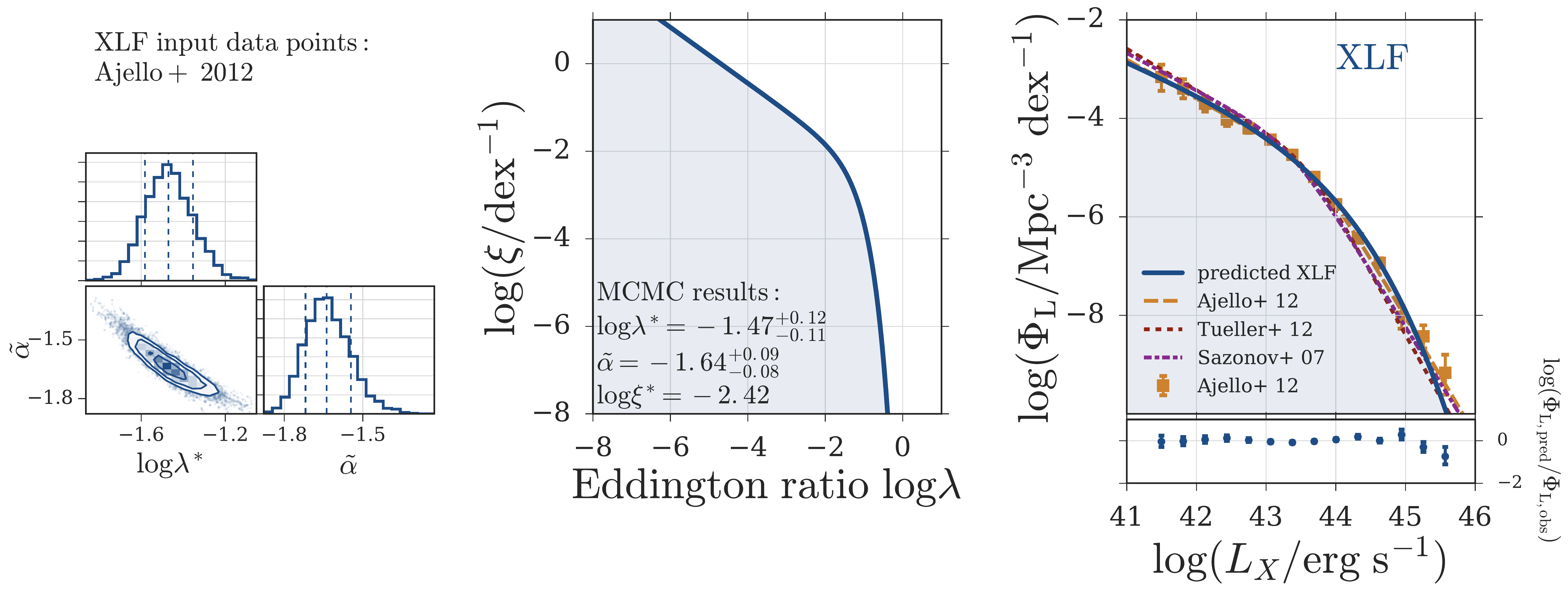}
	\end{minipage}
	\begin{minipage}{\textwidth}
		\includegraphics[width=\textwidth]{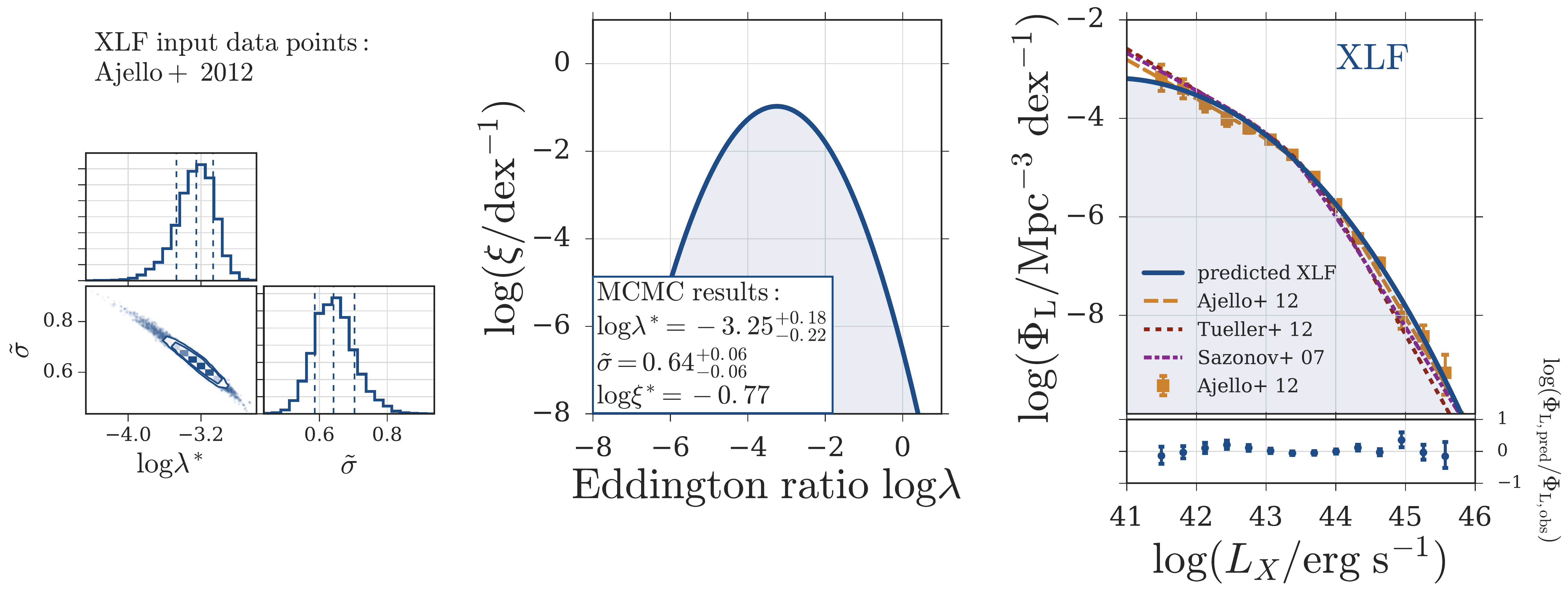}
	\end{minipage}
	\caption{\label{fig:schechter_normal_xray}MCMC results for radiatively efficient AGN assuming a Schechter function (top row) and a log normal shaped (bottom row) ERDF. In our model we assume a broken power law shaped ERDF (see Figure \ref{fig:mcmc_results}). We adjust the MCMC to test additional functional ERDF forms and illustrate the results in this figure. In the upper panel, we show the results for a Schechter function shaped ERDF for which we vary its break $\lambda^{*}$ and its low Eddington ratio slope $\tilde{\alpha}$. In case of the log normal shaped ERDF, which is illustrated in the lower row, the peak $\lambda^{*}$ and the width $\tilde{\sigma}$ are free parameters.}
\end{figure*}

\begin{figure*}
	\centering
	\begin{minipage}{\textwidth}
		\includegraphics[width=\textwidth]{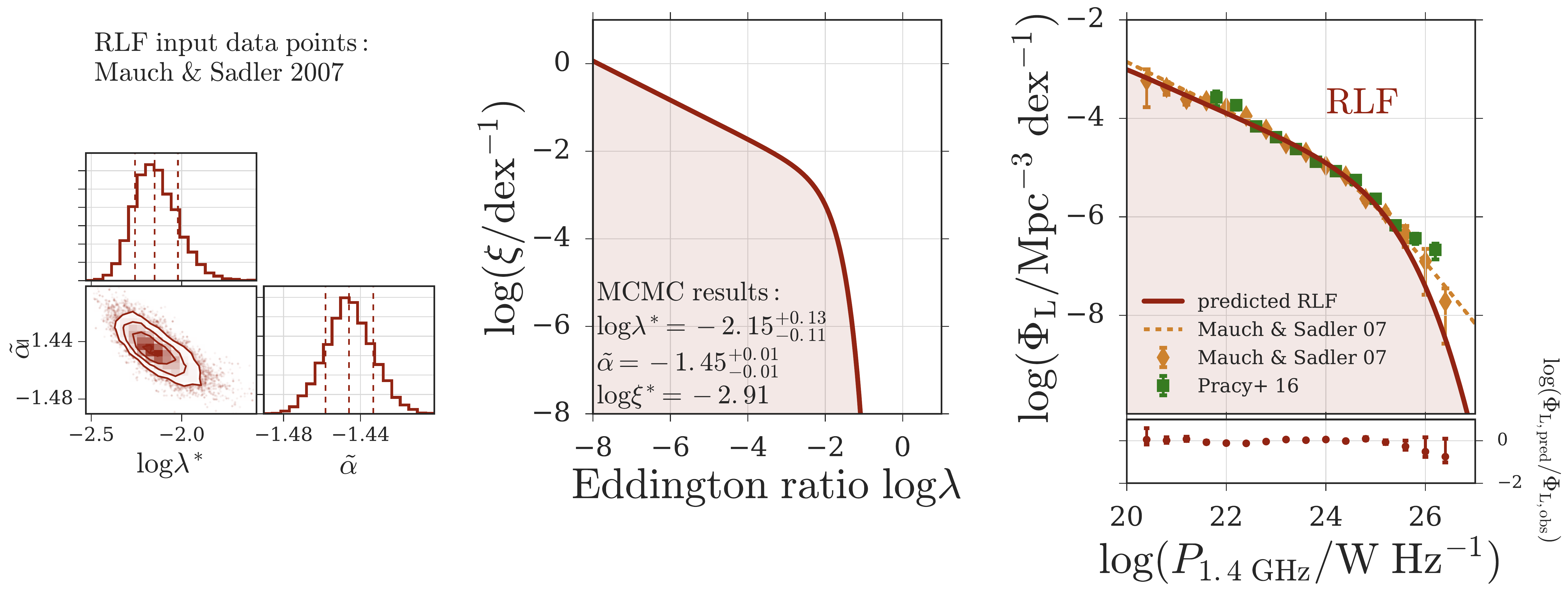}
	\end{minipage}
	\begin{minipage}{\textwidth}
		\includegraphics[width=\textwidth]{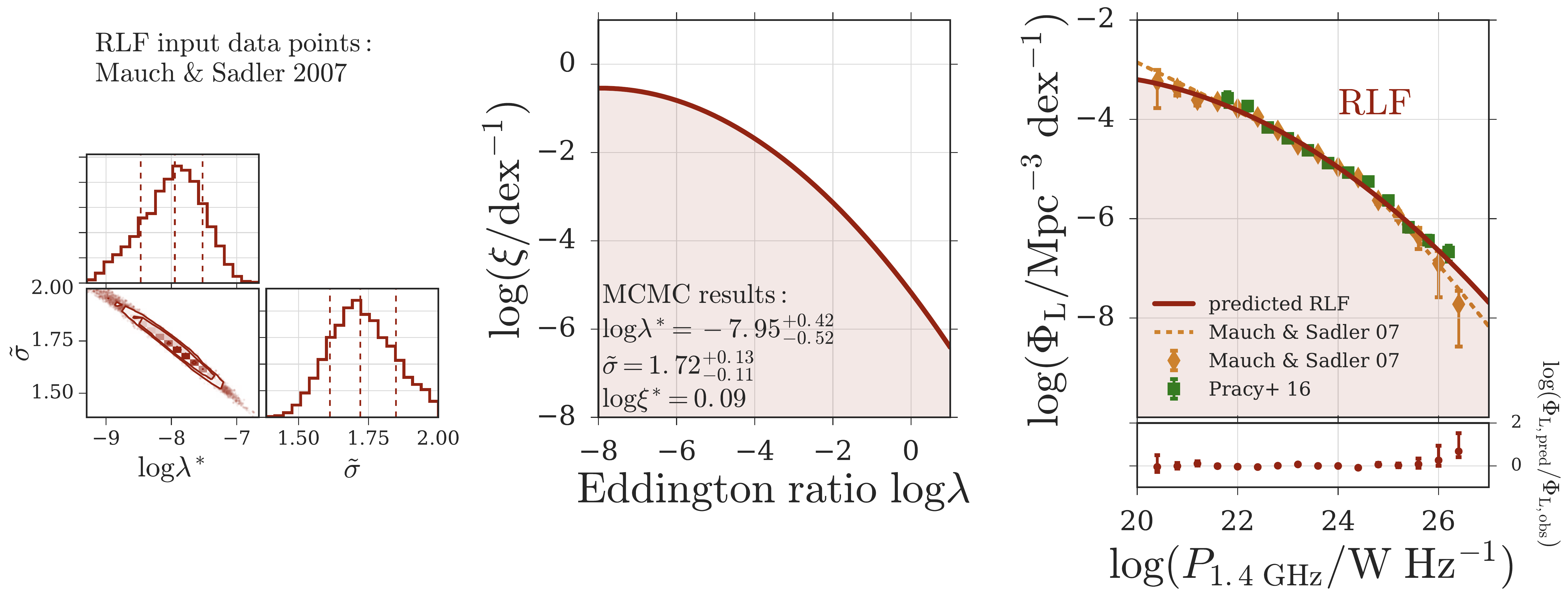}
	\end{minipage}
	\caption{\label{fig:schechter_normal_radio}In analogy to Figure \ref{fig:schechter_normal_xray}, MCMC results for radiatively inefficient AGN assuming a Schechter function (top row) and a log normal shaped (bottom row) ERDF.}
\end{figure*}

Our model is based on a broken power law shaped ERDF (see Section \ref{sec:method}). Yet different ERDF shapes, such as a single power law, a Schechter function or a log normal function, have been proposed. 

A single power law ERDF has, for instance, been used by \cite{Aird:2012aa}. \cite{Bongiorno:2012aa}, who also use a simple power law ERDF, find evidence for a break at high Eddington ratios and \cite{Aird:2013ab} stress the need for a break in the ERDF to match the observed X-ray luminosity function. In recent work, \cite{Bongiorno:2016aa} thus use a broken power law fit to the data. 
\cite{Veale:2014aa} test both, a truncated power law and a log normal shaped ERDF. Furthermore, a log normal ERDF is used by \cite{Kollmeier:2006aa} and \cite{Conroy:2013aa} (`scattered light bulb'). \cite{Kauffmann:2009aa} find a log normal distribution for galaxies with high Eddington ratios and high star formation rates (`feast mode') and a black hole mass and stellar age dependent power law ERDF for galaxies with old stellar populations (`famine mode').

\cite{Hopkins:2009aa}, \cite{Cao:2010aa}, \cite{Hickox:2014aa} and \cite{Trump:2015aa} use a Schechter function ERDF. \cite{Jones:2016aa} show that an intrinsic Schechter function ERDF can be reduced to the log normal ERDF observed by \cite{Kauffmann:2009aa} if the BPT \citep{Baldwin:1981aa} selection effects are taken into account. \cite{Nobuta:2012aa} and \cite{Schulze:2015aa}  consider both, a log normal and a Schechter function ERDF. 

Irrespective of the functional form of the ERDF, we do not distinguish between type-1 and type-2 AGN. We assume unification \citep{Urry:1995aa, Urry:2004aa} and thus expect both groups to have the same intrinsic ERDF. In contrast to our model, \cite{Tucci:2016aa}, for example, stress the difference between the type-1 (log-normal) and the type-2 AGN (power law) ERDF. 

To investigate the differences between a broken power law, a Schechter function and a log normal ERDF, we use our MCMC method (see Section \ref{sec:mcmc}) to fit these additional functional ERDF forms to the observed luminosity functions. For the Schechter function ERDF we vary the low Eddington ratio end slope $\tilde{\alpha}$ and the knee $\log \lambda^{*}$. The log normal ERDF also has two free parameters: the peak $\log \lambda^{*}$ and the width $\tilde{\sigma}$.

In analogy to equation \ref{eq:single_schechter}, we assume the following functional form for the Schechter function ERDF 
\begin{equation}
\begin{aligned}
\xi(\lambda) = &  \frac{dN}{d\log \lambda}\\
= & \ln(10) \xi^{*} \left(10^{\log \lambda - \log \lambda^{*}}\right)^{\tilde{\alpha} + 1} \exp\left(-10^{\log \lambda - \log \lambda^{*}}\right). 
\end{aligned}
\end{equation}
The log normal ERDF is parametrized as:
\begin{equation}
\begin{aligned}
\xi(\lambda) = & \frac{dN}{d \log \lambda}\\
= & \frac{\xi^{*}}{\tilde{\sigma} \sqrt{2\pi}} \times \exp\left(\frac{-(\log \lambda - \log \lambda^{*})^2}{2 \tilde{\sigma}^2}\right).
\end{aligned}
\end{equation}

 In analogy to Section \ref{sec:mcmc_results}, we fix the fitting range to $\fitLminX < \log (L_\mathrm{X}/\mathrm{erg\ s}^{-1}) < \fitLmaxX$ for radiatively efficient AGN and to $\fitLminR < \log(P_\mathrm{1.4\ GHz}/\mathrm{W\ Hz}^{-1}) < \fitLmaxR$ for radiatively inefficient AGN. 

Our priors are given in Table \ref{tab:mcmc_para} and our results are shown in Figure \ref{fig:schechter_normal_xray} and \ref{fig:schechter_normal_radio}. We observe the following: 

\begin{itemize}
	\item a Schechter function ERDF for the XLF: As $\gamma_1 \gtrsim -(\alpha_1 + 1)$, where $\alpha_1$ is the low mass slope of the blue+green stellar mass function, the low $\lambda$ end of the ERDF and not $\alpha_1$ determines $\gamma_1$ (see equation \ref{eq:gamma1}). This is also true for the Schechter function shaped ERDF which we are considering in Figure \ref{fig:schechter_normal_xray}: $\tilde{\alpha}$ is well constrained and steeper than $\alpha_1$. $\log \lambda^{*}$ lies close to our initial guess (see equation \ref{eq:pred_break_X}), but is higher than in the broken power law case (see Section \ref{sec:mcmc_results}). Using a Schechter function shaped ERDF leads to a XLF that is consistent with the majority of the observed XLF. Due to the exponential cut off, it does however fail to reproduce $\Phi_{\rm L}$ in the brightest luminosity bin.
	\item a log normal ERDF for the XLF: When using a log normal shaped ERDF, the MCMC fits the entire observed XLF with the declining part of the ERDF. $\lambda^{*}$ is therefore over one dex lower than in the case of a broken power law or Schechter function shaped ERDF. Due to its broadness, $\xi_{\rm X}(\lambda)$ succeeds in reproducing the XLF within the luminosity range that we are considering. 
	\item a Schechter function ERDF for the RLF: As $\gamma_1 >> -(\alpha_2 + 1)$, where $\alpha_2$ is the high mass end of the red stellar mass function, $\tilde{\alpha}$ for the RLF is well constrained and has smaller uncertainties than in the case of the XLF. As in the XLF case, $\log \lambda^{*}$ for the Schechter function shaped ERDF is higher than the best-fitting value for a broken power law shaped ERDF. The predicted RLF is consistent with the measured $\Phi_{\rm L}(L)$ values. However at the bright end it does deviate significantly  from the best-fitting broken power law RLF by \citetalias{Mauch:2007aa}. 
	\item a log normal ERDF for the RLF: In analogy to the XLF, reproducing the RLF with a log normal shaped ERDF leads to low $\log \lambda^{*}$ values. In the case of the RLF, the best-fitting $\lambda^{*}$ for a log normal $\xi_{\rm R}(\lambda)$ lies over 5 orders of magnitude below the value for a broken power law or Schechter function shaped ERDF. 
	With the broad log normal shaped ERDF we are able to reproduce most of the observed $\Phi_{\rm L}(L)$ values. However the distribution is too shallow to reproduce $\Phi_{\rm L}(L)$ in the brightest luminosity bin.
\end{itemize}

We conclude that in comparison to a broken power law shaped ERDF, reproducing the observed XLF and RLF with a log normal or a Schechter function ERDF is more challenging. This is not unexpected since both functional forms only have two free parameters, whereas a broken power law shaped ERDF allows us to also vary both $\delta_1$ and $\delta_2$. Using a log normal shaped ERDF results in a broad distribution with a low $\log \lambda^{*}$ value which reproduces the shape of a power law or Schechter function shaped ERDF. In the case of the XLF, a Schechter function shaped $\xi_{\rm X}(\lambda)$ fails, a log normal shaped $\xi_{\rm X}(\lambda)$ succeeds in reproducing the observed $\Phi_{\rm L}(L)$ values. Vice versa, Schechter and log normal shaped ERDFs succeed and fail to predict luminosity functions that are consistent with the RLF observations, respectively.
	
To be able to determine if a broken power law, Schechter function or log normal shaped ERDF leads to better agreement between our predictions and the observations, we need luminosity functions that cover a wider range in $\log L$. The faint and the bright end are where Schechter function and log normal shaped ERDFs show their weakness. This can already be seen in the analysis that we presented here, but is likely to become more evident at higher and lower luminosities. If the observed luminosity functions do indeed have a broken power law shape, a Schechter function shaped ERDF is likely to be too steep to produce the bright end. A log normal shaped ERDF is likely to be too shallow to predict a luminosity function that is consistent with the observations at the faint end.

\end{document}